# A Unified Approach to Sparse Signal Processing

F. Marvasti, *Senior Member, IEEE,* A. Amini, F. Haddadi, M. Soltanolkotabi, *Student Member, IEEE,*
B. H. Khalaj, *Member, IEEE,* A. Aldroubi, S. Holm, *Senior Member, IEEE,* S. Sanei, *Senior Member, IEEE,*
J. Chambers, *Senior Member, IEEE, and other invited contributors*

*Abstract*— A unified view of the area of sparse signal processing is presented in tutorial form by bringing together various fields in which the property of sparsity has been successfully exploited. For each of these fields, various algorithms and techniques, which have been developed to leverage sparsity, are described succinctly. The common potential benefits of significant reduction in sampling rate and processing manipulations through sparse signal processing are revealed.

The key application domains of sparse signal processing are sampling, coding, spectral estimation, array processing, component analysis, and multipath channel estimation. In terms of the sampling process and reconstruction algorithms, linkages are made with random sampling, compressed sensing and rate of innovation. The redundancy introduced by channel coding in finite and real Galois fields is then related to over-sampling with similar reconstruction algorithms. The methods of Prony, Pisarenko, and MUltiple SIgnal Classification (MUSIC) are next shown to be targeted at analyzing signals with sparse frequency domain representations. Specifically, the relations of the approach of Prony to an annihilating filter in rate of innovation and Error Locator Polynomials in coding are emphasized; the Pisarenko and MUSIC methods are further improvements of the Prony method. Such narrowband spectral estimation is then related to multi-source location and direction of arrival estimation in array processing. The notions of sparse array beamforming and sparse sensor networks are also introduced. Sparsity in unobservable source signals is also shown to facilitate source separation in Sparse Component Analysis (SCA); the algorithms developed in this area are also widely used in compressed sensing. Finally, the nature of the multipath channel estimation problem is shown to have a sparse formulation; algorithms similar to sampling and coding are used to estimate typical multicarrier communication channels.

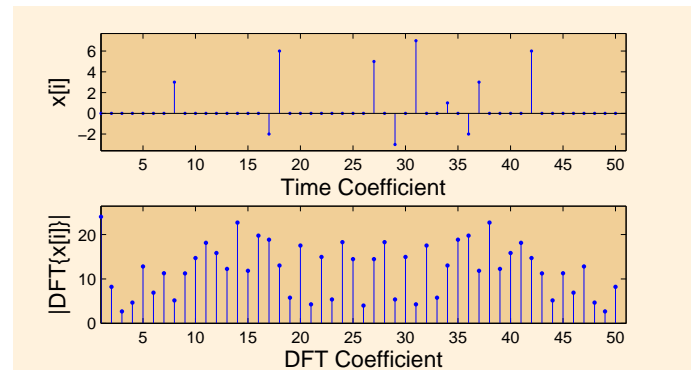

Fig. 1. Sparse discrete time signal with its Discrete Fourier Transform (DFT).

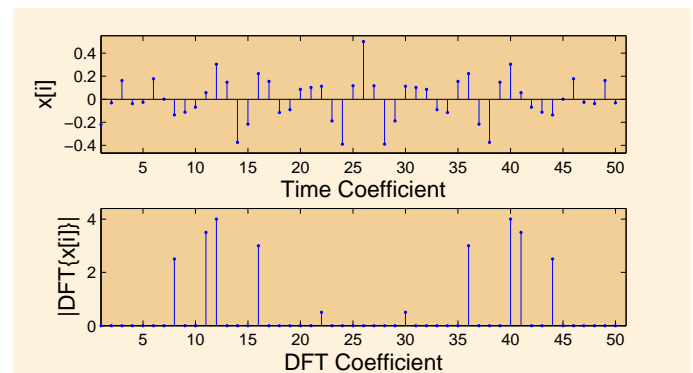

Fig. 2. Sparsity is manifested in the frequency domain.

## I. INTRODUCTION

THERE are many applications in signal processing and communication systems where the discrete signals are sparse in some domain such as time, frequency, or space i.e., most of the samples are zero, or alternatively their transform in another domain (normally called "*frequency coefficients*")

Manuscript received February 05, 2009; revised ??? ??, 2009.

F. Marvasti, A. Amini, F. Haddadi, B. Khalaj and M. Soltanolkotabi are affiliated with Advanced Communication Research Institute (ACRI), Electrical Engineering Department, Sharif University of Technology {marvasti, khalaj}@sharif.edu, {arashsil, farzanhaddadi, msoltan}@ee.sharif.edu

A. Aldroubi is affiliated with Math Department, Vanderbilt University, akram.aldroubi@vanderbilt.edu. The work of Akram Aldroubi was supported in part by grant nsf-dms 0807464.

S. Sanei is affiliated with Centre of Digital Signal Processing, School of Engineering, Cardiff University, Cardiff, UK, SaneiS@cf.ac.uk

J. Chambers is affiliated with Electrical and Electronic Department, Loughborough University, j.a.chambers@lboro.ac.uk

S. Holm is affiliated with the Department of Informatics, University of Oslo, sverre@ifi.uio.no

The invited contributors to specific sections are listed in the acknowledgments.

is sparse (see Figs. 1 and 2). There are trivial sparse transformations where the sparsity is preserved in both "time" and "frequency" domains; the identity transform matrix and its sorted versions are extreme examples. Wavelet transformations that preserve the local characteristics of a sparse signal can be regarded as "almost" sparse in the "frequency" domain; in general, for sparse signals, the more similar the transformation matrix is to an identity matrix, the sparser the signal is in the transform domain. In addition, the transform matrix may be sparse; wavelet transformation matrices are such examples.

In any of these scenarios, sampling and processing can be optimized using sparse signal processing. In other words, the sampling rate and the processing manipulations can be significantly reduced; hence, a combination of data compression and processing time reduction can be achieved[1].

Each field has developed its own tools, algorithms, and reconstruction methods for sparse signal processing. Very few

TABLE I

COMMON NOTATIONS USED THROUGHOUT THE PAPER.

| | |
|---|---|
| $n$ | Length of original vector |
| $k$ | Order of sparsity |
| $m$ | Length of observed vector |
| $\mathbf{x}$ | Original vector |
| $\mathbf{s}$ | Corresponding sparse vector |
| $\mathbf{y}$ | Observed vector |
| $\boldsymbol{\nu}$ | Noise vector |
| $\mathbf{A}$ | Transformation matrix relating $\mathbf{s}$ to $\mathbf{y}$ |
| $\|\mathbf{u}_{n\times 1}\|_{\ell_p}$ | $\left(\sum_{i=1}^{n}|u_i|^p\right)^{\left(\frac{1}{p}\right)}$ |

authors have noticed the similarities of these fields[2]. It is the intention of this tutorial to describe these methods in each field succinctly and show that these methods can be used in other areas and applications often with appreciable improvements. Among these fields are 1-*Sampling*: random sampling of bandlimited signals [2], Compressed Sensing (CS) [3], and sampling with finite rate of innovation [4]; 2- *Coding*: Galois [5], [6] and real-field error correction codes [7]; 3- *Spectral Estimation* [8], [9], [10], [11]; 4- *Array Processing*: Multi-Source Location (MSL) and Direction Of Arrival (DOA) estimation [12], [13], sparse array processing [14], and sensor networks [15]; 5- *Sparse Component Analysis* (SCA): blind source separation [16], [17], [18] and dictionary representation [19], [20], [21]; 6- *Channel Estimation* in Orthogonal Frequency Division Multiplexing (OFDM) [22], [23], [24]. The sparsity properties of these fields are summarized in Table II[3]. The details of each application will be discussed in the next sections but the common traits will be discussed in this introduction.

The columns of Table II consist of 0- category, 1- topics, 2- sparsity domain, 3- type of sparsity, 4- information domain, 5- type of sampling in information domain, 6- minimum sampling rate, 7- reconstruction method, and 8- applications. The first rows (2-7) of column 1 are on sampling techniques. The 8-9th rows are related to channel coding, row 10 is on spectral estimation and rows 11-13 are related to array processing. Rows 14-15 correspond to SCA and finally, row 16 covers multicarrier channel estimation, which is a rather new topic. As shown in column 2 of the table, depending on the topics, sparsity is defined in time, space, or "frequency" domains. In some applications, the sparsity is defined as the number of polynomial coefficients (which in a way could be regarded as "frequency"), the number of sources (which may depend on location or time sparsity for the signal sources), or the number of "words" (signal bases) in a dictionary. The type of sparsity is shown in column 3; for sampling schemes, it is usually low-pass, band-pass, or multiband [25], while for compressed sensing, and most other applications, it is random. Column 4 represents the information domain, where the number of sparsity, locations, and amplitudes can be determined by proper sampling (column 5) of this domain.

The other columns are self explanatory and will be discussed in more details in the following sections.

The rows 2-4 of Table II are related to the sampling (uniform or random) of signals that are bandlimited in the Fourier domain. Band-limitedness is a special case of sparsity where the nonzero coefficients in the frequency domain are consecutive. A better assumption in the frequency domain is to have random sparsity [26], [27], [28] as shown in row 5 and column 3. A generalization of the sparsity in the frequency domain is sparsity in any transform domain such as Discrete Cosine Transform (DCT) and wavelets; this concept is further generalized in compressed sensing (row 6) where sampling is taken by a linear combination of time domain samples [3], [29], [30], [31]. Sampling of signals with finite rate of innovation (row 7) is related to piecewise smooth (polynomial based) signals. The position of discontinuous points is determined by annihilating filters that are equivalent to error locator polynomials in error correction codes and Prony's method [28] as discussed in Sections III and IV, respectively.

Random errors in a Galois field (row 8) and the additive impulsive noise in real-field error correction codes (row 9) are sparse disturbances that need to be detected and removed. For erasure channels, the impulsive noise can be regarded as the negative of the sample value [32]; thus the missing sampling problem, which can also be regarded as a special case of nonuniform sampling, is also a special case of the error correction problem. A subclass of impulsive noise for 2-D signals is salt and pepper noise [33]. The information domain, where the sampling occurs, is called the syndrome which is usually in a transform domain. In addition, for special binary codes such as Low Density Parity Check (LDPC) codes, the parity check matrix is extremely sparse. Sparse matrix inversions and manipulations [34], [35], [36] are utilized in Discrete Wavelet Transform (DWT) and LDPC decoding.

Spectral estimation (row 10) is the dual of error correction codes, i.e., the sparsity is in the frequency domain. MSL (row 11) and multi-target detection in radars are similar to spectral estimation since targets act as spatial sparse mono-tones; each target is mapped to a specific spatial frequency regarding its line of sight direction relative to the receiver. The techniques developed for this branch of science is quite unique; with examples such as MUSIC [8], Prony [9], and Pisarenko [10]. We shall see that the techniques used in real-field error correction codes and SCA can also be used in this area.

The array processing category (rows 11-13) consists of three separate topics. The first one covers MSL in radars and sonars and DOA, which are similar to spectral estimation. The techniques developed for this field are similar to the spectral estimation methods with emphasis on the Minimum Description Length (MDL) [37]. The second topic in the array processing category is related to the design of sparse arrays where some of the array elements are missing; the remaining nodes form a nonuniform sparse grid. In this case, one of the optimization problems is to find the sparsest array (number, location and weight of elements) for a given beampattern. This problem has some resemblance to the missing sampling

TABLE II

Various topics and applications with sparsity properties: the sparsity, which may be in time/space or "frequency" domains, consists of unknown samples/coefficients that need to be determined. The information domain consists of known samples/coefficients in "frequency" or time/space domain (the complement of the sparse domain). A list of acronyms is given in Table XV at the end of the paper; also, a list of common notations is presented in Table I. For definition of ESPRIT on row 11 and column 7, see the footnote on page 18.

| | 0 | 1 | 2 | 3 | 4 | 5 | 6 | 7 | 8 |
|---|---|---|---|---|---|---|---|---|---|
| 1 | Category | Topics | Sparsity Domain | Type of Sparsity | Information Domain | Type of Sampling in Info. Domain | Min Number of Required Samples | Reconstruction Method | Applications |
| 2 | | Uniform sampling | Frequency | Lowpass | Time/Space | Uniform | $2 \times BW - 1$ | Lowpass filtering / Interpolation | A/D |
| 3 | | Nonuniform sampling | Frequency | Lowpass | Time/Space | Missing samples/Jitter/Periodic/Random | $2 \times BW - 1$ (in some cases even BW) | Iterative Methods/Filter banks/ Spline Interp. | Seismic / MRI / CT/ FM / PPM |
| 4 | Sampling | Sampling of multiband signals | Frequency | Union of disjoint intervals | Time/Space | Uniform/Jitter/Periodic/Random | $2 \times \sum BW$ | Iterative methods/Filter banks/ Interpolation | Data Compression/ Radar |
| 5 | | Random sampling | Frequency | Random | Time/Space | Random/Uniform | $2 \times \sum$ #Coeff. | Iterative methods: Adapt. Thresh. RDE / ELP | Missing Samp. Recovery/ Data Comp. |
| 6 | | Compressed sensing | An arbitrary orthonormal transform | Random | Random mapping of Time/Space | Random mixtures of samples | $c \cdot k \cdot \log(\frac{n}{k})$ | Basis pursuit/ Matching pursuit | Data compression |
| 7 | | Finite rate of innovation | Time and polynomial Coeff. | Random and uniform | Filtered time domain | Uniform | # Coeff. + 1 + 2 · (# Discont. Epochs) | Annihilating filter (ELP) | ECG/ OCT/ UWB |
| 8 | Channel coding | Galois field codes | Time | Random | Syndrome | Uniform or random | $2 \times$ # errors | Berlekamp-Massey/Viterbi/ Belief Prop. | Digital communication |
| 9 | | Real field codes | Time | Random | Transform domain | Uniform or random | $2 \times$ # Impulsive noise | Adaptive thresholding RDE / ELP | Fault tolerant system |
| 10 | Spectral estimation | Spectral estimation | Frequency | Random | Time / Autocorrelation | Uniform | $2 \times$ # Tones $-1$ | MUSIC/ Pisarenko/ Prony / MDL | Military/ Radars |
| 11 | Array processing | MSL/ DOA estimation | Space | Random | Space/ Autocorrelation | Uniform | $2\times$ # Sources | MDL+ MUSIC / ESPRIT | Radars/ Sonar/ Ultrasound |
| 12 | | Sparse array beamforming | Space | Random/ Missing elements | Space | Peaks of sidelobes/ [Non]Uniform | $2 \times$ # Desired array elements | Optimization: LP/ SA / GA | Radars/sonar/ Ultrasound/ MSL |
| 13 | | Sensor networks | Space | Random | Space | Uniform | $2\times$ BW of random field | Similar to row 5 | Seismic/ Meteorology/ Environmental |
| 14 | SCA | BSS | Active source/Time | Random | Time | Uniform | $2 \times$ # Active sources | $\ell_1 / \ell_2 /$ SL0 | Biomedical |
| 15 | | SDR | Dictionary | Random | Linear mixture of time samples | Uniform / Random | $2\times$ # Sparse Words | $\ell_1 / \ell_2 /$ SL0 | Data compression |
| 16 | Channel estimation | Multipath channels | Time | Random | Frequency or time | Uniform / Nonuniform | $2 \times$ # Sparse channel components | $\ell_1 /$ MIMAT | Channel equalization/ OFDM |

problem (which is a special case of real-field error correction codes), and may be solved by similar techniques. The third topic is on sensor networks (row 13). Distributed sampling and recovery of a physical field using an array of sparse sensors is a problem of increasing interest in environmental and seismic monitoring applications of sensor networks [38]. Sensor fields may be bandlimited or non-bandlimited. Since the power consumption is the most restricting issue in sensors, it is vital to use the lowest possible number of sensors (sparse sensor networks) with the minimum processing computation.

In Sparse Component Analysis (SCA), the number of observations is much less than the number of sources (signals). However, if the sources are sparse in the time domain, then the active sources and their amplitudes can be determined; this is equivalent to error correction codes. Sparse Dictionary Representation (SDR) is another new area where signals are represented by the sparsest number of words (signal bases) in a dictionary of finite number of words; this sparsity may result in tremendous amount of data compression. When the dictionary is overcomplete, there are many ways to represent the signal; however, we are interested in the sparsest representation. Normally, for extraction of statistically independent sources, Independent Component Analysis (ICA[4]) is used for a complete set of linear mixtures. In the case of a non-complete (underdetermined) set of linear mixtures, ICA can work if the sources are also sparse; for this special case, ICA analysis is synonymous with SCA.

Finally, channel estimation is shown in row 16. In mobile communication systems, multipath reflections create a channel that can be modeled by a sparse FIR filter. For proper decoding of the incoming data, the channel characteristics should be estimated before it can be equalized. For this purpose, a training sequence is inserted within the main data, which enables the receiver to obtain the output of the channel by exploiting this training sequence. The channel estimation problem becomes a deconvolution problem under noisy environments. The sparsity criterion of the channel greatly improves the channel estimation; this is where the algorithms for extraction of a sparse signal could be employed [22], [23], [39].

When sparsity is random, further signal processing is needed. In this case there are three items that need to be considered. 1- Evaluating the number of sparse coefficients (or samples), 2- finding the position of sparse coefficients, and 3- determining the values of these coefficients. In some applications only the first two items are needed; e.g., in spectral estimation. However, in almost all the other cases mentioned in Table II, all the three items should be determined. Various types of Linear Programming (LP) and some iterative algorithms, such as the Iterative Method with Adaptive Thresholding (IMAT), determine the number, positions and values of sparse samples at the same time. On the other hand, the Minimum Description Length (MDL) method, used in DOA/MSL spectral estimation, determines the number of sparse source locations or frequencies. In the subsequent sections, we shall describe, in more details, each algorithm for various areas and applications based on Table II.

Finally, it should be mentioned that the signal model for each topic or application may be deterministic or stochastic. For example, in the sampling category for rows 2-4 and 7, the signal model is typically deterministic although stochastic models could also be envisioned [40]. On the other hand for random sampling and CS (rows 5-6), the signal model is stochastic although deterministic models may also be envisioned [41]. In channel coding and estimation (rows 8-9 and 16), the signal model is normally deterministic. For Spectral and DOA estimation (rows 10-11), stochastic models are assumed; while for array beam-forming (row 12), deterministic models are used. In sensor networks (row 13), both deterministic and stochastic signal models are employed. Finally, in SCA (rows 14-15), statistical independence of sources may be necessary and thus stochastic models are applied.

## II. SAMPLING: UNIFORM, NONUNIFORM, MISSING, RANDOM, COMPRESSED SENSING, RATE OF INNOVATION

Analog signals can be represented by finite rate discrete samples (uniform, nonuniform, or random) if the signal has some sort of redundancies such as band-limitedness, finite polynomial representation (e.g., periodic signals that are represented by a finite number of trigonometric polynomials), and nonlinear functions of such redundant functions [42], [43]. The minimum sampling rate is the Nyquist rate for uniform sampling and its generalizations for nonuniform [2] and multiband signals [44]. When a signal is discrete, the equivalent discrete representation in the "frequency" domain (DFT, DCT, DWT, Discrete Hartley Transform (DHT), Dis-crete Sine Transform (DST)) may be sparse, which is the discrete version of bandlimited or multiband analog signals.

For discrete signals, if the nonzero coefficients ("frequency" sparsity) are consecutive, depending on the location of the zeros, they are called lowpass, bandpass, or multiband discrete signals; otherwise, the "frequency" sparsity is random. The number of discrete time samples needed to represent a frequency-sparse signal follows the law of algebra, that is, the number of time samples should be equal to the number of coefficients in the "frequency" domain; this is equivalent to the Nyquist rate- twice the bandwidth (for discrete signals with DC components it is twice the bandwidth minus one). The dual of frequency-sparsity is time-sparsity, which can happen in a burst or a random fashion. The number of "frequency" coefficients needed follows the Nyquist criterion. This will be further discussed in Section III for sparse additive impulsive noise channels.

### A. Sampling of Sparse Signals

If the sparsity locations of a signal are known in a transform domain, then the number of samples needed in the time (space) domain should be at least equal to the number of sparse coefficients, i.e., the so called Nyquist rate. However, depending on the type of sparsity (lowpass, bandpass, or random) and the type of sampling (uniform, periodic nonuniform, or random), the reconstruction may be unstable and the corresponding reconstruction matrix may be ill-conditioned [45], [46]. Thus in many applications mentioned in Table II, the sampling rate in column 6 is higher than the minimum (Nyquist) rate.

When the location of sparsity is not known, by the law of algebra, the number of samples needed to specify the sparsity is at least twice the number of sparse coefficients. Again for stability reasons, the actual sampling rate is higher than this minimum figure [2], [44]. To guarantee stability, instead of direct sampling of the signal, a combination of the samples can be used. Donoho [30] has recently shown that if we take linear combinations of the samples, the minimum stable sampling rate is of the order $O(k \log(\frac{n}{k}))$, where $n$ and $k$ are the frame size and the sparsity number, respectively.

*1) Reconstruction Algorithms:* There are many reconstruction algorithms that can be used depending on the sparsity pattern, uniform or random sampling, complexity issues, and sensitivity to quantization and additive noise [47], [48]. Among these methods are: Linear Programming (LP), Lagrange interpolation [49], time varying method [50], spline interpolation [51], matrix inversion [52], Error Locator Polynomial (ELP) [53], iterative techniques [1], [46], [54], [55], [56], [57], [58], and Iterative Methods with Adaptive Thresholding (IMAT) [26], [32], [59], [60]. In the following we will only concentrate on the last three methods that have proven to be effective and practical.

*Iterative Methods When the Location of Sparsity is Known:* The reconstruction algorithms have to recover the original sparse signal from the information domain and the type of sparsity in the transform domain. We know the samples (both position and amplitude) and we know the location of sparsity in the transform domain. An iteration between these two

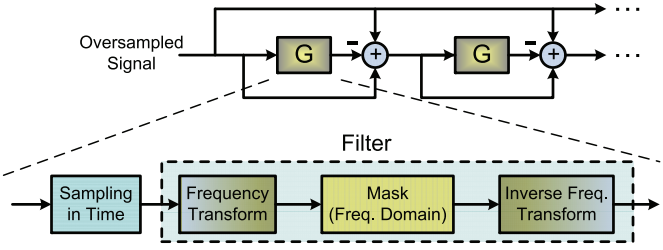

Fig. 3. Block diagram of the iterative reconstruction method. The Mask is an appropriate filter with coefficients of 1's and 0's depending on the type of sparsity in the original signal.

1) Convert the input to the $i^{th}$ iteration ($\mathbf{x}^{(i)}$) into the transform domain (for instance the Fourier domain); $\mathbf{x}^{(0)}$ is normally the initial received signal.
2) Multiply the transformed signal ($X^{(i)}$) by a mask (for instance a band-limiting filter).
3) Take the inverse transform of the result in step 2 to get $\mathbf{r}^{(i)}$.
4) Set the new result as: $\mathbf{x}^{(i+1)} = \mathbf{x}^{(0)} + \mathbf{x}^{(i)} - \mathbf{r}^{(i)}$.
5) Repeat for a given number of iterations.
6) Stop when $\|\mathbf{x}^{(i+1)} - \mathbf{x}^{(i)}\|_{\ell_2} < \epsilon$.

domains (Fig. 3 and Table III) or consecutive Projections Onto Convex Sets (POCS) should yield the original signal [45], [54], [55], [58], [61], [62], [63], [64].

In case of the usual assumption that the sparsity is in the "frequency" domain and for the uniform sampling case of lowpass signals, one projection (bandlimiting in the frequency domain) suffices. However, if the frequency sparsity is random, the time samples are nonuniform, or the "frequency" domain is defined in a domain other than the DFT, then we need several iterations to have a good replica of the original signal. In general, this iterative method converges if the "Nyquist" rate is satisfied, i.e., the number of samples per block is greater than or equal to the number of coefficients. Figure 4 shows the improvement in dB versus the number of iterations for a random sampling set for a bandpass signal. In this figure, besides the standard iterative method, accelerated iterations such as Chebyshev and Conjugate Gradient methods are also used (please see Appendix I for the algorithms) [65].

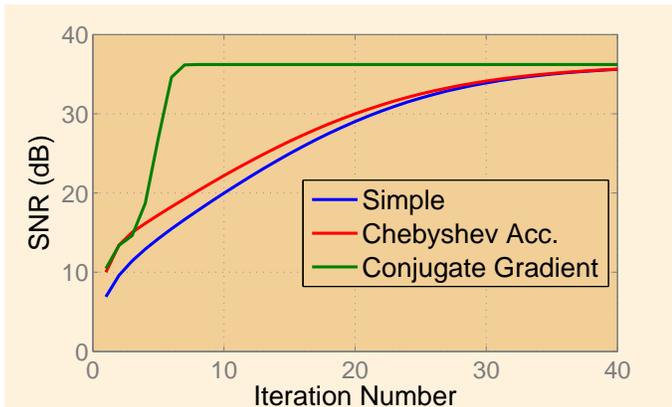

Fig. 4. SNR improvement vs. the no. of iterations for a random sampling set at the Nyquist rate (OSR=1) for a bandpass signal.

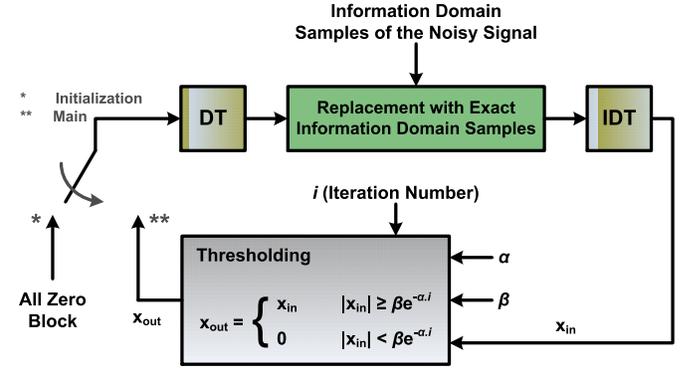

Fig. 5. The Iterative Method with Adaptive Thresholding (IMAT) for detecting the number, location, and values of sparsity.

Iterative methods are quite robust against quantization and additive noise. In fact, we can prove that the iterative methods approach the pseudo-inverse (least squares) solution for a noisy environment; specially, when the matrix is ill-conditioned [44].

*Iterative Method with Adaptive Threshold (IMAT) for Unknown Location of Sparsity:* As expected, when sparsity is assumed to be random, further signal processing is needed. We need to evaluate the number of sparse coefficients (or samples), the position of sparsity, and the values of the coefficients. The above iterative method cannot work since projection (the masking operation in Fig. 3) onto the "frequency" domain is not possible without the knowledge of the positions of sparse coefficients. In this scenario, we need to use the knowledge of sparsity in some way. The introduction of an adaptive nonlinear threshold in the iterative method can do the trick, thus the name: Iterative Method with Adaptive Threshold (IMAT); the block diagram and the algorithm are depicted in Fig. 5 and Table IV, respectively. The algorithms in [66], [32], [26], [24] are variations of this method. Figure 5 shows that by alternate projections between information and sparsity domains (adaptively lowering or raising the threshold levels in the sparsity domain), the sparse coefficients are gradually picked up after several iterations. This method can be considered as a modified version of Matching Pursuit as described in Section VI-D.1; the results are shown in Fig. 6. The sampling rate in the time domain is twice the number of unknown sparse coefficients. This is called the full capacity rate; this figure shows that after about 15 iterations, the SNR reaches its peak value. In general, the higher the sampling rate relative to the full capacity, the faster is the convergence rate and the better is the SNR value.

*Matrix Solutions:* When the sparse nonzero locations are known, matrix approaches can be utilized to determine the values of sparse coefficients [52]. Although these approaches are rather straight forward, they may not be robust against quantization or additive noise.

There are other approaches such as Spline interpolation [51], nonlinear/time varying methods [52], Lagrange interpolation

TABLE IV

GENERIC IMAT OF FIG. 5 FOR ANY SPARSITY IN THE DISCRETE
TRANSFORM (DT), WHICH IS TYPICALLY THE FAST FOURIER
TRANSFORM (FFT).

1) Use the all-zero block as the initial value of the sparse domain signal ($0^{th}$ iteration)
2) Convert the current estimate of the signal in the sparse domain into the information domain (for instance the time domain into the Fourier domain)
3) Where possible, replace the values with the known samples of the signal in the information domain.
4) Convert the signal back to the sparse domain.
5) Use adaptive hard thresholding to distinguish the original nonzero samples.
6) If neither the maximum number of iterations has past nor a given stopping condition is fulfilled, return to the 2nd step.

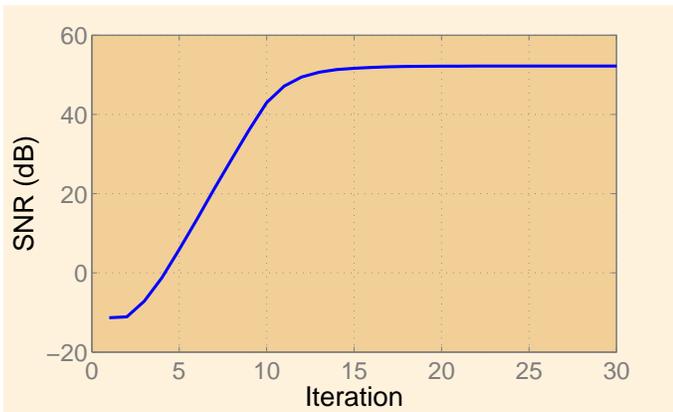

Fig. 6. SNR vs. the no. of iterations for sparse signal recovery using the IMAT (Table IV).

[49] and Error Locator Polynomial (ELP) [67] that will not be discussed here. These methods work quite well in the absence of additive noise but they may not be robust in the presence of noise. However the ELP approach will be discussed in Sec. III-A; variations of this method are called the annihilating filter in sampling with finite rate of innovation (Sec. II-C) and the Prony's method in spectral and DOA estimation (Sec. IV-A).

### B. Compressed Sensing (CS)

The relatively new topic of Compressed (Compressive) Sensing (CS) which was originally introduced in [30] and further extended in [68], [69] and [31] deals with sampling of sparse signals, in general. The idea is to introduce sampling schemes with low number of required samples which uniquely represent the original sparse signal; these methods have lower computational complexities than the traditional techniques that employ oversampling and then apply compression. In other words, compression is achieved exactly at the time of sampling. Unlike the classical sampling theorem [70], the signals are assumed to be sparse in an arbitrary transform domain, not necessarily the Fourier transform. Furthermore, there is no restricting assumption for the location of nonzero coefficients in the sparsity domain; i.e., the locations should not follow a specific pattern such as lowpass or multiband structure. Clearly, this assumption includes a more general class of signals than the ones previously studied.

Since the concept of sparsity in a transform domain is easier to study for discrete signals, most of the research in this field is focused along discrete type signals [71]; however, recent results [72] show that most of the work can be generalized to continuous signals that have a sparse representation in a Riesz basis[5] [73]. We first study discrete signals and then briefly discuss the extension to the continuous case.

*1) CS Mathematical Modeling:* Let the vector $\mathbf{x} \in \mathbb{R}^n$ be a finite length discrete signal in the time domain which has to be under-sampled. We assume that $\mathbf{x}$ has a sparse representation in a transform domain denoted by a unitary matrix $\mathbf{\Psi}_{n \times n}$; i.e., we have:

$$\mathbf{x} = \mathbf{\Psi} \cdot \mathbf{s} \tag{2}$$

where $\mathbf{s}$ is an $n \times 1$ vector which has at most $k$ non-zero elements ($k$-sparse vectors). In practical cases, $\mathbf{s}$ has at most $k$ significant elements and the insignificant elements are set to zero which means $\mathbf{s}$ is an almost $k$-sparse vector. For example, $\mathbf{x}$ can be the pixels of an image and $\mathbf{\Psi}$ can be the corresponding IDCT matrix. In this case, most of the DCT coefficients are insignificant and if they are set to zero, the quality of the image will not degrade significantly. In fact, this is the main concept behind some of the lossy compression methods such as JPEG2000. Since the inverse transform on $\mathbf{x}$ yields $\mathbf{s}$, the vector $\mathbf{s}$ can be used instead of $\mathbf{x}$, which can be succinctly represented by the location and values of the nonzero elements of $\mathbf{s}$. Although this method efficiently compresses $\mathbf{x}$, it initially requires all the samples of $\mathbf{x}$ to produce $\mathbf{s}$, which undermines the whole purpose of CS.

Now let us assume that instead of samples of $\mathbf{x}$, we take $m$ linear combinations of the samples (called generalized samples). If we represent these linear combinations by the matrix $\mathbf{\Phi}_{m \times n}$ and the resultant vector of samples by $\mathbf{y}_{m \times 1}$, we have:

$$\mathbf{y}_{m \times 1} = \mathbf{\Phi}_{m \times n} \cdot \mathbf{x}_{n \times 1} = \mathbf{\Phi}_{m \times n} \cdot \mathbf{\Psi}_{n \times n} \cdot \mathbf{s}_{n \times 1} \tag{3}$$

The question is how the matrix $\mathbf{\Phi}$ and the size $m$ should be chosen to ensure that these samples uniquely represent the original signal $\mathbf{x}$. Apparently, the case of $\mathbf{\Phi} = \mathbf{I}_{n \times n}$ for $m = n$ yields a trivial solution (keeping all the samples of $\mathbf{x}$) that does not employ the sparsity characteristic. We look for $\mathbf{\Phi}$ matrices with as few rows as possible which can guarantee the invertibility of the sampling process for the class of sparse inputs.

To solve this problem, we introduce probabilistic measures; i.e., instead of exact recovery of signals, we focus on the probability that a random sparse signal (according to a given probability density function) fails to be reconstructed using its generalized samples. If the probability of failure approaches

---

[5]The sequence of vectors $\{\mathbf{v}_n\}$ is called a Riesz basis if there exist scalars $0 < A \leq B < \infty$ such that for every absolutely summable sequence of scalars $\{a_n\}$, we have the following inequalities:

$$A\left(\sum_n |a_n|^2\right) \leq \left\|\sum_n a_n \mathbf{v}_n\right\|_{\ell_2}^2 \leq B\left(\sum_n |a_n|^2\right) \tag{1}$$

zero, we can state that the sampling scheme (the joint pair of $\boldsymbol{\Psi}, \boldsymbol{\Phi}$) is successful in recovering $\mathbf{x}$ with probability 1.

Let us assume that $\boldsymbol{\Phi}^{(m)}$ represents the first $m$ rows of an invertible matrix $\boldsymbol{\Phi}_{n \times n}$. It is apparent that if we use $\{\boldsymbol{\Phi}^{(m)}\}_{m=0}^{n}$ as the sampling matrices for a given sparsity domain, the failure probabilities for $\boldsymbol{\Phi}^{(0)}$ and $\boldsymbol{\Phi}^{(n)}$ are one and zero respectively, and as the index $m$ increases, the failure probability decreases. The important point is that the decreasing rate of the failure probability is exponential with respect to $\frac{m}{k}$ [74]. Therefore, we expect to reach an almost zero failure probability much earlier than $m = n$ despite the fact that the exact rate highly depends on the mutual behavior of the two matrices $\boldsymbol{\Psi}, \boldsymbol{\Phi}$. More precisely, it is shown in [74] that:

$$P_{failure} < n \cdot e^{-\frac{c}{\mu^2(\boldsymbol{\Psi}, \boldsymbol{\Phi}^{(m)})} \cdot \frac{m}{k}} \tag{4}$$

where $c$ is a positive constant and $\mu(\boldsymbol{\Psi}, \boldsymbol{\Phi}^{(m)})$ is the maximum coherence between the rows of $\boldsymbol{\Psi}$ and $\boldsymbol{\Phi}^{(m)}$ defined by [75]:

$$\mu(\boldsymbol{\Psi}, \boldsymbol{\Phi}^{(m)}) = \sqrt{n} \cdot \max_{1 \le a \le n, 1 \le b \le m} \left| \langle \psi_a, \phi_b \rangle \right| \tag{5}$$

where $\psi_a, \phi_b$ are the $a^{th}$ and $b^{th}$ rows of the matrices $\boldsymbol{\Psi}$ and $\boldsymbol{\Phi}$, respectively. The above result implies that, the probability of reconstruction is close to one for:

$$m \ge \underbrace{\mu^2(\boldsymbol{\Psi}, \boldsymbol{\Phi})}_{\ge \mu^2(\boldsymbol{\Psi}, \boldsymbol{\Phi}^{(m)})} \frac{k \cdot \ln n}{c} \tag{6}$$

The above derivation implies that, the lower the maximum coherence between the two matrices, the lower is the number of required samples. Thus, to decrease the number of samples, we should look for matrices $\boldsymbol{\Phi}$ with low coherence with $\boldsymbol{\Psi}$. For this purpose, we use a random $\boldsymbol{\Phi}$. It is shown that the coherence of a random matrix with i.i.d. Gaussian distribution with any unitary $\boldsymbol{\Psi}$ is considerably small [30], which makes it a proper candidate for the sampling matrix. Investigation on the probability distribution has shown that the Gaussian PDF is not the only solution (for example a binary distribution is considered in [76]) but may be the simplest to analyze.

The inequality shown in (6) can be simplified in terms of $m, n$ and $k$; it can be shown [3] and [71] that a sparse signal can be reconstructed from its compressed samples with a probability of almost one if:

$$m \ge c \, k \log(\frac{n}{k}) \tag{7}$$

*2) Reconstruction from Compressed Measurements:* In this subsection, we would like to consider the reconstruction algorithms and stability issues. Essentially, there are three methods: A- Geometric, B- Combinatorial, and C- Information Theoretic. We would like to briefly discuss these three methods.

*Geometric Methods:* The oldest methods for reconstruction from compressed sampling are geometric, i.e., $\ell_1$ minimization techniques for finding a $k$-sparse vector $\mathbf{s} \in \mathbb{R}^n$ from a set of $m = O(k \log(\frac{n}{k}))$ measurements $(y_i)$; see e.g., [30], [74], [77], [78], [79]. Let us assume that we have applied a suitable $\boldsymbol{\Phi}$ which guarantees the invertibility of the sampling process. The reconstruction method should be a technique to recover a $k$-sparse vector $\mathbf{s}_{n \times 1}$ from the observed samples $\mathbf{y}_{m \times 1} =$ $\boldsymbol{\Phi}_{m \times n} \cdot \boldsymbol{\Psi}_{n \times n} \cdot \mathbf{s}_{n \times 1}$ or possibly $\mathbf{y}_{m \times 1} = \boldsymbol{\Phi}_{m \times n} \cdot \mathbf{s}_{n \times 1} + \boldsymbol{\nu}_{m \times 1}$, where $\boldsymbol{\nu}$ denotes the noise vector. Suitability of $\boldsymbol{\Phi}$ implies that $\mathbf{s}_{n \times 1}$ is the only $k$-sparse vector that produces the observed samples; therefore, $\mathbf{s}_{n \times 1}$ is also the sparsest solution for $\mathbf{y} = \boldsymbol{\Phi} \cdot \boldsymbol{\Psi} \cdot \mathbf{s}$. Consequently, $\mathbf{s}$ can be found using:

$$\text{minimize } \|\mathbf{s}\|_{\ell_0} \quad \text{subject to } \mathbf{y} = \boldsymbol{\Phi} \cdot \boldsymbol{\Psi} \cdot \mathbf{s} \tag{8}$$

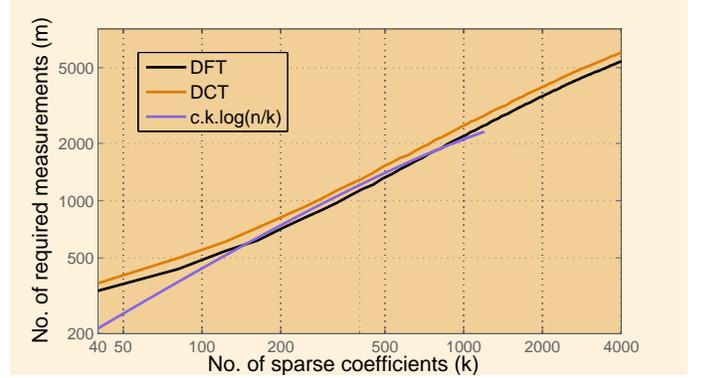

Fig. 7. Relation between $m$ and $k$ for sparse DFT and DCT signals; the frame size is $n = 2^{13}$.

Minimization with respect to $\ell_0$-norm (sparsity) is an NP-complete problem in general. However, it is shown in [80] that minimization with $\ell_1$-norm results in the same vector $\mathbf{s}$ for many cases. The interesting part is that the number of required samples to replace $\ell_0$ with $\ell_1$-minimization has the same order of magnitude as the one for the invertibility of the sampling scheme. Hence, $\mathbf{s}$ can be derived from (8) using $\ell_1$-minimization. It is worthwhile to mention that replacement of $\ell_1$-norm with $\ell_2$-norm, which is faster to implement, does not necessarily produce reasonable solutions. However, there are greedy methods (Matching Pursuit as discussed in Sec. VI on SCA [81], [82]) which iteratively approach the best solution faster than $\ell_1$-norm optimization (Basis Pursuit as discussed in Sec. VI on SCA).

The technique of IMAT, discussed in random sampling and simulated in Fig. 6 for sparse DFT signals, can also be used for the recovery of a sparse signal in other transform domains such as DCT. It should be noted that this sampling process is a special case of CS. Instead of a linear combination of samples, the actual random samples are used [27]. Simulation results shown in Fig. 7 confirm the relation represented in (7). In our simulation results, perfect reconstruction corresponds to an SNR value of $100 \ dB$ with a reliability of at least 80%. It is interesting to see that (7) closely tracks the DFT sparse signal for a range of $k$ when $c = 1$. When $k$ increases for a given $n$ (i.e., the signal becomes less sparse), the relative number of needed measurements to specify the signal is decreased and the relation (7) is no longer valid.

A sufficient condition for these methods to work is that the matrix $\boldsymbol{\Phi} \cdot \boldsymbol{\Psi}$ must satisfy the so called *Restricted Isometric Property* (RIP) [83], [84], [76]; which will be discussed in the following subsection:

*RIP:* It is important that in the presence of noise, the algorithms produce the best approximate solution to within a precision; in other words, small perturbations in the signal caused by noise result in small distortions in the output solution. This characteristic is usually defined as stability. In compressed sensing, the stability of the reconstruction is determined by the characteristics of the sampling matrix $\boldsymbol{\Phi}$. We say that the matrix $\boldsymbol{\Phi}$ has RIP of order $k$, when for all $k$-sparse vectors $\mathbf{s}$, we have [31]:

$$1 - \delta_k \leq \frac{\|\boldsymbol{\Phi} \cdot \mathbf{s}\|_{\ell_2}^2}{\|\mathbf{s}\|_{\ell_2}^2} \leq 1 + \delta_k \qquad (9)$$

where $0 \leq \delta_k < 1$ (isometry constant). The RIP is a sufficient condition that provides us with the maximum and minimum power of the samples with respect to the input power and ensures that none of the $k$-sparse inputs fall in the null space of the sampling matrix. The RIP property essentially states that every $4k$ columns of the matrix $\boldsymbol{\Phi}_{m \times n}$ must be almost orthonormal. The explicit construction of a matrix with such a property is difficult for any given $n \gg m$; however, the problem has been studied in some cases [41], [85]. Moreover, given such a matrix $\boldsymbol{\Phi}$, finding $\mathbf{s}$ (or alternatively $\mathbf{x}$) via the minimization problem involves linear programming with $n$ variables and $m$ constraints which can be computationally expensive.

Among the matrices that satisfy the RIP condition are Gaussian random matrices. If $\boldsymbol{\Phi}$ is a Gaussian random matrix with the number of rows satisfying (6), $\boldsymbol{\Phi} \cdot \boldsymbol{\Psi}$ is also a Gaussian random matrix with the same number of rows and thus it satisfies RIP, which guarantees a stable recovery. Assume that instead of $\boldsymbol{\Phi} \cdot \mathbf{s}$, we have $\boldsymbol{\Phi} \cdot \mathbf{s} + \boldsymbol{\nu}$, where $\boldsymbol{\nu}$ represents the additive noise vector. Since $\boldsymbol{\Phi} \cdot \mathbf{s} + \boldsymbol{\nu}$ may not belong to the range space of $\boldsymbol{\Phi}$ over $k$-sparse vectors, the $\ell_1$ minimization of (8) may not produce a solution. Thus we employ the following minimization instead:

$$\text{minimize } \|\mathbf{s}\|_{\ell_1} \qquad \text{subject to } \|\mathbf{y} - \boldsymbol{\Phi} \cdot \mathbf{s}\|_{\ell_2} < \epsilon \qquad (10)$$

where $\epsilon^2$ is the maximum noise power. Let us denote the result of the above minimization for $\mathbf{y} = \boldsymbol{\Phi} \cdot \mathbf{s} + \boldsymbol{\nu}$ by $\hat{\mathbf{s}}$, which is also a $k$-sparse vector. Now we have:

$$\|\mathbf{y} - \boldsymbol{\Phi} \cdot \mathbf{s}\|_{\ell_2} = \|\boldsymbol{\Phi} \cdot (\hat{\mathbf{s}} - \mathbf{s}) + \boldsymbol{\nu}\|_{\ell_2} < \epsilon$$
$$\Rightarrow \|\boldsymbol{\Phi} \cdot (\hat{\mathbf{s}} - \mathbf{s})\|_{\ell_2} < \|\boldsymbol{\nu}\|_{\ell_2} + \epsilon \qquad (11)$$

Since both $\mathbf{s}$ and $\hat{\mathbf{s}}$ are $k$-sparse, $\hat{\mathbf{s}} - \mathbf{s}$ is $2k$-sparse, and by using the RIP, we get:

$$(1 - \delta_{2k}) \|\hat{\mathbf{s}} - \mathbf{s}\|_{\ell_2} < \|\boldsymbol{\Phi} \cdot (\hat{\mathbf{s}} - \mathbf{s})\|_{\ell_2} < \|\boldsymbol{\nu}\|_{\ell_2} + \epsilon \qquad (12)$$

Or equivalently,

$$\|\hat{\mathbf{s}} - \mathbf{s}\|_{\ell_2} < \frac{\|\boldsymbol{\nu}\|_{\ell_2} + \epsilon}{1 - \delta_{2k}} \qquad (13)$$

This shows that small perturbations in the input cause small perturbations in the output (stability). Moreover, as $\delta_{2k}$ approaches unity, the distortion in the output caused by the input additive noise becomes more significant; the ideal case is when $\delta_{2k} = 0$.

*Combinatorial:* Another standard approach for reconstruction of compressed sampling is combinatorial. The sampling matrix $\boldsymbol{\Phi}$ is found using bipartite graphs, and consists of binary entries, i.e., entries that are either 1 or 0. Binary search methods are then used to find an unknown $k$-sparse vector $\mathbf{s} \in \mathbb{R}^n$, see e.g., [77], [86], [87], [88], [89], [90], [91], [92] and the references therein. Typically, the binary matrix $\boldsymbol{\Phi}$ has $m = O(k \log n)$ rows, and there exist fast algorithms for finding the solution $\mathbf{x}$ from the $m$ measurements (typically a linear combination). However, the construction of $\boldsymbol{\Phi}$ is also difficult.

*Information Theoretic:* A more recent approach is adaptive and information theoretic [93]. In this method, the signal $\mathbf{s} \in \mathbb{R}^n$ is assumed to be an instance of a vector random variable $\mathbf{x} = (x_1, \ldots, x_n)^t$, and the $i^{th}$ row of $\boldsymbol{\Phi}$ is constructed using the value of the previous sample $y_{i-1}$. Tools from the theory of Huffman coding are used to develop a deterministic construction of a sequence of binary sampling vectors $\phi_\sigma$ (i.e., the components of $\phi_\sigma$ consist of 0 or 1) in such a way as to minimize the average number of sampling (rows of $\boldsymbol{\Phi}$) needed to determine a signal.

*3) Almost Sparse Signals and Noisy Measurements:* An important issue in compressed sampling is the robustness to noisy measurements. Specifically, the reconstruction of a $k$-sparse vector $\mathbf{s}$ from noisy $y_{i_1} = \langle \phi_{i_1}, \mathbf{x} \rangle + \eta_i$ should produce a $k$-sparse vector $\hat{\mathbf{s}}$, which is close to $\mathbf{s}$.

Another important aspect is the behavior of compressive sampling algorithms on almost sparse signals which are more likely to occur in applications than exactly $k$-sparse vectors. For example a $k$-sparse vector $\mathbf{s}$ may be corrupted by noise $\eta \in \mathbb{R}^n$, producing the vector $\tilde{\mathbf{s}} = \mathbf{s} + \eta$. Another example is the wavelet transform of an image which consists mostly of small coefficients and a few large coefficients. Obviously, any method for the sampling and reconstruction of sparse signals must also be well adapted to almost sparse signals, i.e., if a sampling and reconstruction method is applied to an almost $k$-sparse signal $\mathbf{s}$, it must produce a $k$-sparse signal $\hat{\mathbf{s}}$ that includes the $k$ most significant coefficients of $\mathbf{s}$ (up to a small error).

*4) CS for Analog Signals:* Recently, there have been efforts to extend the concept of CS to analog signals [72]. The sparse signals are assumed to be the elements of a Shift-Invariant (SI) space generated by $n$ kernels with period $T$. For instance, assume $\{a_l(t)\}_{l=1}^n$ form a Riesz basis (see the footnote in page 6) [73] for $L_2$; the respective generated SI space is

$$SI = \Big\{ \sum_{l=1}^n \sum_{k \in \mathbb{Z}} d_l[k] \cdot a_l(t - kT) \ \Big| \ d_l[k] \in \mathbb{R}$$
$$, \sum_{l=1}^n \sum_{k \in \mathbb{Z}} \|d_l[k]\|^2 < \infty \Big\} \qquad (14)$$

For example, the set of all lowpass signals with bandwidth $B$ form an SI space with a single kernel ($sinc(2Bt)$). Similarly, the set of multiband signals with the frequency support defined as $n$ fixed disjoint intervals of equal length form an $n$-kernel SI space. The classical sampling theorems suggest that the sampling rate of $\frac{n}{T}$ is sufficient for the signal recovery for the space given in (14).

Now, a signal $x(t) \in SI$ is called $k$-sparse if in its basis representation, at most $k$ generators among the total $n$ are active; i.e., $d_l[k] = 0$ for $l \notin \{l_1, l_2, \ldots, l_k\}$ where $\{l_1, l_2, \ldots, l_k\}$ is an arbitrary subset of $\{1, 2, \ldots, n\}$. Similar to the discrete case, we are looking for sampling schemes that employ the inherent sparsity in order to decrease the sampling rate. In [72], it is shown that the rate could be as low as $\frac{2k}{T}$ for $k$-sparse signals; this is the theoretical lower bound for invertible sampling rates. Instead of the sampling matrix $\boldsymbol{\Phi}$, a filter bank is used for sampling; the signal is passed through $p$ filters ($2k \leq p < n$) followed by samplers, each sampling at rate $\frac{1}{T}$. The analog continuous signal $x(t)$ is thus transformed to an infinite length vector; generalization of the CS for finite length vectors has also been studied in [72].

### C. Sampling with Finite Rate of Innovation

The classical sampling theorem states that:

$$x(t) = \sum_{i \in \mathbb{Z}} x\left(\frac{i}{2B}\right) \cdot \mathrm{sinc}(2Bt - i) \tag{15}$$

where $B$ is the bandwidth of $x(t)$ with the Nyquist interval $T_s = \frac{1}{2B}$. These uniform samples can be regarded as the degrees of freedom of the signal; i.e., a lowpass signal with bandwidth $B$ has one degree of freedom in each Nyquist interval $T_s$. Replacing the $sinc$ function with other kernels in (15), we can generalize the sparsity (bandlimitedness) in the Fourier domain to a wider class of signals known as the SI spaces:

$$x(t) = \sum_{i \in \mathbb{Z}} c_i \cdot \varphi\left(\frac{t}{T_s} - i\right) \tag{16}$$

Similarly, the above signals have one degree of freedom in each $T_s$ period of time (the coefficients $c_i$). A more general definition for the degree of freedom is introduced in [4] and is named the *Rate of Innovation*. For a given signal model, if we denote the degree of freedom in the time interval of $[t_1, t_2]$ by $C_x(t_1, t_2)$, the local rate of innovation is defined by $\frac{1}{t_2 - t_1} C_x(t_1, t_2)$ and the global rate of innovation ($\rho$) is defined as

$$\rho = \lim_{\tau \to \infty} \frac{1}{2\tau} C_x(t - \tau, t + \tau) \tag{17}$$

provided that the limit exists; in this case we say that the signal has finite rate of innovation [4], [28], [94], [95]. As an example, for the lowpass signals with bandwidth $B$ we have $\rho = 2B$, which is the same as the Nyquist rate. In fact by proper choice of the sampling process, we are extracting the innovations of the signal. Now the question that arises is that whether the uniform sampling theorems can be generalized to the signals with finite rate of innovation. The answer is positive for a class of non-bandlimited signals including the SI spaces. Consider the following signals:

$$x(t) = \sum_{i \in \mathbb{Z}} \sum_{r=1}^{R} c_{i,r} \cdot \varphi_r\left(\frac{t - t_i}{T_s}\right) \tag{18}$$

where $\{\varphi_r(t)\}_{r=1}^{k}$ are arbitrary but known functions and $\{t_i\}_{i \in \mathbb{Z}}$ is a realization of a point process with mean $\mu$. The free parameters of the above signal model are $\{c_{i,r}\}$ and $\{t_i\}$. Therefore, for this class of signals we have $\rho = \frac{2}{\mu}$; however, the classical sampling methods cannot reconstruct these kinds of signals with the sampling rate predicted by $\rho$. There are many variations for the possible choices of the functions $\varphi_r(t)$; nonetheless, we just describe the simplest version. Let the signal $x(t)$ be a finite mixture of sparse Dirac functions:

$$x(t) = \sum_{i=1}^{k} c_i \cdot \delta(t - t_i) \tag{19}$$

where $\{t_i\}$ is assumed to be an increasing sequence. We intend to show that the samples generated by proper sampling kernels $\varphi(t)$ (shown in Fig. 8) can be used to reconstruct the sparse Dirac functions. In fact we choose the $\varphi(t)$ to satisfy the so called Strang-Fix condition of order $2k$:

$$
\forall\, 0 \leq r \leq 2k - 1,\ \exists\, \{\alpha_{r,i}\}_{i \in \mathbb{Z}}:\\
\sum_{i \in \mathbb{Z}} \alpha_{r,i} \varphi(t - i) = t^r \tag{20}
$$

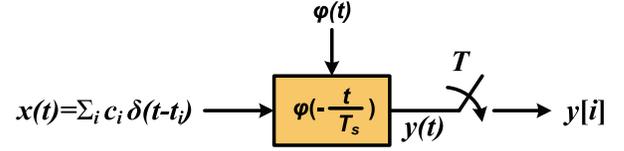

Fig. 8. Sampling with the kernel $\varphi(t)$

The above condition for the Fourier domain becomes:

$$
\begin{cases}
\Phi(\Omega = 0) \neq 0 \\
\Phi^{(r)}(\Omega = 2\pi i) = 0, & \forall\, i \neq 0 \in \mathbb{Z} \\
& r = 0, \ldots, 2k - 1
\end{cases} \tag{21}
$$

where $\Phi(\Omega)$ denotes the Fourier transform of $\varphi(t)$, and the superscript $(r)$ represents the $r^{th}$ derivative. It is also shown that such functions are of the form $\varphi(t) = f(t) * \beta_{2k}(t)$, where $\beta_{2k}(t)$ is the B-spline of order $2k^{th}$ and $f(t)$ is an arbitrary function with nonzero DC frequency [94]. Therefore, the function $\beta_{2k}(t)$ is itself among the possible options for the choice of $\varphi(t)$.

We can show that for the sampling kernels which satisfy the Strang-Fix condition (20), the innovations of the signal $x(t)$ (19) can be extracted from the samples ($y[j]$):

$$y[j] = \left(x(t) * \varphi\left(-\frac{t}{T_s}\right)\right)\Big|_{t = j \cdot T_s} = \sum_{i=1}^{k} c_i \varphi(t_i - j) \tag{22}$$

Thus,

$$
\begin{aligned}
\tau_r &\triangleq \sum_{j \in \mathbb{Z}} \alpha_{r,j} y[j] \\
&= \sum_{i=1}^{k} c_i \sum_{j \in \mathbb{Z}} \alpha_{r,j} \varphi(t_i - j) \\
&= \sum_{i=1}^{k} c_i t_i^r
\end{aligned} \tag{23}
$$

In other words, we have filtered the discrete samples ($y[j]$) in order to obtain the values $\tau_r$; (23) shows that these values are only a function of the innovation parameters (amplitudes $c_i$ and time instants $t_i$). However, the values $\tau_r$ are nonlinearly related to the time instants and therefore, the innovations cannot be extracted from $\tau_r$ using linear algebra[6]. However, these nonlinear equations form a well-known system which was studied by *Prony* in the field of spectral estimation (see Sec. IV-A) and its discrete version is also employed in both real and Galois field versions of Reed-Solomon codes (see Sec. III-A). This method which is called the *annihilating filter* is as follows:

The sequence $\{\tau_r\}$ can be viewed as the solution of a recursive equation. In fact if we define $H(z) = \sum_{i=0}^{k} h_i z^i = \prod_{i=1}^{k} (z - t_i)$, we will have (see Sec. III-A and Appendices II, III for the proof of a similar theorem):

$$\forall \; r : \quad \tau_{r+k} = -\sum_{i=1}^{k} h_i \cdot \tau_{r+i-1} \qquad (24)$$

In order to find the time instants $t_i$, we find the polynomial $H(z)$ (or the coefficients $h_i$) and we look for its roots. A recursive relation for $\tau_r$ becomes:

$$\begin{bmatrix} \tau_1 & \tau_2 & \ldots & \tau_k \\ \tau_2 & \tau_3 & \ldots & \tau_{k+1} \\ \vdots & \vdots & \ddots & \vdots \\ \tau_k & \tau_{k+1} & \ldots & \tau_{2k-1} \end{bmatrix} \cdot \begin{bmatrix} h_1 \\ h_2 \\ \vdots \\ h_k \end{bmatrix} = - \begin{bmatrix} \tau_{k+1} \\ \tau_{k+2} \\ \vdots \\ \tau_{2k} \end{bmatrix} \qquad (25)$$

By solving the above linear system of equations, we obtain coefficients $h_i$ (for a discussion on invertibility of the left side matrix see [94], [96]) and consequently, by finding the root of $H(z)$, the time instants will be revealed. It should be mentioned that the choice of $\tau_1, \ldots, \tau_{2k}$ in (25), can be replaced with any $2k$ consecutive terms of $\{\tau_i\}$. After determining $\{t_i\}$, (23) becomes a linear system of equations with respect to the values $\{c_i\}$ which could be easily solved.

This reconstruction method can be used for other types of signals satisfying (18) such as the signals represented by piecewise polynomials [94]. An important issue in nonlinear reconstruction is the noise analysis; for the purpose of denoising and performance under additive noise the reader is encouraged to see [28].

## III. ERROR CORRECTION CODES: GALOIS AND REAL/COMPLEX FIELDS

The relation between sampling and channel coding is the result of the fact that over-sampling creates redundancy [97]. This redundancy can be used to correct for "sparse" impulsive noise. Normally, the channel encoding is performed in finite Galois fields as opposed to real/complex fields; The reason is the simplicity of logic circuit implementation and insensitivity to the pattern of errors. On the other hand, the real/complex field implementation of error correction codes has stability problems with respect to the pattern of impulsive, quantization and additive noise [46], [53], [67], [98], [99], [100], [101]. Nevertheless, such implementation has found applications in fault tolerant computer systems [102], [103], [104], [105], [106] and impulsive noise removal from 1-D and 2-D signals [32], [33]. Similar to finite Galois fields, real/complex field codes can be implemented in both block and convolutional fashions.

A discrete real-field block code is an oversampled signal with $n$ samples such that, in the transform domain (e.g., DFT), a contiguous number of high frequency components are zero. In general, the zeros do not have to be the high frequency components or contiguous. However, if they are contiguous, the resultant $m$ equations (from the syndrome information domain) and $m$ unknown erasures form a Vandermonde matrix, which ensures invertibility and consequently erasure recovery. The DFT block codes are thus a special case of Reed-Solomon (RS) codes in the field of real/complex numbers [97]. A more general condition to have a Vandermonde matrix is that the indices of the zero frequencies form an arithmetic progression using $mod(n)$. This is equivalent to having contiguous zeros in the Sorted DFT (SDFT[7]) domain [2], [53], [107].

Figure 9 represents a convolutional encoder of rate $\frac{1}{2}$ of finite constraint length [97] and infinite precision per symbol. Fig. 9(a) is a systematic convolutional encoder and resembles an oversampled signal discussed in Section II if the FIR filter acts as an ideal lowpass filter. Fig. 9(b) is a non-systematic encoder used in the simulations to be discussed subsequently. In case of additive impulsive noise, errors could be detected based on the side information that there are frequency gaps in the original oversampled signal (syndrome). In the following subsections, various algorithms for decoding along with simulation results are given for both block and convolutional codes. Some of these algorithms can be used in other applications such as spectral and channel estimation.

### A. Decoding of Block Codes- ELP Method

Iterative reconstruction for an erasure channel is identical to the missing sampling problem [108] discussed in Section II-A.1 and therefore, will not be discussed here. Let us assume that we have a finite discrete signal $x_{orig}[i]$, where $i = 1, \ldots, l$. The DFT of this sequence yields $l$ complex coefficients in the frequency domain ($X_{orig}[j]$, $j = 1, \ldots, l$). If we insert $p$ consecutive zeros[8] to get $n = l + p$ samples ($X[j]$, $j = 1, \ldots, n$) and take its inverse DFT, we end up with an oversampled version of the original signal with $n$ complex samples ($x[i]$, $i = 1, \ldots, n$). This oversampled signal is real iff Hermitian symmetry (complex conjugate symmetry) is preserved in the frequency domain, e.g., the set $\Theta$ of $p$ zeros are centered at $\frac{n}{2}$. For erasure channels, the sparse missing samples are denoted by $e[i_m] = x[i_m]$, where $i_m$'s denote the positions of the lost samples; consequently, for $i \neq i_m$, $e[i] = 0$. The Fourier transform of $e[i]$ (called

---

[6]Note that the Strang-Fix condition can be also used for an exponential polynomial assuming the delta functions are non-uniformly periodic; in that case $\tau_r$ in equation (23) is similar to $E$, the DFT of the impulses, as defined in Appendices II and III.

[7]The kernel of SDFT is $\exp(\frac{2\pi j}{n} i \, q)$, where $q$ is relatively prime w.r.t. $n$; this is equivalent to a sorted version of DFT coefficients according to a $mod$ rule.

[8]We call the set of indices of consecutive zeros syndrome positions and denote it by $\Theta$; this set includes the complex conjugate part in a block of $n$ samples.

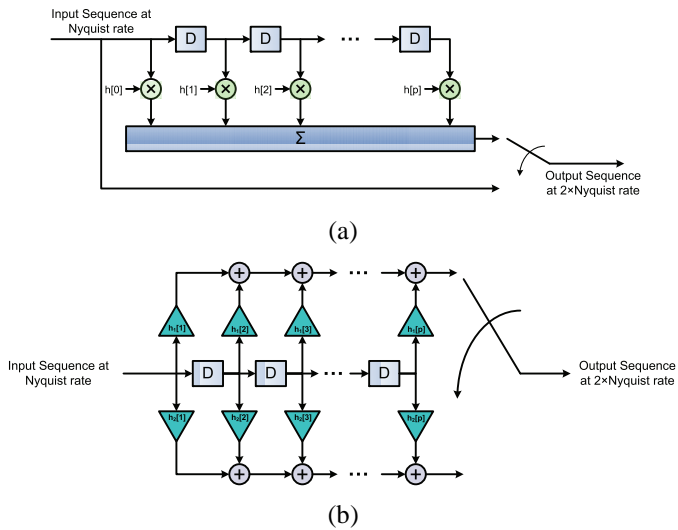

**(a)**

**(b)**

Fig. 9. Convolutional Encoders (a) A real-field systematic convolutional encoder of rate $\frac{1}{2}$; $h[i]$'s are the taps of an FIR filter; (b) A non-systematic convolutional encoder of rate $\frac{1}{2}$, $h_1[i]$'s and $h_2[i]$'s are the taps of 2 FIR filters.

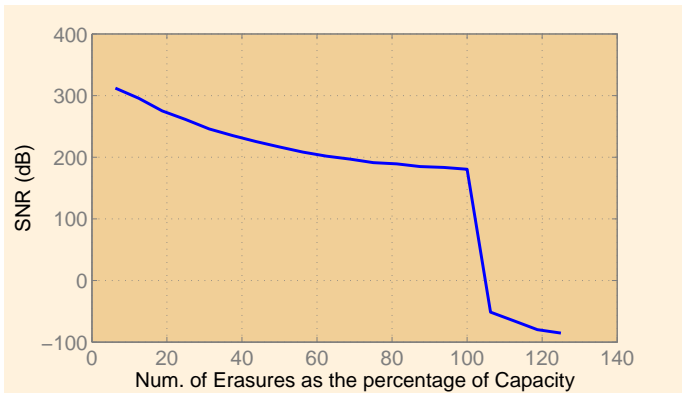

Fig. 10. Recovery of a burst of 16 sample losses.

$E[j], \; j = 1, \ldots, n)$ is known for the syndrome positions $\Theta$. The remaining values of $E[j]$ can be found from the following recursion (see Appendix II):

$$E[r] = -\frac{1}{h_k} \sum_{t=1}^{k} E[r+t] h_{k-t} \qquad (26)$$

where $r$ is a member of the complement of $\Theta$ and the index additions are in $mod(n)$. After finding $E[j]$ values, the spectrum of the recovered oversampled signal $X[j]$ can be found by removing $E[j]$ from the received signal (see (97) in Appendix II). Hence the original signal can be recovered by removing the inserted zeros at the syndrome positions of $X[j]$. The above algorithm, called Error Locator Polynomial (ELP) algorithm, is capable of correcting any combination of erasures. However, if the erasures are bursty, the above algorithm may become unstable. To combat bursty erasures, we can use the SDFT [2], [53], [109], [110], [107] instead of DFT. The simulation results for block codes with erasure and impulsive noise channels are given in the following two subsections.

*1) Simulation Results for Erasure Channels:* The simulation results for the ELP decoding implementation for $n = 32$, $p = 16$, and $k = 16$ erasures (a burst of 16 consecutive missing samples from position 1 to 16) are shown in Fig. 10.

Since consecutive sample losses represent the worst case [53], [110], the proposed method works better for random samples. In practice, the error recovery capability of this technique degrades with the increase of the block and/or burst size due to the accumulation of round-off errors. In order to reduce the round-off error, instead of the DFT, a transform based on the SDFT, or Sorted DCT (SDCT) can be used [2], [53], [110]. These types of transformations act as an interleaver to break down the bursty erasures.

*2) Simulation Results for Random Impulsive Noise Channel:* There are several methods to determine the number,

locations and values of the impulsive noise samples, namely: Modified Berlekamp-Massey for real fields [111], [112], ELP, IMAT, and CFAR-RDE. The Berlekamp-Massey method for real numbers is sensitive to noise and will not be discussed here [111]. The ELP method is described in the next subsection, while the other two methods are discussed below.

*ELP Method [96]:* When the number and positions of the impulsive noise samples are not known, $h_t$ in (26) is not known for any $t$; therefore, we assume the maximum possible number of impulsive noise samples per block, i.e., $k = \lfloor \frac{n-l}{2} \rfloor$ as given in (94) in Appendix II. To solve for $h_t$, we need to know only $n - l$ samples of $E$ in the positions where zeros are added in the encoding procedure. Once the values of $h_t$ are determined from the pseudo-inverse [96], the number and positions of impulsive noise can be found from (96) in Appendix II. The actual values of the impulsive noise can be determined from (26) as in the erasure channel case. For the actual algorithm, please refer to Appendix III. As we are using the above method in the field of real numbers, exact zeros of $\{H_k\}$, which are the DFT of $\{h_i\}$, are rarely observed; consequently, the zeros can be found by thresholding the magnitudes of $H_k$. Alternatively, the magnitudes of $H_k$ can be used as a mask for soft-decision; in this case, thresholding is not needed.

*CFAR-RDE and IMAT Methods [32]:* The Constant False Alarm Rate with Recursive Detection Estimation (CFAR-RDE) method is similar to the Iteration Method with Adaptive Thresholding (IMAT) with the additional inclusion of the CFAR module to estimate the impulsive noise; CFAR is extensively used in radars to detect and remove clutter noise from data. In CFAR, we compare the noisy signal with its neighbors and determine if an impulsive (sparse) noise is present or not (using soft decision)[9]. After removing the impulsive noise in a "soft" fashion, we estimate the signal using the iterative method for an erasure channel as described in Section II-A.1 for random sampling or using the ELP method. The impulsive noise and signal detection and estimation go through several iterations in a recursive fashion as shown in Fig. 11. As the number of recursions increases, the certainty about the detection of impulsive noise locations also increases; thus, the soft decision is designed to act more like the hard decision at the latter parts of the iteration steps, which yields the error locations. Meanwhile, further iterations are performed to enhance the quality of the original signal since suppression of the impulsive noise also suppresses the original signal samples at the location of the impulsive noise. The improvement of using CFAR-RDE over a simple soft decision RDE is shown in Fig. 12.

---

[9]This has some resemblance to soft decision iteration for turbo codes[101].

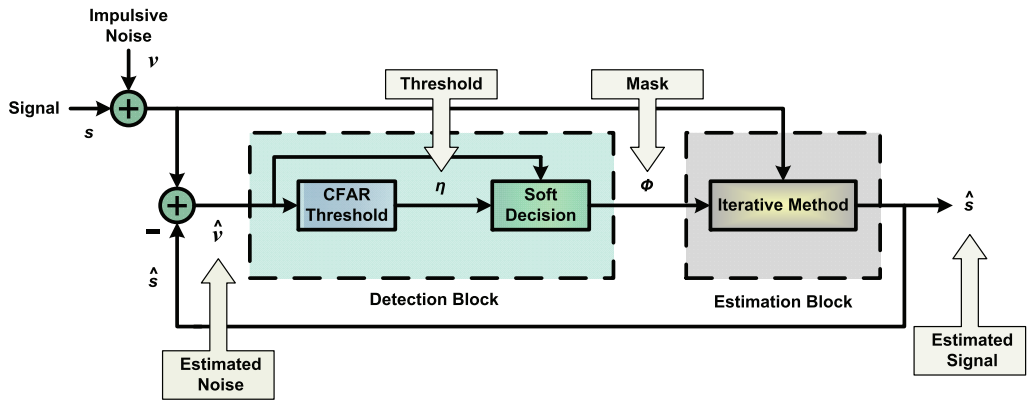

Fig. 11. CFAR-RDE method with the use of adaptive soft thresholding and an iterative method for signal reconstruction.

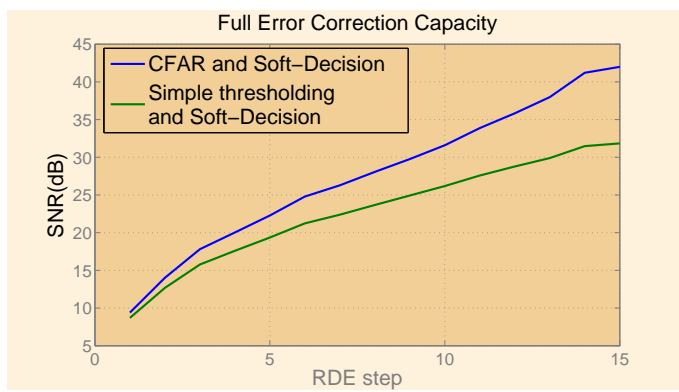

Fig. 12. Comparison of CFAR-RDE and a simple soft decision RDE for DFT block codes.

### B. Decoding for Convolutional Codes

The performance of convolutional decoders depends on the coding rate, the number and values of FIR taps for the encoders, and the type of the decoder. Let us take the convolutional encoder of rate $\frac{1}{2}$ of Fig. 9(b) as our platform for simulations. For simulation results, the taps of the encoder of Fig. 9(b) are

$$h_1 = [1, 2, 3, 4, 5, 16],$$
$$h_2 = [16, 5, 4, 3, 2, 1] \tag{27}$$

The input signal is taken from a uniform random distribution of size 50 and the simulations are run 1000 times and then averaged. The following subsections describe the simulation results for erasure and impulsive noise channels.

#### 1) Decoding for Erasure Channels:
For the erasure channels, we employ two methods as described below:

*Iterations with Averaging:* An iterative method to decode for erasures in the convolutional code of Fig. 9(b) is shown in Fig. 13. This figure is designed for the rate $\frac{1}{2}$ convolutional encoder. At each stage of decoding, the results of the two branches are averaged. For the rate $\frac{1}{2}$ and specific FIR structure, the SNR improvement versus the relative rate of erasures with respect to the theoretical maximum rate of correction capability (full capacity) is shown in Fig. 14. This figure shows that the SNR values gradually decrease as the sampling rate reaches the full capacity.

*Decoding Using the Generator Matrix:* The generator matrix of a convolutional encoder of the type depicted in Fig. 9(b) with taps given in (27) can be shown to be [5]

$$
G =
\begin{bmatrix}
1 & 0 & 0 & 0 & 0 & \cdots \\
16 & 0 & 0 & 0 & 0 & \cdots \\
2 & 1 & 0 & 0 & 0 & \cdots \\
5 & 16 & 0 & 0 & 0 & \cdots \\
3 & 2 & 1 & 0 & 0 & \cdots \\
4 & 5 & 16 & 0 & 0 & \cdots \\
4 & 3 & 2 & 1 & 0 & \cdots \\
3 & 4 & 5 & 16 & 0 & \cdots \\
5 & 4 & 3 & 2 & 1 & \cdots \\
2 & 3 & 4 & 5 & 16 & \cdots \\
16 & 5 & 4 & 3 & 2 & \cdots \\
1 & 2 & 3 & 4 & 5 & \cdots \\
0 & 16 & 5 & 4 & 3 & \cdots \\
0 & 1 & 2 & 3 & 4 & \cdots \\
0 & 0 & 16 & 5 & 4 & \cdots \\
\vdots & \vdots & \vdots & \vdots & \vdots & \ddots
\end{bmatrix}^T
\tag{28}
$$

An iterative decoding scheme for this matrix representation is similar to that of Fig. 3 except that the operator $G$ consists of the generator matrix, a mask (erasure operation) and the transpose of the generator matrix. If the rate of erasure does not exceed the encoder full capacity, the matrix form of the operator $\mathbf{G}$ can be shown to be a nonnegative definite square matrix and therefore its inverse exists [1], [45]. Thus, with a proper choice of the relaxation parameter, the iteration represented in Fig. 13 converges to the actual signal.

By using the above operator $G$ in our iterative simulations,

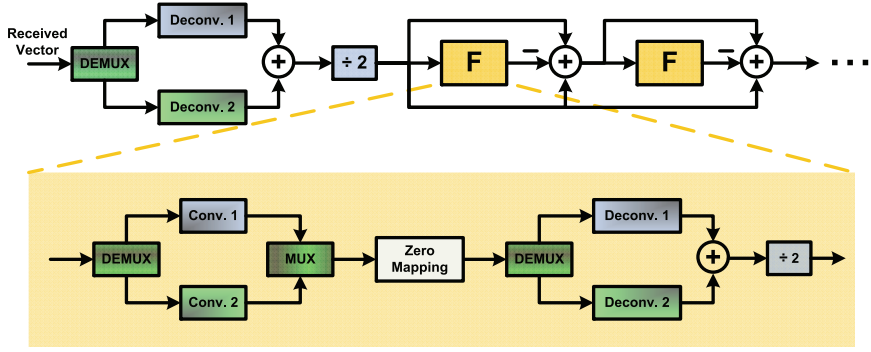

Fig. 13. Iterative decoding for a rate $\frac{1}{2}$ convolutional encoder as shown in Fig. 9(b).

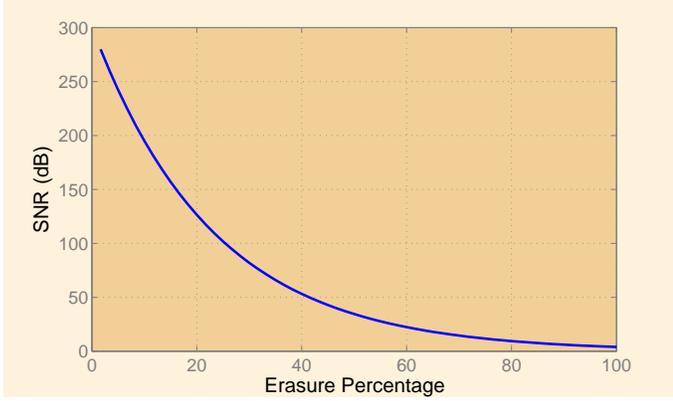

Fig. 14. SNR vs. the percentage of erasure for the convolutional decoder of Fig. 13 after 50 iterations.

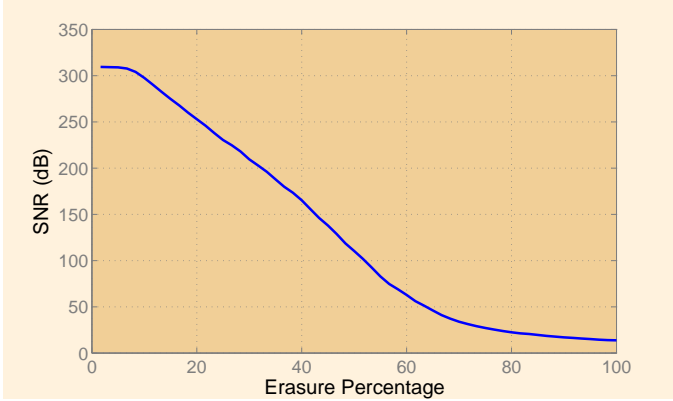

Fig. 15. Simulation results of a convolutional decoder, using the iterative method with the generator matrix, after 30 CG iterations (see Appendix I); SNR vs. the relative rate of erasures (w.r.t. full capacity) in an erasure channel.

better results can be obtained in comparison with the averaging method of Fig. 14. Figure 15 shows that the SNR values gradually decrease as the rate of erasure reaches its maximum (capacity). This figure shows that the generator matrix approach of decoding using the iteration matrix performs much better than the averaging method represented in Figs. 13 and 14. However, the complexity of the matrix approach is higher than the averaging method.

### 2) Decoding for Impulsive Noise Channels:

Let us consider $\mathbf{x}$ and $\mathbf{y}$ as the input and the output streams of the encoder, respectively, related to each other through the generator matrix $\mathbf{G}$ as $\mathbf{y} = \mathbf{Gx}$.

Denoting the observation vector at the receiver by $\hat{\mathbf{y}}$, we have $\hat{\mathbf{y}} = \mathbf{y} + \boldsymbol{\nu}$, where $\boldsymbol{\nu}$ is the impulsive noise vector. Multiplying $\hat{\mathbf{y}}$ by the transpose of the parity check matrix $\mathbf{H}^T$, we get

$$\mathbf{H}^T \hat{\mathbf{y}} = \mathbf{H}^T (\mathbf{y} + \boldsymbol{\nu}) = \underbrace{\mathbf{H}^T \mathbf{Gx}}_{0} + \mathbf{H}^T \boldsymbol{\nu}$$

$$\Rightarrow \mathbf{H}^T \hat{\mathbf{y}} = \mathbf{H}^T \boldsymbol{\nu} \tag{29}$$

Multiplying the resultant by the right pseudo-inverse of the $\mathbf{H}^T$, we derive:

$$\mathbf{H}(\mathbf{H}^T\mathbf{H})^{-1}\mathbf{H}^T \hat{\mathbf{y}} = \mathbf{H}(\mathbf{H}^T\mathbf{H})^{-1}\mathbf{H}^T \boldsymbol{\nu}$$

$$= \tilde{\boldsymbol{\nu}} \tag{30}$$

Thus by multiplying the received vector by $\mathbf{H}(\mathbf{H}^T\mathbf{H})^{-1}\mathbf{H}^T$, we obtain an approximation of the impulsive noise. In the IMAT method, we apply the operator $\mathbf{H}(\mathbf{H}^T\mathbf{H})^{-1}\mathbf{H}^T$ in the iteration of Fig. 5; the threshold level is reduced exponentially at each iteration step. The block diagram of IMAT in Fig. 5 is modified as shown in Fig. 16.

For simulation results, we use the generator matrix shown in (28); its generator matrix can be calculated from [5] and is given below:

$$H = \begin{bmatrix} -1 & 0.063 & 0 & 0 & \cdots \\ -0.313 & 0.125 & -1 & 0.063 & \cdots \\ -0.25 & 0.188 & -0.313 & 0.125 & \cdots \\ -0.188 & 0.25 & -0.25 & 0.188 & \cdots \\ -0.125 & 0.313 & -0.188 & 0.25 & \cdots \\ -0.063 & 1 & -0.125 & 0.313 & \cdots \\ 0 & 0 & -0.063 & 1 & \cdots \\ 0 & 0 & 0 & 0 & \cdots \\ 0 & 0 & 0 & 0 & \cdots \\ \vdots & \vdots & \vdots & \vdots & \ddots \end{bmatrix} \tag{31}$$

In our simulations, the locations of the impulsive noise samples are generated randomly and their amplitudes have Gaussian distributions with zero mean and variance equal to 1, 2, 5 and 10 times the variance of the encoder output. The results are shown in Fig. 17 after 300 iterations. This figure shows that the high variance impulsive noise has a better performance.

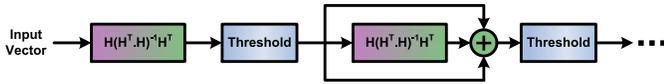

Fig. 16. The modified diagram of IMAT method from Fig. 5.

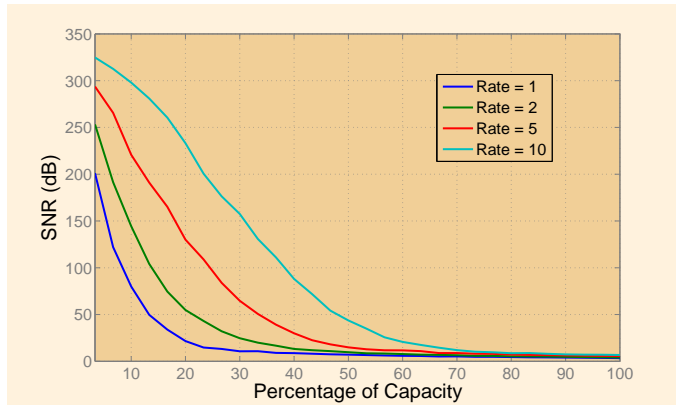

Fig. 17. Simulation results by using the IMAT method for detecting the location and amplitude of the impulsive noise, $\lambda = 1.9$.

### C. LDPC Codes

Low Density Parity Check (LDPC) codes are another example of the application of sparse matrices in reducing complexity [113], [114], [115], [116]. LDPC codes are linear block codes whose parity check matrix, **H** is sparse[10]. The term "low-density" refers to this fact. A code with a sparse **H** matrix which has equal number of 1's in each row and equal number of 1's in each column is called a regular LDPC code. The iterative decoding algorithm that is used to decode a given code, with either a high or low density of ones in the **H** matrix, is such that the decoding complexity grows almost linearly with the block length. The coefficient of this increase is directly dependent on the number of 1's in the **H** matrix as is explained below.

For decoding LDPC codes, we use Tanner graphs [117]; these graphs may also be useful in decoding real-field codes when the parity matrix is sparse and possibly binary. Each iteration module of the decoding algorithm consists of two steps. In the first step, for each row, certain operations are performed for each 1 on that row. Similarly for each column, certain operations are performed for each 1 on that column in the second step. Therefore, the sparser the matrix is, the lower the decoding complexity will be.

As an example, consider an **H** matrix of a regular LDPC code which has 6 and 3 ones in its rows and columns, respectively. Consider another code whose **H** matrix has similarly 8 and 4 ones. The decoding complexity of the former code is 75% of the latter. Note that in general (assuming that

**H** matrix has full rank), for a given code rate $R$, the ratio of the number of ones in each column to the number of ones in each row is a constant which is equal to $1 - R$. Therefore, if the number of ones in the rows is scaled by a factor of $c$, the number of ones in the columns has to be scaled with the same factor to keep the code rate fixed. Consequently, the decoding complexity is scaled by $c$.

*1) LDPC Decoding Using Linear Programming (LP):* [11]

The linear programming decoding of block codes was initially suggested by Feldman in 2003 [118], [119]. Consider Maximum Likelihood (ML) decoding of block codes; this is an optimization problem where we want to minimize a cost function by finding a codeword that has the minimum Euclidean distance to the received vector from the channel. In the logarithm domain, the cost function can be transformed into a linear function of the coded bits (see [120]). Note that the codeword has to satisfy a linear set of parity check equations. Therefore, the problem of decoding a block code can be considered as an LP optimization, and the computational complexity grows exponentially with the block length. However, a sub-optimal solution to this problem significantly reduces the complexity [120]. This modification however makes the overall performance and complexity of the method comparable to that of the conventional decoding method of LDPC codes, the Maximum A Posteriori (MAP) decoding on a bit level.

The performance of the LP decoding method is usually inferior to that of MAP decoding but it has been shown to be superior to that of the MIN-SUM, which is a simplification of the original MAP decoding algorithm [121].

## IV. Spectral Estimation

In this section, we review some of the methods which are used to evaluate the frequency content of data [8], [9], [10], [11]. In the field of signal spectrum estimation, there are several methods which are appropriate for different types of signals. Some methods are known to better estimate the spectrum of wideband signals, where some others are proposed for the extraction of narrow-band components. Since our focus is on sparse signals, it would be reasonable to assume sparsity in the frequency domain, i.e., we assume the signal to be a combination of several sinusoids plus white noise.

Conventional methods for spectrum analysis are non-parametric methods in the sense that they do not assume any model (statistical or deterministic) for the data, except that it is zero or periodic outside the observation interval. As a well known non-parametric method, we can name the periodogram $\hat{P}_{per}(f)$ that can be computed via the FFT:

$$\hat{P}_{per}(f) = \frac{1}{mT_s} \left| T_s \sum_{r=0}^{m-1} x_r e^{-j2\pi f r T_s} \right|^2 \qquad (32)$$

where $m$ is the number of observations, $T_s$ is the sampling interval (usually assumed as unity), and $x_r$ is the signal. Although non-parametric methods are robust with low computational complexity, they suffer from fundamental limitations.

---

[10]Note that the generator matrix G of an LDPC code is not necessarily sparse and the above discussions are only valid for decoding and not encoding.

[11]For further discussion on LP and its relation to Basis Pursuit, please refer to Sec. VI-D.2 and Table IX.

The most important limitation is its resolution; too closely spaced harmonics cannot be distinguished if the spacing is smaller than the inverse of the observation period.

To overcome this resolution problem, parametric methods are devised. Assuming a statistical model with some unknown parameters, we can get more resolution by estimating the parameters from the data at the cost of more computational complexity. Theoretically, in parametric methods we can resolve two closely spaced harmonics with limited data length if the SNR goes to infinity[12].

In this section, we shall discuss three parametric approaches for spectral estimation: the Pisarenko, the Prony, and the well known MUSIC algorithms. The first two are mainly used in spectral estimation, while the MUSIC method that was first developed for array processing has been extended to spectral estimation (see the next section). It should be noted that the parametric methods unlike the non-parametric approaches, require prior knowledge of the model order (the number of tones). This can be decided from the data using the Minimum Discription Length (MDL) method discussed in the next section.

### A. Prony Method

The Prony method was originally proposed for modeling the expansion of gases [122]; however, now it is known as a general spectral estimation method. In fact, Prony tried to fit a weighted mixture of $k$ damped complex exponentials to $2k$ data measurements. The original approach is related to the noiseless measurements; however, it has been extended to produce the least squared solutions for noisy measurements. We focus only on the noiseless case here. The signal is modeled as a weighted mixture of $k$ complex exponentials with complex amplitudes and frequencies:

$$x_r = \sum_{i=1}^{k} b_i z_i^r \tag{33}$$

where $x_r$ is the noiseless discrete sparse signal consisting of $k$ exponentials with parameters

$$\begin{aligned} b_i &= a_i e^{j\theta_i} \\ z_i &= e^{j2\pi f_i T_s} \end{aligned} \tag{34}$$

where $a_i, \theta_i, f_i$ represent the amplitude, phase and the frequency ($f_i$ is a complex number in general), respectively. Let us define the polynomial $H(z)$ such that its roots represent the complex exponential functions related to the sparse tones (see section II-C on FRI, (26) on ELP and Appendix II):

$$H(z) = \prod_{i=1}^{k} (z - z_i) = \sum_{i=0}^{k} h_i z^{k-i} \tag{35}$$

By shifting the index of (33) and multiplying by the parameter $h_j$ and summing over $j$ we get:

$$\sum_{j=0}^{k} h_j x_{r-j} = \sum_{i=1}^{k} b_i z_i^{r-k} \sum_{j=0}^{k} h_j z_i^{k-j} = 0 \tag{36}$$

TABLE V
BASIC PRONY ALGORITHM

| |
|---|
| 1) Solve the recursive equation in (36) to evaluate $h_i$'s. |
| 2) Find the roots of the polynomial represented in (35); these roots are the complex exponentials defined as $z_i$ in (33). |
| 3) Solve (33) to obtain the amplitudes of the exponentials ($b_i$'s). |

where $r$ is indexed in the range $k + 1 \le r \le 2k$. This formula implies a recursive equation to solve for $h_i$ [9]. After the evaluation of $h_i$'s, the roots of (35) yield the frequency components. Hence the amplitudes of the exponentials can be evaluated from a set of linear equations given in (33). The basic Prony algorithm is given in Table V.

The Prony method is sensitive to noise, which was also observed in the ELP and the annihilating filter methods discussed in sections III-A and II-C. There are extended Prony methods that are better suited for noisy measurements.

### B. Pisarenko Harmonic Decomposition (PHD)

The PHD method is based on the polynomial of the Prony method that utilizes the eigen-decomposition of the data covariance matrix [11]. Assume $k$ complex tones are present in the spectrum of the signal. Then, decompose the covariance matrix of $k + 1$ dimensions into a $k$-dimensional signal subspace and a 1-dimensional noise subspace that are orthogonal to each other. The noise subspace is spanned by the eigenvector $\mathbf{v}$ which satisfies $\mathbf{R}\mathbf{v} = \sigma^2 \mathbf{v}$.

By including the additive noise, the observations are given by:

$$y_r = x_r + \nu_r \tag{37}$$

where $y$ is the observation sample and $\boldsymbol{\nu}$ is a zero-mean noise term that satisfies $E\{\nu_r \nu_{r+i}\} = \sigma^2 \delta[i]$. By replacing $x_r = y_r - \nu_r$ in the difference equation (36), we get

$$\sum_{i=0}^{k} h_i y_{r-i} = \sum_{i=0}^{k} h_i \nu_{r-i} \tag{38}$$

which reveals the Auto-Regressive Moving Average (ARMA) structure (order $(k, k)$) of the observations $y_r$ as a random process. To benefit from the tools in linear algebra, let us define the following vectors

$$\begin{aligned} \mathbf{y} &= [y_r, \ \ldots, \ y_{r-k}]^T \\ \mathbf{h} &= [1, \ h_1, \ \ldots, \ h_k]^T \\ \boldsymbol{\nu} &= [\nu_r, \ \ldots, \ \nu_{r-k}]^T \end{aligned} \tag{39}$$

Now (38) can be written as

$$\mathbf{y}^H \mathbf{h} = \boldsymbol{\nu}^H \mathbf{h} \tag{40}$$

Multiplying both sides of (40) by $\mathbf{y}$ and taking the expected value, we get $E\{\mathbf{y}\mathbf{y}^H\}\mathbf{h} = E\{\mathbf{y}\boldsymbol{\nu}^H\}\mathbf{h}$. Note that

$$E\{\mathbf{y}\mathbf{y}^H\} = \mathbf{R}_{\mathbf{y}\mathbf{y}} \triangleq \begin{bmatrix} R_{\mathbf{y}\mathbf{y}}(0) & \ldots & R_{\mathbf{y}\mathbf{y}}^*(k) \\ \vdots & \ddots & \vdots \\ R_{\mathbf{y}\mathbf{y}}(k) & \ldots & R_{\mathbf{y}\mathbf{y}}(0) \end{bmatrix} \tag{41}$$

TABLE VI
PHD ALGORITHM

1) Given the model order $k$ (number of sinusoids), find the autocorrelation matrix of the noisy observations with dimension $k+1$ ($\mathbf{R_{yy}}$).
2) Find the smallest eigenvalue ($\sigma^2$) of $\mathbf{R_{yy}}$ and the corresponding eigenvector ($\mathbf{h}$).
3) Set the elements of the obtained vector as the coefficients of the polynomial in (35). The roots of this polynomial are the estimated frequencies.

$$E\{\mathbf{y}\boldsymbol{\nu}^H\} = E\{(\mathbf{x}+\boldsymbol{\nu})\boldsymbol{\nu}^H\} = E\{\boldsymbol{\nu}\boldsymbol{\nu}^H\} = \sigma^2\mathbf{I} \qquad (42)$$

We thus have an eigen-equation

$$\mathbf{R_{yy}}\mathbf{h} = \sigma^2\mathbf{h} \qquad (43)$$

which is the key equation of the Pisarenko method. The eigen-equation of (43) states that the elements of the eigenvector of the covariance matrix, corresponding to the smallest eigenvalue ($\sigma^2$), are the same as the coefficients in the recursive equation of $x[r]$ (coefficients of the ARMA model in (38)). Therefore, by evaluating the roots of the polynomial with coefficients that are the elements of this vector, we can find the tones in the spectrum.

Although we started by eigen-decomposition of $\mathbf{R_{yy}}$, we observed that only one of the eigenvectors is required; the one that corresponds to the smallest eigenvalue. This eigenvector can be found using simple approaches (in contrast to eigen-decomposition). The PHD method is briefly shown in Table VI.

A different formulation of the PHD method with linear programming approach (refer to Sec. VI-D.2 for description of linear programming) for array processing is studied in [123]. The PHD method is shown to be equivalent to a geometrical projection problem which can be solved using $\ell_1$-norm optimization. Let us convert the autocorrelation matrix $\mathbf{R}_{m\times m}$ into an $m^2 \times 1$ vector. If $\mathbf{R_1}$ and $\mathbf{R_2}$ are the spatial autocorrelation matrices of two uncorrelated sources, the overall autocorrelation matrix is $\mathbf{R_1}+\mathbf{R_2}$ when both sources are active, which is translated to a similar summation of the corresponding vectors. Although it seems that the vectorized notation generates a subspace for uncorrelated sources, only linear combinations of $m^2 \times 1$ vectors with positive coefficients are acceptable (resulting in a hyper-cone) since $\mathbf{R}$ is positive definite. When the number of sources is less than the number of sensors, the respective vector is restricted to lie on the surface of the cone. Due to the existence of noise, the observed vector falls within the cone. It is shown [123] that the PHD method projects the measured vectors into the surface. For the purpose of the projection, $\ell_1$-norm optimization can be employed instead of the common Pisarenko method.

### C. MUSIC

MUltiple SIgnal Classification (MUSIC), is a method originally devised for high resolution source direction estimation in the context of array processing that will be discussed in the next section [124]. The inherent equivalence of the array processing and time series analysis paves the way for the employment of this method in spectral estimation. MUSIC can be understood as a generalization and improvement of the Pisarenko method. It is known that in the context of array processing, MUSIC attains the statistical efficiency[13] in the limit of asymptotically large number of observations [12]. This is a valuable property since MUSIC contains only a 1-D search while efficient ML methods require $k$-dimensional searches.[14]

MUSIC estimates the frequencies by finding $k$ local maxima of a 1-D spectrum function while ML exhaustively searches a $k$-dimensional parameter space to find the global maximum of a $k$-variable spectrum function.

In the PHD method, we construct an autocorrelation matrix of dimension $k+1$ under the assumption that its smallest eigenvalue ($\sigma^2$) belongs to the noise subspace. Then we use the Hermitian property of the covariance matrix to conclude that the noise eigenvector should be orthogonal to the signal eigenvectors. In MUSIC, we extend this method, using a noise subspace of dimension greater than one to improve the performance. We also use some kind of averaging over noise eigenvectors to obtain a more reliable signal estimator.

The data model for the sum of exponentials plus noise can be written in the matrix form as

$$\mathbf{y} = \mathbf{Ab} + \boldsymbol{\nu} \qquad (44)$$

where the length of data is taken as $m > k$ and

$$\mathbf{A} \triangleq \begin{bmatrix} 1 & \cdots & 1 \\ e^{j\omega_1} & \cdots & e^{j\omega_k} \\ \vdots & \ddots & \vdots \\ e^{j(m-1)\omega_1} & \cdots & e^{j(m-1)\omega_k} \end{bmatrix}$$

$$\mathbf{b} \triangleq [b_1, \ldots, b_k]^T$$
$$\boldsymbol{\nu} \triangleq [\nu_1, \ldots, \nu_m]^T$$
$$\mathbf{y} \triangleq [y_1, \ldots, y_m]^T \qquad (45)$$

where $\boldsymbol{\nu}$ represents the noise vector. Since the frequencies are different, $\mathbf{A}$ is of rank $k$ and the first term in (44) forms a $k$-dimensional signal subspace, while the second term is randomly distributed in both signal and noise subspaces; i.e., unlike the first term, it is not confined to a subspace of lower dimension. The correlation matrix of the observations is given by

$$\mathbf{R} = \mathbf{Abb}^H\mathbf{A}^H + \sigma^2\mathbf{I} \qquad (46)$$

where the noise is assumed to be white with variance $\sigma^2$. If we decompose $\mathbf{R}$ into its eigenvectors, $k$ eigenvalues corresponding to the $k$-dimensional subspace of the first term of (46) are essentially greater than the remaining $m - k$ values, $\sigma^2$, corresponding to the noise subspace; thus, by sorting the eigenvalues, the noise and signal subspaces can be determined. Assume $\omega$ is an arbitrary frequency and $\mathbf{e}(\omega) = [1, e^{j\omega}, \ldots, e^{j(m-1)\omega}]$. The MUSIC method estimates the spectrum content of the signal at frequency $\omega$ by projecting the vector $\mathbf{e}(\omega)$ into the noise subspace. When the projected vector is zero, the vector $\mathbf{e}(\omega)$ falls in the signal subspace and most likely, $\omega$ is among the spectral tones. In fact, the frequency content of the spectrum is inversely proportional to the $\ell_2$-norm of the projected vector:

$$P_{MU}(\omega) = \frac{1}{\mathbf{e}^H(\omega)\mathbf{\Pi}^{\perp}\mathbf{e}(\omega)} \tag{47}$$

$$\mathbf{\Pi}^{\perp} = \sum_{i=k+1}^{m} \mathbf{v}_i \mathbf{v}_i^H \tag{48}$$

where $\mathbf{v}_i$'s are eigenvectors of $\mathbf{R}$ corresponding to the noise subspace.

---

[13] Statistical efficiency of an estimator means that it is asymptotically unbiased and its variance goes to zero.

[14] The error function is non-convex and has multiple local minima; thus gradient descent methods cannot be used to achieve a global minimum.

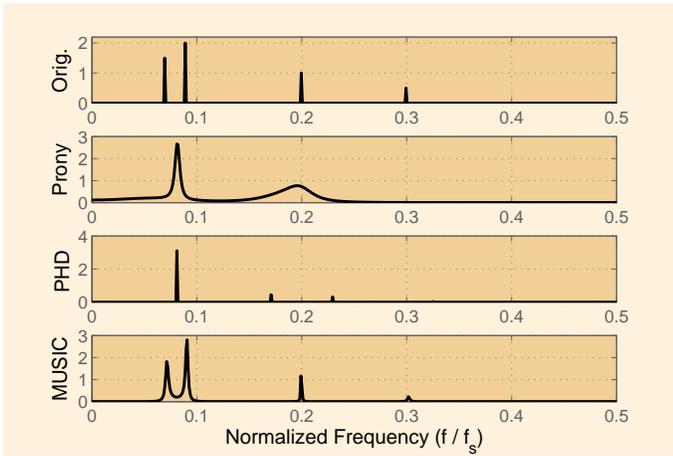

Fig. 18. Spectral estimation of a sparse mixture of sinusoids using Prony, Pisarenko and MUSIC methods; input SNR is $5dB$ and 1024 time samples are used.

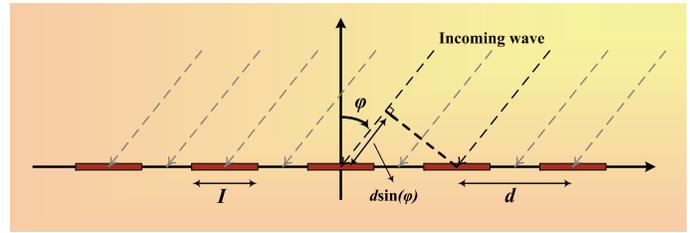

Fig. 19. Uniform linear array with element distance $d$, element length $I$, and a wave arriving from direction $\varphi$.

The $k$ peaks of $P_{MU}(\omega)$ are selected as the frequencies of the sparse signal. The determination of the number of frequencies (model order) in MUSIC is based on the MDL and Akaike Information Criterion (AIC) methods to be discussed in the next section.

Figure 18 shows results for various spectral line estimation methods discussed above. The first upper figure shows the original spectral lines, and the three other figures belong to the results of the estimation methods. We can see that the Prony method (which is similar to ELP and annihilating filter of Sec. II-C and (26)) does not yield good results, while the PHD method is better.

## V. SPARSE ARRAY PROCESSING

We address three types of array processing: 1- Estimation of Multi-Source Location (MSL) and Direction of Arrival (DOA), 2- Sparse Array Beam-forming and Design, and 3- Sparse Sensor Networks. The first topic is related to estimating the direction and/or the location of multiple targets; this problem is very similar to the problem of spectral estimation dealt with in the previous section. The second topic is related to the design of sparse arrays with some missing and/or random array sensors. The last topic, depending on the type of sparsity, is either similar to the second topic or related to CS of sparse signal fields in a network. Below each topic will be briefly described.

### A. Array Processing for MSL and DOA Estimation

Among the important fields of active research in array processing are MSL and DOA estimation [124], [125], [126]. In such schemes, a passive or active array of sensors is used to locate the sources of narrow-band signals. Some applications may assume far-field sources (e.g. radar signal processing) where the array is only capable of DOA estimation, while other applications (e.g. biomedical imaging systems) assume near-field sources where the array is capable of locating the sources of radiation. A closely related field of study is spectral estimation due to similar linear statistical models. The stochastic sparse signals pass through a partially known linear transform (e.g., array response or inverse Fourier transform) and are observed in a noisy environment.

In the array processing context, the common temporal frequency of the source signals are known. Spatial sampling of the signal is used to extract direction of the signal (spatial frequency). As a far-field approximation, the signal wavefronts are assumed to be planar. Consider a signal arriving with angle $\varphi$ as in Fig. 19. Simultaneous sampling of this wavefront on the array will exhibit a phase change of the signal from sensor to sensor. In this way, discrete samples of a complex exponential are obtained, where its frequency can be translated to the direction of the signal source. The response of a Uniform Linear Array (ULA) to a wavefront impinging on the array from direction $\varphi$ is

$$\mathbf{a}(\varphi) = [1, \ e^{j2\pi\frac{d}{\lambda}\sin(\varphi)}, \ \dots, \ e^{j(n-1)2\pi\frac{d}{\lambda}\sin(\phi)}] \tag{49}$$

where $d$ is the inter-element spacing of the array, $\lambda$ is the wavelength, and $n$ is the number of sensors in the array. When multiple sources are present, the observed vector is the sum of the response(sweep) vectors and noise. This resembles the spectral estimation problem with the difference that sampling of the array elements is not limited in time. In fact, in array processing an additional degree of freedom (the number of elements) is present; thus, array processing is more general than spectral estimation.

Two main fields in array processing are MSL and DOA for estimating the source locations and directions, respectively; for both purposes, the angle of arrival (azimuth and elevation) should be estimated while for MSL an extra parameter of range is also needed. The simplest case is the 1-D ULA (azimuth-only) for DOA estimation.

For the general case of $k$ sources with angles $\varphi_1, \ldots, \varphi_k$ with respect to the array, the ULA response is given by the matrix $\mathbf{A}(\boldsymbol{\varphi}) = [\mathbf{a}(\varphi_1), \ldots, \mathbf{a}(\varphi_k)]$, where the vector $\boldsymbol{\varphi}$ of DOA's is defined as $\boldsymbol{\varphi} = [\varphi_1, \ldots, \varphi_k]$. In the above notation, $\mathbf{A}$ is a matrix of size $n \times k$ and $a(\varphi_k)$'s are column vectors. Now, the vector of observations at array elements ($\mathbf{x}[i]$) is given by

$$\mathbf{x}[i] = \mathbf{As}[i] + \boldsymbol{\nu}[i] \tag{50}$$

where the vector $\mathbf{s}[i]$ represents the multi-source signals and $\boldsymbol{\nu}[i]$ is the white Gaussian noise. Source signals and additive noise are assumed to be zero-mean and i.i.d. normal processes with covariance matrices $\mathbf{P}$ and $\sigma^2\mathbf{I}$, respectively. With these assumptions, the observation vector $\mathbf{x}[i]$ will also follow an $n$-dimensional zero-mean normal distribution with the covariance matrix

$$\mathbf{R} = E\{\mathbf{xx}^H\} = \mathbf{APA}^H + \sigma^2\mathbf{I} \tag{51}$$

In the field of DOA estimation, extensive research has been accomplished in 1) source enumeration, and 2) DOA estimation methods. Both of the subjects correspond to the determination of parameters $k$ and $\boldsymbol{\varphi}$.

Although some methods are proposed for simultaneous detection and estimation of the model statistical characteristics [127], most of the literature is devoted to two-stage approaches; first, the number of active sources is detected and then their directions are estimated by techniques such as Estimation of Signal Parameters via Rotational Invariance Techniques (ESPRIT)[15] [128], [129], [130], [131], [132]. Usually the joint detection-estimation methods outperform the two-stage approaches with the cost of higher computational complexity. Source enumeration techniques often rely on the eigenvalues of the covariance matrix obtained from the received data. These eigenvalues are indicators of the energy level in certain principal components of the data. In the MSL estimation step, similar to the MUSIC method, the signal is decomposed into two orthogonal subspaces of signal and noise. The signal subspace corresponds to the column space of the sweep matrix $\mathbf{A}$; thus, proper estimation of the signal subspace leads to estimation of the matrix $\mathbf{A}$ and consequently $\boldsymbol{\varphi}$ and range [12], [13]. In the sequel, we take a closer look at these two steps.

*1) Minimum Description Length (MDL):* One of the most successful methods in array processing for source enumeration is the use of MDL criterion [133]. This technique is very powerful and outperforms its older versions including AIC [134], [135], [136]. Hence, we confine our discussion to MDL algorithms.

*Preliminaries:* MDL is an optimum method of finding the model order and parameters for the most compressed representation of the observed data. For the purpose of statistical modeling, the MAP probability or, the suboptimal criterion of ML is used; more precisely, conditioned on the observed data, the maximum probability among the possible options is found (hypotheses testing) [137]. When the model parameters are not known, the MAP and ML criteria result in the most complex approach; consider fitting a finite sequence of data to a polynomial of unknown degree [37]:

$$x(t_i) = P(t_i) + \nu(t_i), \quad i = 1, \ldots, m \tag{52}$$

where $P(t) = a_0 + a_1t + \cdots + a_kt^k$, $\nu(t)$ is the observed Gaussian noise and $k$ is the unknown model order (degree of the polynomial $P(t)$) which determines the complexity. Clearly, $m - 1$ is the maximum required order for unique description of the data ($m$ observed samples) and the ML criterion, always selects this maximum value ($\hat{k}_{ML} = m - 1$); i.e., the ML method forces the polynomial $P(t)$ to pass through all the points. MDL, on the other hand, yields a sparser solution ($\hat{k}_{MDL} < m - 1$).

Due to the existence of additive noise, it is quite rational to look for a polynomial with degree less than $m$ which also takes the complexity order into account. In MDL, the idea of how to consider the complexity order is borrowed from information theory: Given a specific statistical distribution, we can find an optimum source coding scheme (e.g., Huffman coding) which attains the lowest average code length for the symbols. Furthermore, if $p_x$ is the distribution of the source $x$ and $q_x$ is another distribution, we have [138]:

$$H(x) = E_{p_x}(-\log p_x) \leq E_{p_x}(-\log q_x) \tag{53}$$

where $H(x)$ is the entropy of the signal. This implies that the minimum average code length is obtained only for the correct source distribution (model parameters); in other words, the choice of wrong model parameters (distribution function) leads to larger code lengths. When a particular model with the set of parameters $\boldsymbol{\theta}$ is assumed for the data a priori, each time a sequence $x$ is received, the parameters first should be estimated. The optimum estimation method is usually the ML estimator which results in $\hat{\boldsymbol{\theta}}_{ML}$. Now, the probability distribution for a received sequence $x$ becomes $p(x|\hat{\boldsymbol{\theta}}_{ML})$ which according to information theory, requires an average code length of $-\log\left(p(x|\hat{\boldsymbol{\theta}}_{ML}(x))\right)$ bits. In addition to the data, the model parameters should also be encoded which in turn require $\frac{\kappa}{2}\log(m)$ bits where $\kappa$ is the number of independent parameters to be encoded in the model and $m$ is the number of data points[16]. Thus, the *two part MDL* selects the model that minimizes the whole required code length which is given by [139]:

$$-\log\left(p(x|\hat{\boldsymbol{\theta}}_{ML})\right) + \frac{\kappa}{2}\log(m) \tag{54}$$

The first term is the ML term for data encoding and the second term is a penalty function that inhibits the number of free parameters of the model to get very large.

*MDL Source Enumeration:* In the source enumeration problem, our model is a multivariate Gaussian random process with zero mean and covariance of the type shown in (51), where the number of active sources is unknown. In some enumeration methods (other than MDL), the exact form of (51) is employed which results in high computational complexity. In the conventional MDL method, it is assumed that the model is a covariance matrix with a spherical subspace[17] of dimension $n - k$. Suppose the sample covariance matrix is

$$\hat{\mathbf{R}} = \frac{1}{m} \sum_{i=1}^{m} \mathbf{x}_i \mathbf{x_i}^H \tag{55}$$

and assume the ordered eigenvalues of $\hat{\mathbf{R}}$ are $\hat{\lambda}_1 \geq \hat{\lambda}_2 \geq \cdots \geq \hat{\lambda}_n$, while the ordered eigenvalues of the exact covariance matrix $\mathbf{R}$ are $\lambda_1 \geq \cdots \geq \lambda_k \geq \lambda_{k+1} = \cdots = \lambda_n = \sigma^2$. The normal distribution function of the received complex data $\mathbf{x}$ is [129]

$$p(\mathbf{x}; \mathbf{R}) = \frac{1}{det(\pi \mathbf{R})^m} e^{-tr\{\mathbf{R}^{-1}\hat{\mathbf{R}}\}} \tag{56}$$

where $tr(.)$ stands for the trace operator. The ML estimate of signal eigenvalues in $\mathbf{R}$ are $\hat{\lambda}_i$, $i = 1, \ldots, k$ with the respective eigenvectors $\{\hat{\mathbf{v}}_i\}_{i=1}^{k}$. Since $\lambda_{k+1} = \cdots = \lambda_n = \sigma^2$, the ML estimate of the noise eigenvalue is $\hat{\sigma}_{ML}^2 = \frac{1}{n-k} \sum_{i=k+1}^{n} \hat{\lambda}_i$ and $\{\hat{\mathbf{v}}_i\}_{i=k+1}^{n}$ are all noise eigenvectors. Thus, the ML estimate of $\mathbf{R}$ given $\hat{\mathbf{R}}$ is

$$\mathbf{R}_{ML} = \sum_{i=1}^{k} \hat{\lambda}_i \hat{\mathbf{v}}_i \hat{\mathbf{v}}_i^H + \hat{\sigma}_{ML}^2 \sum_{i=k+1}^{n} \hat{\mathbf{v}}_i \hat{\mathbf{v}}_i^H \tag{57}$$

---

[15]The array in ESPRIT is composed of sensor doublets with the same displacement. The parameters of the impinging signals can be estimated via a rotational invariant property of the signal subspace. The complexity and storage of ESPRIT is less than MUSIC; it is also less vulnerable to array imperfections. ESPRIT, unlike MUSIC results in an unbiased DOA estimate; nontheless, MUSIC outperforms ESPRIT, in general.

[16]For a video introduction to these concepts, please refer to *http://videolectures.net/icml08_grunwald_mdl*

In fact, since we know that $\mathbf{R}$ has a spherical subspace of dimension $n - k$, we correct the observed $\hat{\mathbf{R}}$ to obtain $\mathbf{R}_{ML}$.

Now, we calculate $-\log\big(p(\mathbf{x}|\mathbf{R}_{ML})\big)$; from Appendix IV, we have:

$$tr\{\mathbf{R}_{ML}^{-1}\hat{\mathbf{R}}\} = n \tag{58}$$

which is independent of $k$ and can be omitted in the minimization of (54). Thus, for the first term of (54) we only need the determinant $|\mathbf{R}_{ML}|$ which is the product of the eigenvalues, and the MDL criterion becomes (see Appendix IV):

$$m \sum_{i=1}^{k} \log(\hat{\lambda}_i) \quad + \quad m(n-k) \log\left(\frac{1}{n-k} \sum_{i=k+1}^{n} \hat{\lambda}_i\right)$$
$$+ \quad \frac{\kappa}{2}\log(m) \tag{59}$$

where $\kappa$ is the number of free parameters in the distribution. This expression should be computed for different values of $0 \leq k \leq n - 1$ and its minimum point should be $\hat{k}_{MDL}$. Note that we can subtract the term $m \sum_{i=1}^{n} \log(\hat{\lambda}_i)$ from the expression, which is not dependent on $k$ to get the well known MDL criterion [129] (also see Appendix IV):

$$m(n-k) \log\left[\frac{\frac{1}{n-k}\sum_{i=k+1}^{n} \hat{\lambda}_i}{\prod_{i=k+1}^{n} \hat{\lambda}_i^{\frac{1}{n-k}}}\right] + \frac{\kappa}{2}\log(m) \tag{60}$$

where the first term is the likelihood ratio for the sphericity test of the covariance matrix. This likelihood ratio is a function of arithmetic and geometric means of the noise subspace eigenvalues [140]. Figure 20 is an example of MDL performance in determining the number of sources in array processing. It is evident that in low SNR's, the MDL has a strong tendency to underestimate the number of sources, while as SNR increases, it gives a consistent estimate. Also at high SNR's, underestimation is more probable than overestimation.

[17]Spherical subspace implies the eigenvalues of the cross-correlation matrix are equal in that subspace.

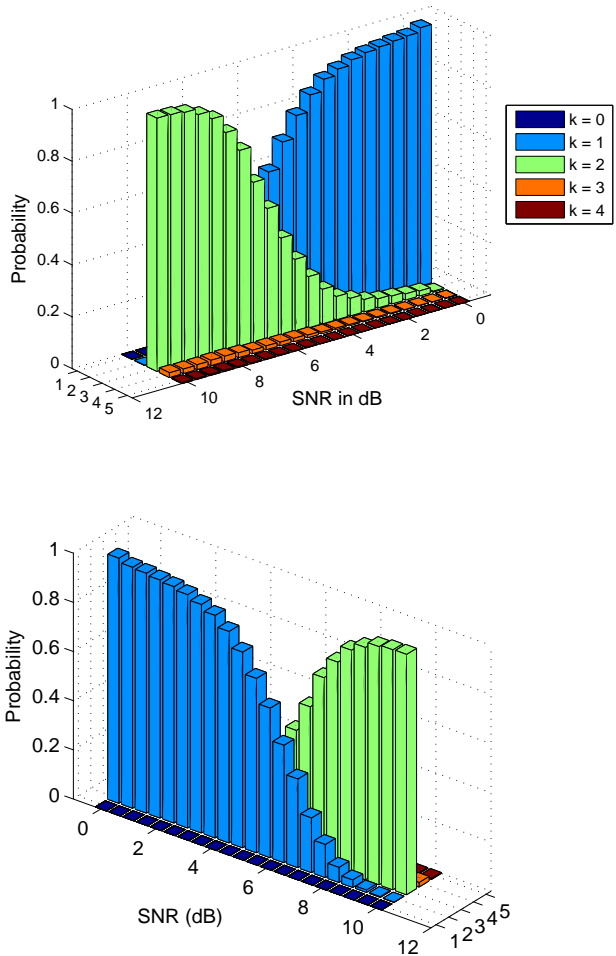

Fig. 20. An example of the performance of the MDL criterion in array processing. The MDL estimates the number of active sources, which is 2.

Now we compute the number of independent parameters ($\kappa$) in the model. Since the noise subspace is spherical, the choice of eigenvectors in this subspace can accept any arbitrary orthonormal set; i.e., no information is revealed when these vectors are known. Thus, the set of parameters is $\{\lambda_1, \ldots, \lambda_k, \sigma^2, \mathbf{v}_1, \ldots, \mathbf{v}_k\}$. The eigenvalues of a hermitian matrix (correlation matrix) are all real while the eigenvectors are normal complex vectors. Therefore, the eigenvalues (including $\sigma^2$) introduce $k+1$ degrees of freedom. The first eigenvector has $2n - 2$ degrees of freedom (since its first nonzero element can be adjusted to unity); while the second, due to its orthogonality to the first eigenvector, has $2n - 4$ degrees of freedom. With the same argument, it can be shown that there are $2(n - i)$ free parameters in the $i^{th}$ eigenvector;

hence

$$\kappa = 1 + k + \sum_{i=1}^{k} 2(n-i) = n(2n-k) + 1 \qquad (61)$$

where the last integer 1 can be omitted since it is independent of $k$.

The two-part MDL, despite its very low computational complexity, is among the most successful methods for source enumeration in array processing. Nonetheless, this method does not reach the best attainable performance for finite number of measurements [141]. The new version of MDL, called *one-part* or *Refined MDL* has improved the performance for the cases of finite measurements which has not been applied to the array processing problem [37].

### B. Sparse Array Beam-forming and Design

Sparsely sampled irregular and random arrays have been used in several fields such as radar, sonar, ultrasound imaging, and seismology. The objective is to improve economy of design, e.g., achieve better resolution for a given maximum number of array elements. The whole idea of sparse arrays depends on having a feature which is better than needed, e.g., sidelobe suppression which can be traded for resolution.

In array signal processing, the aperture smoothing function plays the same role as the transfer function of an LTI filter. Assume that $n$ elements are spaced as a 1-D uniform linear array with a distance $d$ and are located at $x_i = i \cdot d$ for $i = 0, 1, \ldots, n-1$, as shown in Fig. 19. The aperture smoothing function when each element is weighted by a scalar $w_i$ is

$$W(u) = \sum_{i=0}^{n-1} w_i e^{-2\pi j i \frac{u}{\lambda} d} \qquad (62)$$

The variable $u$ is defined by $u = \sin \varphi$ where $\varphi$ is called the azimuth angle, $\lambda$ the wavelength, and the weights, $w_i$, are a standard window function [14]. The weights, $w_i$, could also be angle-dependent if the individual elements are directional (dipoles). The aperture smoothing function determines how the wavefield Fourier transform is filtered by a finite aperture, just as the filter transfer function determines how the received signal spectrum is shaped by the filtering operation. The condition for avoiding aliasing is that the argument in the exponent satisfies

$$2\pi \frac{|u|}{\lambda} d = |\beta_x| \cdot d \leq \pi \qquad (63)$$

where $\beta_x = 2\pi \frac{u}{\lambda}$ is the $x$ component of the wave-number. The relationship between the array pattern for a regular 1-D array and a filter transfer function is

$$\begin{aligned} \omega &\leftrightarrow \beta_x = 2\pi \frac{u}{\lambda} \\ T &\leftrightarrow d \\ h_i &\leftrightarrow w_i \end{aligned} \qquad (64)$$

By using these parallels, the time-frequency sampling theorem $T \leq \frac{\pi}{\omega_{max}}$ translates into the spatial sampling theorem $d \leq \frac{\lambda_{min}}{2}$. The aperture smoothing function at the Nyquist rate of a uniform linear array is depicted in Fig. 21.

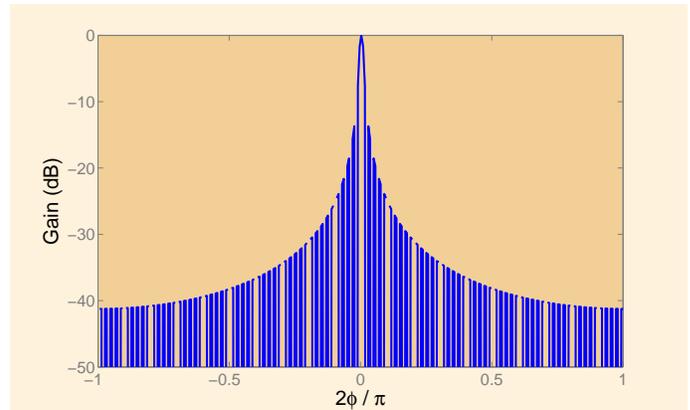

Fig. 21. The element response for an array with ($d = \lambda/2$) and omni-directional elements ($1 \ll \lambda$).

In addition, array processing has some additional degrees of freedom that are not so common in time-frequency processing, such as irregular sampling, and things that are not found at all such as elements along curved surfaces.

A sparse array has elements removed from the aperture (thinning), and there are basically two approaches to finding the best thinning. The first is through optimization; This can be performed in 1-D and 2-D, and also with elements on an underlying regular grid or with freely chosen positions in the aperture (avoiding overlapping elements). It also makes a difference whether the array is flat or curved. The most popular optimization criterion is to minimize the peak sidelobe in the beampattern with a condition on the maximum mainlobe width.

The second approach is more heuristic, and is based on the experience that it is often not possible to formulate optimality in a sense that is compatible with standard optimization algorithms. Also, in an imaging system, more degrees of freedom can be obtained in the optimization, if one allows the layouts of the transmitter and the receiver to be different. We will still assume that the receiver and transmitter arrays are located in the same position (monostatic), but now there is partial or no overlap between the selected elements.

In the optimization approach, the challenge is the combinatorial problem of finding the best layout of sparse elements for one and two dimensions. The optimization problem is creation of beampatterns with low mainlobe width and low sidelobes. The layout optimization methods consist of LP, Genetic Algorithms (GA) and Simulated Annealing (SA) [142].

With a sparse optimized random array, we can get an array pattern as shown in Fig. 22. Comparing this figure to Fig. 21, we see that with even $\frac{1}{4}$ of the elements, we can get comparable results for the peak sidelobe value. It is also evident that thinning has a price in that the overall energy in the sidelobe region has increased.

Another example is shown in Fig. 23 where enhanced simulated annealing has been used. This approach can handle arbitrary layouts where the elements are not locked to a Cartesian grid and where each individual element can have different directivity. Also, it can easily be extended to handle wideband excitations and the response optimization in the near

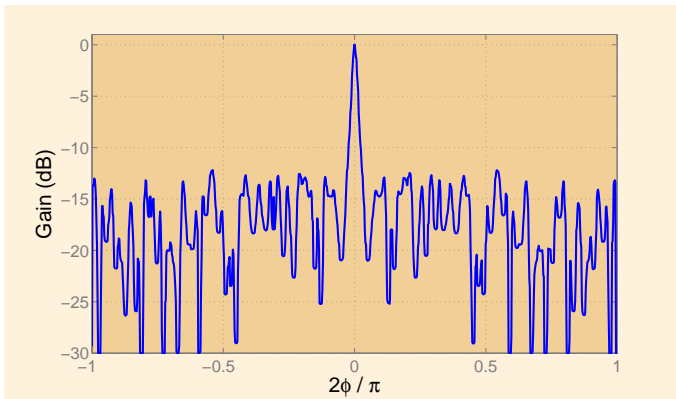

Fig. 22. Array pattern for optimized array with 25 elements out of 101.

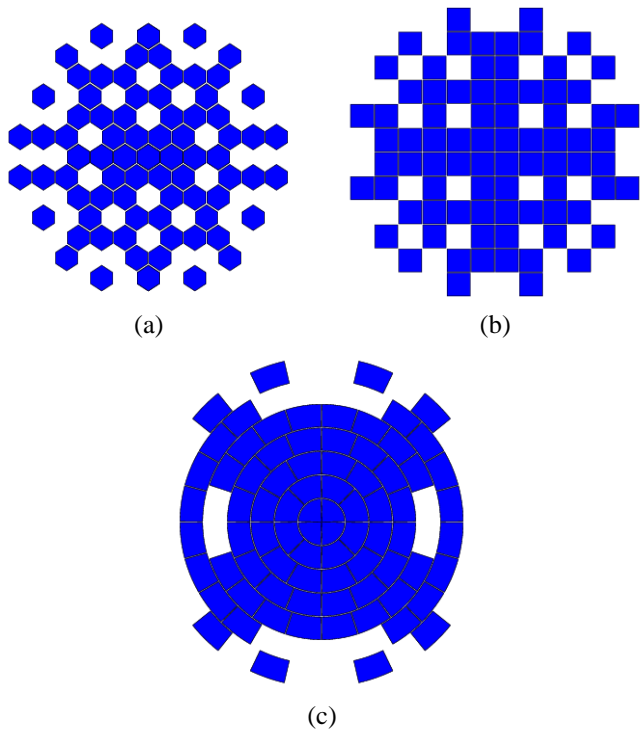

Fig. 23. Symmetric non-cartesian sparse arrays, (a) sparse hex-grid array thinned from 112 to 80 elements, (b) a square element array thinned from 121 to 77 elements, and (c) a ring array thinned from 112 to 80 elements. Optimization is by simulated annealing [144].

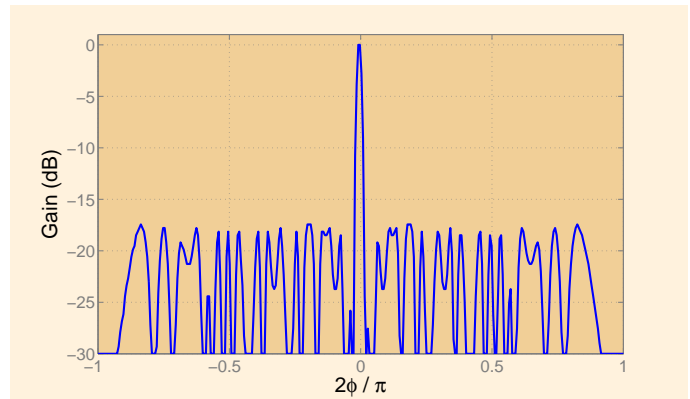

Fig. 24. One-way array pattern after optimization for 64-element array randomly thinned to 48 elements. Thinning and weights are shown in Fig. 25.

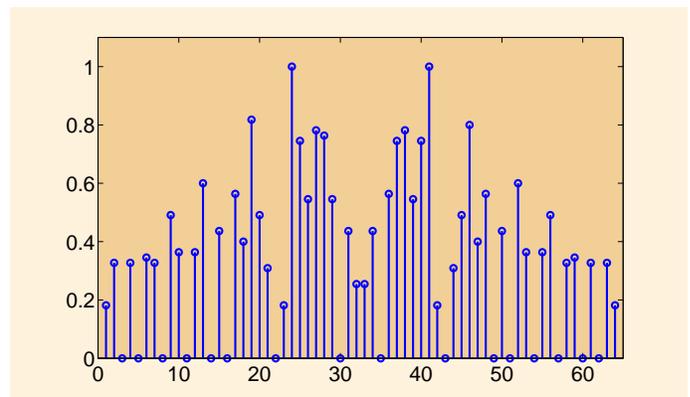

Fig. 25. Weights found after optimization for beam-pattern of the random layout as shown in Fig. 24.

field (focused arrays) [143].

The simple examples of Fig. 23 illustrate the geometries that the algorithm can handle and shows a sparse hex-grid optimized array with hexagonally shaped elements, a linear grid design with square elements, and a ring array having increasing element sizes for increasing ring radii.

One of the optimization problems in array processing is to design the sparsest array for a given beam pattern [142], [145]. In this case, as shown in Table II in row 12, the random sparsity is in the space domain and the information domain is a desirable aperture array pattern; we conjecture that, for a given beampattern, the IMAT used for random sampling and impulsive noise removal (Section II-A.1) can be used in synthesizing the number, locations and weights of the sparse array elements. Another optimization issue for a given number of array elements is equivalent to designing the weights and locations of sparse random arrays to yield an optimum array/beam-pattern [146], see also Figs. 24 and 25; this kind of optimization is unique for this application and has no parallel in other applications. The weights and the array locations can be optimized separately or jointly; joint optimization of thinning pattern and weights has been reported in sonar arrays. Joint optimization of positions and weights is also possible, usually by iterating over a sequence of position optimization followed by weight optimization. Optimization of the element positions of a sparse array is considerably more difficult than weight optimization. The reason is that for an array with $n$ elements, the number of combinations for selecting $k$ array elements is

$$\binom{n}{k} = \frac{n!}{k!(n-k)!} \tag{65}$$

When $n$, $k$ are large or for 2-D arrays, an exhaustive search is out of the question. There are several ways that this number can be reduced. The optimization criteria are the same for the layout problem as for the weight problem, i.e.,

- Minimize main sidelobe of the pattern while restricting the width of the main-lobe below a given limit.

- Minimize the total energy of the pattern wasted in the sidelobes while keeping the peak value of the main lobe above a given threshold.

The second heuristic approach builds on the properties of some of the basic building blocks of sparse arrays: random arrays, binned arrays, and periodic arrays.

Assume an array with $n$ elements where only $k$ elements are kept after random thinning. The average array pattern is equal to that of an aperture weighted with the probability density function of the thinning. This also determines the shape of the main lobe. For a uniform thinning, it turns out that the ratio of the average sidelobe power to the main lobe power is $\frac{1}{k}$. But the variance is about $k$ for $|u| = |\sin(\varphi)| > \frac{\lambda}{L}$, where $L$ is the aperture. The relative peak level of a 1-D random array is $\sqrt{k}\ln(k)$ [147] which gives, in our experience, a fairly good estimate of the peak level.

In the binned array, the aperture is divided into $k$ equal size bins and one element is chosen at random in each bin. This resembles a nearest neighbor restriction as there can be no more than two neighbor elements in a 1-D binned array. With a uniform distribution in each bin, it can be shown that the binned array has the same properties as the random array except that the variance does not reach the value of $k$ until the angle reaches $|u| = |\sin(\varphi)| > k\frac{\lambda}{L}$. Thus the binned array has random sidelobes close to the mainlobe that are much lower than the random array [147].

Periodic arrays are thinned with a periodicity, e.g., 1001001001, 101010101, or 11001100. This means that grating lobes are formed whose position and size are easily predicted. Periodic arrays have turned out to be particularly useful in imaging systems where one can use different periodicities for the transmitter and the receiver. As the two-way beampattern is the product of the transmitter and the receiver beampatterns, by proper design, grating lobes formed by the transmitter can be suppressed by the receiver and vice versa. A special case is the Vernier array which has periodicities $p$ and $p-1$ [148].

The simplest way to utilize the above properties is to use the properties of periodic arrays and variants where grating lobes in the transmitter are used to cancel the receiver grating lobes and vice versa. References [149], [150] describe simulations and experiments with a 2-D array for medical ultrasound of size $48 \times 48$ elements where several variations and combinations are utilized for minimizing sidelobes. The different variants investigated with partial overlap between transmitter and receiver elements are

- Periodic 2-D arrays with different periodicity on transmission and reception
- Arrays with diagonal periodicity combined with each other and with periodic arrays, see for example Fig. 26 [149]. It gives a two-way maximum sidelobe of $-49\ dB$ when un-steered and $-40\ dB$ when steered to $(30, 30)$ degrees.
- Arrays with periodicity along only a single axis and combinations with diagonal periodic arrays
- Arrays with periodicity along radius vectors rather than the $x$- and $y$-axes

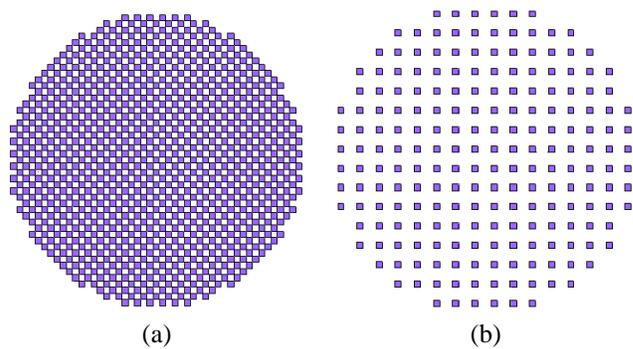

Fig. 26. (a) Periodic diagonal array with 877 out of 2304 elements used for transmission and (b) periodic array with 208 elements for reception-[149].

Arrays which have no overlap between the transmitter and receiver elements are

- Binned arrays with different layouts for transmission and reception
- Polar binned arrays where the bins are along rays emanating from the center of the array

The proposed configurations have been tested on real data for an array that has been built. Good correspondence with simulations have been obtained.

*Curved Arrays:* More recently, there has been some research on optimizations of curved arrays. The beam pattern can be found by projecting the elements onto a flat surface (Fourier projection-slice theorem), therefore, the beampattern is equivalent to that of an array with unequal spacing between the elements. This reduces grating lobes. There is an optimal radius of the curvature for the array, which minimizes the peak sidelobe in the two-way response [151]. The optimal value varies with the thinning method. Similar work for the optimization of one-way response has also been applied to cylindrical arrays for sonar applications [152].

### C. Sparse Sensor Networks

Wireless sensor networks typically consist of a large number of sensor nodes, spatially distributed over a region of interest, that observe some physical environment including acoustic, seismic, and thermal fields with applications in a wide range of areas such as health care, geographical monitoring, homeland security, and hazard detection. The way sensor networks are used in practical applications can be divided into two general categories:

1) There exists a central node known as the Fusion Center (FC) that retrieves relevant field information from the sensor nodes and communication from the sensor nodes to FC generally takes place over a power- and bandwidth-constrained wireless channel.
2) Such a central node does not exist and the nodes take specific decisions based on the information they obtain and exchange among themselves. Issues such as distributed computing and processing are of high importance in such scenarios.

In general, there are three main tasks that should be implemented efficiently in a wireless sensor network: sensing,

communication, and processing. The main challenge in design of practical sensor networks is to find an efficient way of jointly performing these tasks, while using the minimum amount of system resources (computation, power, bandwidth) and satisfying the required system design parameters (such as distortion levels). For example, one such metric is the so-called energy-distortion tradeoff which determines how much energy the sensor network consume in extracting and delivering relevant information up to a given distortion level. Although many theoretical results are already available in the case of point-to-point links in which separation between source and channel coding can be assumed, the problem of efficiently transmitting or sharing information among a vast number of distributed nodes remains a great challenge. This is due to the fact that well-developed theories and tools for distributed signal processing, communications, and information theory in large-scale networked systems are still under development. However, recent results on distributed estimation or detection indicate that joint optimization through some form of source-channel matching and local node cooperation can result in significant system performance improvement [153], [154], [155], [156], [157].

*1) How sparsity can be exploited in a sensor network:* Sparsity appears in many applications for which sensor networks are deployed, e.g., localization of targets in a large region or estimation of physical phenomena such as temperature fields that are sparse under a suitable transformation. For example, in radar applications, under a far-field assumption, the observation system is linear and can be expressed as a matrix of steering vectors [158], [159]. In general, sparsity can arise in a sensor network from two main perspectives:

1) Sparsity of node distribution in spatial terms
2) Sparsity of the field to be estimated

Although nodes in a sensor network can be assumed to be regularly deployed in a given environment, such an assumption is not valid in many practical scenarios. Therefore, the non-uniform distribution of nodes can lead to some type of sparsity in spatial domain that can be exploited to reduce the amount of sensing, processing, and/or communication. This issue is subsequently related to extensions of the nonuniform sampling techniques to two-dimensional domains through proper interpolation and data recovery when samples are spatially sparse [38], [160]. The second scenario that provides a proper basis for exploiting the sparsity concepts arises when the field to be estimated is a sparse multi-dimensional signal. From this point of view, ideas such as those presented earlier in the context of compressed sensing (Sec. II-B) provide the proper framework to address the sparsity in such fields.

*Spatial Sparsity and Interpolation in Sensor Networks:* Although general two-dimensional interpolation techniques are well-known in various branches of statistics and signal processing, the main issue in a sensor network is exploring proper spatio/temporal interpolation such that communication and processing are also efficiently accomplished. While there is a wide range of interpolation schemes (polynomial, Fourier, and least squares [161]), many of these schemes are not directly applicable for spatial interpolation in sensor networks due to their communication complexity.

Another characteristic of many sensor networks is the non-uniformity of node distribution in the measurement field. Although non-uniformity has been dealt with extensively in contexts such as signal processing, geo-spatial data processing, and computational geometry [2], the combination of irregular sensor data sampling and intra-network processing is a main challenge in sensor networks. For example, reference [162] addresses the issue of spatio-temporal non-uniformity in sensor networks and how it impacts performance aspects of a sensor network such as compression efficiency and routing overhead. In order to reduce the impact of non-uniformity, the authors in [162] propose using a combination of spatial data interpolation and temporal signal segmentation. A simple interpolation wavelet transform for irregular sampling which is an extension of the 2-D irregular grid transform to 3-D spatio-temporal sampling grids is also proposed in [163]. Such a multi-scale transform extends the approach in [164] and removes the dependence on building a distributed mesh within the network. It should be noted that although wavelet compression allows the network to trade reconstruction quality for communication energy and bandwidth usage, such energy savings are naturally offset by the overhead cost of computing the wavelet coefficients.

Distributed wavelet processing within sensor networks is yet another approach to reduce communication energy and wireless bandwidth usage. Use of such distributed processing makes it possible to trade long-haul transmission of raw data to the FC for less costly local communication and processing among neighboring nodes [163]. In addition, local collaboration among nodes decorrelates measurements and results in a sparser data set.

*Compressive Sensing in Sensor Networks:* Most natural phenomena in SN's are compressible through representation in a natural basis [79]. Some examples of these applications are imaging in a scattering medium [158], MIMO radar [159], and geo-exploration via underground seismic data. In such cases, it is possible to construct a highly compressed version of a given field, in a decentralized fashion. If the correlations between data at different nodes are known a-priori, it is possible to use schemes that have very favorable power-distortion-latency tradeoffs ([153], [165], [166]). In such cases, distributed source coding techniques, such as Slepian-Wolf coding, can be used to design compression schemes without collaboration between nodes (see [165] and the references therein). Since prior knowledge of such correlations is not available in many applications, collaborative, intra-network processing and compression are used to determine unknown correlations and dependencies through information exchange between network nodes. In this regard, the concept of compressive wireless sensing has been introduced in [157] for energy-efficient estimation at the FC of sensor data, based on ideas from wireless communications [153], [155], [166], [167], [168] and compressive sampling theory [30], [84], [169]. The main objective in such an approach is to combine processing and communications in a single distributed operation [170], [171], [172].

*Methods to obtain the required sparsity in a SN:* While transform-based compression is well-developed in traditional signal and image processing domains, the understanding of sparse transforms for networked data is not as trivial [173]. There are methods such as associating a graph with a given network, where the vertices of the graph represent the nodes of the network, and edges between vertices represent relationships among data at adjacent nodes. The structure of the connectivity is the key to obtaining effective sparse transformations for networked data [173]. For example, in the case of uniformly distributed nodes, tools such as DFT or DCT can be adopted to exploit the sparsity in the frequency domain. In more general settings, wavelet techniques can be extended to handle the irregular distribution of sampling locations [163]. There are also scenarios in which standard signal transforms may not be directly applicable. For example, network monitoring applications rely on the analysis of communication traffic levels at the network nodes where network topology affects the nature of node relationships in complex ways. Graph wavelets [174] and diffusion wavelets [175] are two classes of transforms that have been proposed to address such complexities. In the former case, the wavelet coefficients are obtained by computing the digital differences of the data at different scales. The coefficients at the first scale are differences between neighboring data points, and those at subsequent spatial scales are computed by first aggregating data in neighborhoods and then computing differences between neighboring aggregations. The resulting graph wavelet coefficients are then defined by aggregated data at different scales, and computing differences between the aggregated data [174]. In the latter scheme, diffusion wavelets are based on construction of an orthonormal basis for functions supported on a graph and obtaining a custom-designed basis by analyzing eigenvectors of a diffusion matrix derived from the graph adjacency matrix. The resulting basis vectors are generally localized to neighborhoods of varying size and may also lead to sparse representations of data on a graph [175]. One example of such an approach is where the node data correspond to traffic rates of routers in a computer network.

*Implementation of CS in a wireless SN:* Two main approaches to implement random projections in a SN are discussed in the literature [173]. In the first approach, the CS projections are simultaneously calculated through superposition of radio waves and communicated using amplitude-modulated coherent transmissions of randomly-weighted values directly from the nodes in the network to the FC (Fig. 27). This scheme, introduced in [157], [167] and further refined in [176], is based on the notion of so-called *matched source-channel communication* [166], [167]. Although the need for complex routing, intra-network communications, and processing are alleviated, local phase synchronization among nodes is an issue to be addressed properly in this approach.

In the second approach, the projections can be computed and delivered to every subset of nodes in the network using gossip/consensus techniques, or be delivered to a single point using clustering and aggregation. This approach is typically used for networked data storage and retrieval applications. In this method, computation and distribution of each CS sample is accomplished through two simple steps [173]. In the first step, each of the sensors multiplies its data with the corresponding element of the compressing matrix. Then, in the second step, the resulting local terms are simultaneously aggregated and distributed across the network using randomized gossip [177], which is a simple iterative decentralized algorithm for computing linear functions. Because each node only exchanges information with its immediate neighbors in the network, gossip algorithms are more robust to failures or changes in the network topology and cannot be easily compromised by eliminating a single server or fusion center [178].

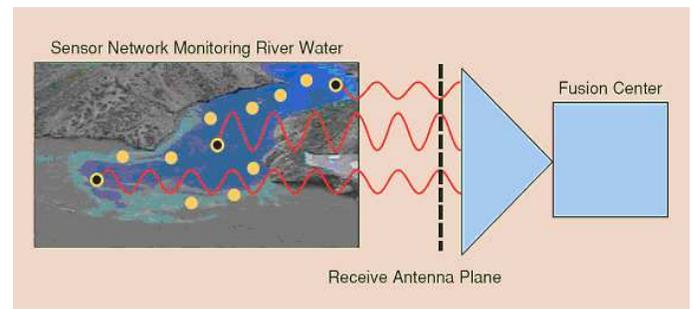

Fig. 27. Computation of CS projections through superposition of radio waves of randomly weighted values directly from the nodes in the network to the FC (from [173]).

Finally, it should be noted that in addition to the encoding process, the overall system performance is significantly affected by the decoding process [82], [179], [180]; this study and its extensions to sparse SN's remain as challenging tasks.

*2) Sensing Capacity:* Despite wide-spread development of SN ideas in recent years, understanding of fundamental performance limits of sensing and communication between sensors is still under development. One of the issues that has recently attracted attention in theoretical analysis of sensor networks, is the concept of sensor capacity. The sensing capacity was initially introduced for discrete alphabets in applications such as target detection [181], and later extended in [15], [182], [183] to the continuous case. The questions in this area are related to the problem of sampling of sparse signals, [30], [69], [169] and sampling with finite rate of innovation [4], [95]. In the context of the CS, sensing capacity provides bounds on the maximum signal dimension or complexity per sensor measurement that can be recovered to a pre-defined degree of accuracy. Alternatively, it can be interpreted as the minimum number of sensors necessary to monitor a given region to a desired degree of fidelity based on noisy sensor measurements. The inverse of sensing capacity is the compression rate; i.e., the ratio of the number of measurements to the number of signal dimensions which characterizes the minimum rate to which the source can be compressed. As shown in [15], sensing capacity is a function of SNR, the inherent dimensionality of the information space, sensing diversity, and the desired distortion level.

Another issue to be noted with respect to the sensing capacity is the inherent difference between sensor network and CS scenarios in the way in which the SNR is handled [15], [183]. In sensor networks composed of many sensors, fixed SNR can be imposed for each individual sensor. Thus, the sensed SNR per location is spread across the field of view leading to a row-wise normalization of the observation matrix. On the other hand, in CS, the vector-valued observation corresponding to each signal component is normalized by each column. This difference has led to different regimes of compression rate [183]. In SN, in contrast to the CS setting, sensing capacity is generally small and correspondingly the number of sensors required does not scale linearly with the target sparsity. Specifically, the number of measurements is generally proportional to the signal dimension and is weakly dependent on target density sparsity. This issue has raised questions on compressive gains in power-limited SN applications based on sparsity of the underlying source domain.

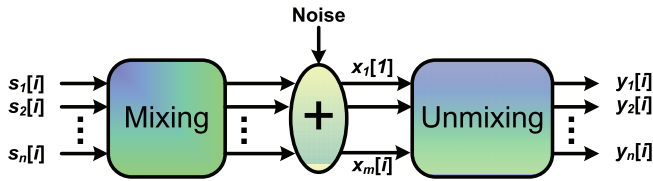

Fig. 28. The BSS concept; the unobservable sources $s_1[i]$, ..., $s_n[i]$ are mixed and corrupted by additive zero mean noise to generate the observations $x_1[i]$, ..., $x_m[i]$. The target of BSS is to estimate an unmixing system to recover the original sources in $y_1[i]$, ..., $y_n[i]$.

## VI. Sparse Component Analysis: BSS and SDR

### A. Introduction

Recovery of the original source signals from their mixtures, without having a priori information about the sources and the way they are mixed, is called Blind Source Separation (BSS). This process is impossible if no assumption about the sources can be made. Such an assumption on the sources may be uncorrelatedness, statistical independence, lack of mutual information, or disjointness in some space [19], [20], [43].

The signal mixtures are often decomposed into their constituent principal components, independent components or are separated based on their disjoint characteristics described in a suitable domain. In the latter case the original sources should be sparse in that domain. Independent Component Analysis (ICA) is often used for separation of the sources in the former case whereas Sparse Component Analysis (SCA) is employed for the latter case. These two mathematical tools are described in the following sections followed by some results and illustrations of their applications.

### B. Independent Component Analysis (ICA)

The main assumption in ICA is the statistical independence of the constituent sources. Based on this assumption, ICA can play a crucial role in the separation and denoising of signals (BSS).

There has been recent research interest in the field of BSS due to its practicality in a wide range of problems. For example, BSS of acoustic signals measured in a room is often referred to as the Cocktail Party problem, which means separation of individual sounds from a number of recordings in an echoic and noisy environment. Figure 28 illustrates the BSS concept, wherein the mixing block represents the multipath propagation model between the original sources and the microphone measurements.

Generally, BSS algorithms make assumptions about the environment in order to make the problem more tractable. There are typically three assumptions about the mixing medium. The most simple but widely used case is the instantaneous case, where the source signals arrive at the sensors at the same time. This has been considered for separation of biological signals such as the EEG where the signals have narrow bandwidths and the sampling frequency is normally low [184]. The generative model for BSS in this case can be easily formulated as:

$$\mathbf{x}[i] = \mathbf{H} \cdot \mathbf{s}[i] + \boldsymbol{\nu}[i] \tag{66}$$

where $\mathbf{s}[i]$, $\mathbf{x}[i]$, and $\boldsymbol{\nu}[i]$ denote respectively the vector of source signals, size $n \times 1$, observed signals size $m \times 1$, and noise signals size $m \times 1$. $\mathbf{H}$ is the mixing matrix of size $m \times n$. Generally, the mixing process can be nonlinear (due to inhomogenity of the environment and that the medium can change with respect to the source signal variations; e.g. stronger vibration of a drum as a medium, with louder sound). However, in an instantaneous linear case where the above problems can be avoided or ignored, the separation is performed by means of a separating matrix, $\mathbf{W}$ of size $n \times m$, which uses only the information contained in $\mathbf{x}[i]$ to reconstruct the original source signals (or the independent components) as:

$$\mathbf{y}[i] = \mathbf{W} \cdot \mathbf{x}[i] \tag{67}$$

where $\mathbf{y}[i]$ is the estimate for the source signal $s[i]$. The early approaches in instantaneous BSS started from the work by Herault and Jutten [185] in 1986. In their approach, they considered non-Gaussian sources with equal number of independent sources and mixtures. They proposed a solution based on a recurrent artificial neural network for separation of the sources.

In the cases where the number of sources is known any ambiguity caused by false estimation of the number of sources can be avoided. If the number of sources is unknown, a criterion may be established to estimate the number of sources beforehand. In the context of model identification this is referred to as *Model Order Selection* and methods such as the Final Prediction Error (FPE), AIC, Residual Variance (RV), MDL and Hannan and Quinn (HNQ) methods [186] may be considered to solve this problem.

In acoustic applications, however, there are usually time lags between the arrival times of the signals at the sensors. The signals also may arrive through multiple paths. This type of mixing model is called a convolutive model [187]. The convolutive mixing model can also be classified into two subcategories: anechoic and echoic. In both cases the vector representations of mixing and separating processes are modified as $\mathbf{x}[i] = \mathbf{H}[i] * \mathbf{s}[i] + \boldsymbol{\nu}[i]$ and $\mathbf{y}[i] = \mathbf{W}[i] * \mathbf{x}[i]$, respectively, where $*$ denotes the convolution operation. In an anechoic model, however, the expansion of the mixing process

may be given as:

$$x_r[i] = \sum_{j=1}^{n} h_{r,j} s_j[i - \delta_{r,j}] + \nu_r[i], \quad \text{for } r = 1, \ldots, m \quad (68)$$

where the attenuation, $h_{r,j}$, and delay $\delta_{r,j}$ of source $j$ to sensor $r$ would be determined by the physical position of the source relative to the sensors. Then the unmixing process to estimate the sources will be given as:

$$y_j[i] = \sum_{r=1}^{m} w_{j,r} x_r[i - \delta_{j,r}], \quad \text{for } j = 1, \ldots, n \quad (69)$$

where the $w_{j,r}$'s are the elements of $\mathbf{W}$. In an echoic mixing environment it is expected that the signals from the same sources reach the sensors through multiple paths. Therefore the expansion of the mixing and separating models will be changed to

$$x_r[i] = \sum_{j=1}^{n} \sum_{l=1}^{L} h_{r,j}^l s_j[i - \delta_{r,j}^l] + \nu_r[i], \quad r = 1, \ldots, m \quad (70)$$

where $L$ denotes the maximum number of paths for the sources, $\nu_r[i]$ is the accumulated noise at sensor $r$, and $(.)^l$ refers to the $l^{th}$ path. The unmixing process will be formulated similarly to the anechoic one. For a known number of sources an accurate result may be expected if the number of paths is known; otherwise, the overall number of observations in an echoic case is infinite.

The aim of BSS using ICA is to estimate an unmixing matrix $\mathbf{W}$ such that $\mathbf{Y} = \mathbf{WX}$ best approximates the independent sources $\mathbf{S}$, where $\mathbf{Y}$ and $\mathbf{X}$ are respectively matrices with columns $\mathbf{y}[i] = [y_1[i], \ y_2[i], \ \ldots, \ y_n[i]]^T$ and $\mathbf{x}[i] = [x_1[i], \ x_2[i], \ \ldots, \ x_m[i]]^T$. Thus the ICA separation algorithms are subject to permutation and scaling ambiguities in the output components, i.e. $\mathbf{W} = \mathbf{PDH}^{-1}$, where $\mathbf{P}$ and $\mathbf{D}$ are the permutation and scaling (diagonal) matrices, respectively. Permutation of the outputs is troublesome in places where either the separated segments of the signals are to be joined together or when a frequency-domain BSS is performed.

Mutual information is a measure of independence and maximizing the non-Gaussianity of the source signals is equivalent to minimizing the mutual information between them [188].

In those cases where the number of sources is more than the number of mixtures (underdetermined systems), the above BSS schemes cannot be applied simply because the mixing matrix is not invertible, and generally the original sources cannot be extracted. However, when the signals are sparse, the methods based on disjointness of the sources in some domain may be utilized. Separation of the mixtures of sparse signals is potentially possible in the situation where, at each sample instant, the number of nonzero sources is not more than a fraction of the number of sensors (see Table II, row and column 6). The mixtures of sparse signals can also be instantaneous or convolutive.

### C. Sparse Component Analysis (SCA)

While the independence assumption for the sources is widely exploited in the design of BSS algorithms, the possible disjointness of the sources in some domain has not been considered. In SCA, this property is directly employed. Blind source separation by sparse decomposition has been addressed by Zibulevsky and Pearlmutter [189] for both overdetermined/exactly-determined and underdetermined systems using the maximum a posteriori approach. One way of formulating SCA is by representing the sources using a proper signal dictionary:

$$s_r[i] = \sum_{l=1}^{n} c_{r,l} \phi_l[i] \quad (71)$$

where $r = 1, \ldots, m$ and $n$ is the number of basis functions in the dictionary. The functions $\phi_l[i]$ are called atoms or elements of the dictionary. These atoms do not have to be linearly independent and may form an overcomplete dictionary. The sparsity property requires that only a small number of the coefficients $c_{r,l}$ differ significantly from zero. Based on this definition, the mixing and unmixing systems are modeled as follows:

$$\begin{aligned} \mathbf{x}[i] &= \mathbf{As}[i] + \boldsymbol{\nu}[i] \\ \mathbf{s}[i] &= \mathbf{C\Phi}[i] \end{aligned} \quad (72)$$

where $\boldsymbol{\nu}[i]$ is an $m \times 1$ vector. $\mathbf{A}$ and $\mathbf{C}$ can be determined by optimization of a cost function based on an exponential distribution for $c_{i,j}$ [189]. Figure 29 shows three separated sources using the above approach.

In places where the sources are sparse and at each time instant, at most one of the sources has significant nonzero value, the columns of the mixing matrix may be calculated individually, which makes the solution to the underdetermined case possible.

The SCA problem can be stated as a clustering problem since the lines in the scatter plot can be separated based on their directionalities by means of clustering. A number of works on this method have been reported [19], [191], [192]. In the work by Li et al. [192], the separation has been performed in two different stages. First, the unknown mixing matrix is estimated using the k-means clustering method. Then, the source matrix is estimated using a standard linear programming algorithm. The line orientation of a data set may be thought of as the direction of its greatest variance. One way is to perform eigenvector decomposition on the covariance matrix of the data, the resultant principal eigenvector, i.e., the eigenvector with the largest eigenvalue, indicates the direction of the data, since it has the maximum variance. In [191], GAP statistics as a metric which measures the distance between the total variance and cluster variances, has been used to estimate the number of sources followed by a similar method to Li's algorithm explained above. In line with this approach, Bofill and Zibulevsky [16] developed a potential function method for estimating the mixing matrix followed by $\ell_1$-norm decomposition for the source estimation. Local maxima of the potential function correspond to the estimated directions of the basis vectors. After the mixing matrix is identified, the sources have to be estimated. Even when $\mathbf{A}$ is known the solution is not unique. So, a solution is found for which the $\ell_1$-norm is minimized. Therefore, for $\mathbf{x}[i] = \sum \mathbf{a}_j s_j[i]$, $\sum_j |s_j|$ is minimized using linear programming.

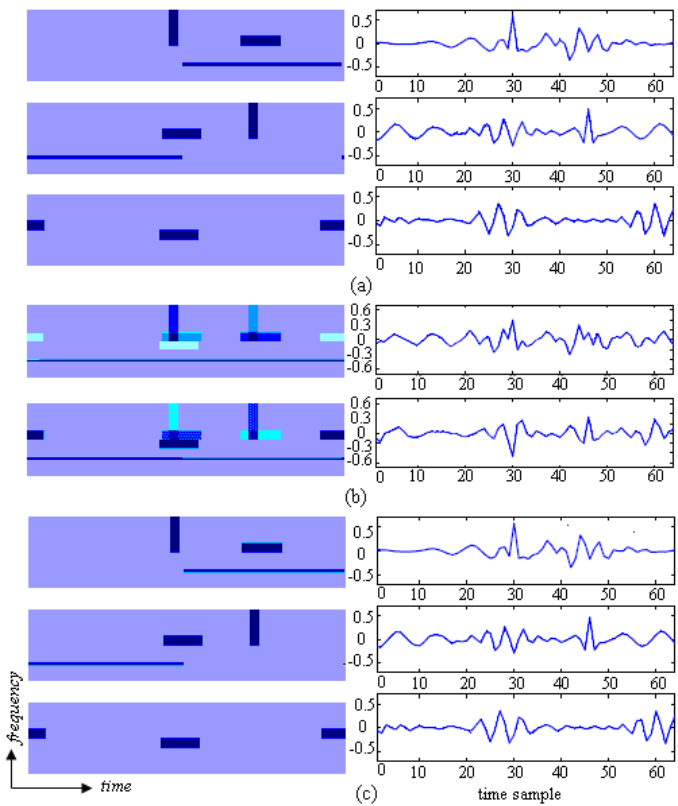

Fig. 29. Separation of sparse signals using the algorithm given in [190], left (a) sources, (b) mixtures, and (c) reconstructed sources, in both time-frequency left, and time, right.

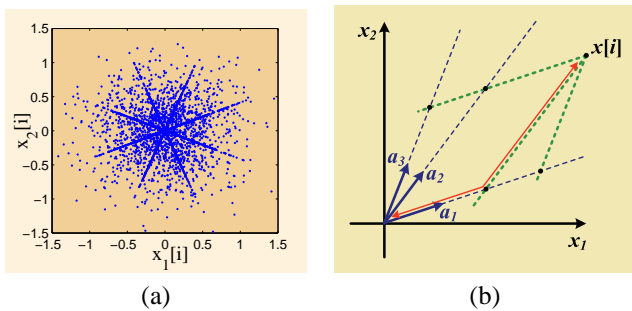

Fig. 30. (a) the scatter plot and (b) the shortest path from the origin to the data point, $x[i]$, extracted from [16].

Geometrically, for a given feasible solution, each source component is a segment of length $|s_j|$ in the direction of the corresponding $\mathbf{a}_j$ and, by concatenation, their sum defines a path from the origin to $x[i]$. Minimizing $\sum_j |s_j|$ amounts therefore to finding the shortest path to $x[i]$ over all feasible solutions $j = 1, \ldots, n$, where $n$ is the dimension of space of the independent basis vectors [19]. Figure 30 shows the scatter plot, polar plot, and the shortest path from the origin to the data point $x[i]$.

There are many cases for which the sources are disjoint in other domains, rather than the time-domain, or when they can be represented as sum of the members of a dictionary which can consist for example of wavelets or wavelet packets. In these cases the sparse component analysis can be performed in those domains more efficiently. Such methods often include transformation to time-frequency domain followed by a binary masking [193] or a BSS followed by binary masking [187]. One such approach, called DUET [193], transforms the anechoic convolutive observations into the time-frequency domain using a short-time Fourier transform and the relative attenuation and delay values between the two observations are calculated from the ratio of corresponding time-frequency points. The regions of significant amplitudes (atoms) are then considered to be the source components in the time-frequency domain. In this method only two mixtures have been considered and as a major limit of this method, only one source has been considered active at each time instant.

For instantaneous separation of sparse sources, the common approach used by most researchers is to attempt to maximize the sparsity of the extracted signals at the output of the separator. The columns of the mixing matrix $\mathbf{A}$ assign each observed data point to only one source based on some measure of proximity to those columns [194], i.e., at each instant only one source is considered active. Therefore the mixing system can be presented as:

$$x_r[i] = \sum_{j=1}^{n} a_{j,r} s_j[i] \quad , \quad r = 1, \ldots, m \qquad (73)$$

where in an ideal case $a_{j,r} = 0$ for $r \neq j$. Minimization of the $\ell_1$-norm is one of the most logical methods for estimation of sources as long as the signals can be considered sparse. $\ell_1$-norm minimization is a piecewise linear operation that partially assigns the energy of $\mathbf{x}[i]$ to the $m$ columns of $\mathbf{A}$ around $\mathbf{x}[i]$ in $\mathbb{R}^n$ space. The remaining $n - m$ columns are assigned zero coefficients, therefore the $\ell_1$-norm minimization can be manifested as:

$$\min \|\mathbf{s}[i]\|_{\ell_1} \quad \text{subject to} \quad \mathbf{A} \cdot \mathbf{s}[i] = \mathbf{x}[i] \qquad (74)$$

A detailed discussion of signal recovery using $\ell_1$-norm minimization is presented by Takigawa *et. al.* [195] and described below. As mentioned above, it is important to choose a domain that sparsely represents the signals.

On the other hand, in the method developed by Pederson *et. al.*, as applied to stereo signals, the binary masks are estimated after BSS of the mixtures and then applied to the microphone signals. The same technique has been used for convolutive sparse mixtures after the signals are transformed to the frequency domain.

In another approach [196] the effect of outlier noise has been reduced using median filtering then hybrid fast ICA filtering, and $\ell_1$-norm minimization have been used for separation of temporomandibular joint sounds. It has been shown that for such sources, this method outperforms both the Degenerate Unmixing Estimation Technique (DUET) and Li's algorithms. The authors of [197] have recently extended the DUET algorithm to separation of more than two sources in an echoic mixing scenario in the time-frequency domain.

TABLE VII

SCA STEPS

| |
|---|
| 1) Consider the model $\mathbf{x} = \mathbf{A} \cdot \mathbf{s}$, we need a linear transformation that applies to both sides of the equation to yield a new sparse source vector. |
| 2) Estimate the mixing matrix $\mathbf{A}$. Several approaches are presented for this step, such as natural gradient ICA approaches, and clustering techniques with variants of k-means algorithm [199], [19]. |
| 3) Estimate the source representation based on the sparsity assumption. A majority of proposed methods are primarily based on minimizing some norm or pseudo-norm of the source representation vector. The most effective approaches are Matching Pursuit [199], [200], Basis Pursuit, [189], [201], [78], [202]], FOCUSS [203], IDE [66] and Smoothed $\ell_0$-norm [204]. |

In a very recent approach it has been considered that brain signal sources in the space-time-frequency domain are disjoint. Therefore, clustering the observation points in the space-time-frequency-domain can be effectively used for separation of brain sources [198].

As can be seen, generally, BSS exploits independence of the source signals whereas SCA benefits from the disjointness property of the source signals in some domain. While the BSS algorithms mostly rely on ICA with statistical properties of the signals, SCA uses their geometrical and behavioral properties. Therefore, in SCA, either a clustering approach or a masking procedure can result in estimation of the mixing matrix. Often, an $\ell_1$-norm is used to recover the source signals. Generally, in places where the source signals are sparse, the SCA methods often result in more accurate estimation of the signals with less ambiguities in the estimation.

### D. SCA Algorithms

There are three main steps for the solution of an SCA problem as shown in Table VII [199]. The first step of Table VII shows a linear model for the SCA problem, the second step consists of estimating the mixing matrix $\mathbf{A}$ using sparsity information, and finally the third step is to estimate the sparse source representation based on the estimate of $\mathbf{A}$.

In the following, we present a survey of major approaches that are suggested for the third step.

*1) Matching Pursuit:* Mallat and Zhang have developed a general iterative method for approximating sparse decomposition [200]. When the dictionary is orthogonal and the signal $x$ is composed of $k \ll n$ atoms, the algorithm recovers the sparse decomposition exactly after $n$ steps. As we will see, the algorithm is greedy [205]. Since the algorithm is myopic, in some certain cases, wrong atoms are chosen in the first few iterations, and thus the remaining iterations are spent on correcting the first few mistakes. The algorithm is shown in Table VIII.

*2) Basis Pursuit:* The mathematical representation of counting the number of sparse components is denoted by $\ell_0$. However, $\ell_0$ is not a proper norm and is not computationally tractable. The closest convex norm to $\ell_0$ is $\ell_1$. The $\ell_1$ optimization of an over complete dictionary is called Basis Pursuit.

TABLE VIII

MATCHING PURSUIT ALGORITHM

| |
|---|
| 1) Let $\mathbf{x}^{(0)} = \mathbf{0}_{n \times 1}$, $\mathbf{r}^{(0)} = \mathbf{x}$ and $i = 1$. |
| 2) Find index $\gamma_i$ such that $\langle \mathbf{r}^{(i-1)}, \mathbf{a}_{\gamma_i} \rangle$ is maximum where $\mathbf{a}_j$ corresponds to columns of the mixing matrix $\mathbf{A}$ (atoms). |
| 3) Let $s_i = \langle \mathbf{r}^{(i-1)}, \mathbf{a}_{\gamma_i} \rangle$, $\mathbf{x}^{(i)} = \mathbf{x}^{(i-1)} + s_i \mathbf{a}_{\gamma_i}$ and $\mathbf{r}^{(i)} = \mathbf{x} - \mathbf{x}^{(i)}$. |
| 4) If the last condition is not satisfied, increase $i$ and return to step 2. |

TABLE IX

RELATION BETWEEN LP AND BASIS PURSUIT (THE NOTATION FOR LINEAR PROGRAMMING IS FROM [207].)

| Basis Pursuit | Linear Programming |
|:---:|:---:|
| $m$ | $2p$ |
| $\mathbf{s}$ | $\mathbf{x}$ |
| $(1, \ldots, 1)_{1 \times m}$ | $\mathbf{C}$ |
| $\pm \mathbf{A}$ | $\mathbf{A}$ |
| $\mathbf{x}$ | $\mathbf{b}$ |

However the $\ell_1$-norm is non-differentiable and we cannot use gradient methods for optimal solutions [206]. On the other hand, the $\ell_1$ solution is stable due to its convexity (the global optimum is the same as the local one) [21].

Formally, the Basis Pursuit can be formulated as:

$$\min \|\mathbf{s}\|_{\ell_1} \quad \text{s.t.} \quad \mathbf{x} = \mathbf{A} \cdot \mathbf{s} \qquad (75)$$

We now explain how the Basis Pursuit is related to Linear Programming (LP). The standard form of linear programming is a constrained optimization problem defined in terms of variable $\mathbf{x} \in \mathbb{R}^n$ by:

$$\min \mathbf{C}^T \mathbf{x} \quad \text{s.t.} \quad \mathbf{A}\mathbf{x} = \mathbf{b}, \quad \forall i : \ x_i \geq 0 \qquad (76)$$

where $\mathbf{C}^T \mathbf{x}$ is the objective function, $\mathbf{A}\mathbf{x} = \mathbf{b}$ is a set of equality constraints and $\forall i : \ x_i \geq 0$ is a set of bounds. Table IX shows this relationship. Thus, the solution of (75) can be obtained by solving the equivalent LP. There are two major approaches to solve LP: 1) Interior Point methods and 2) Simplex algorithms, depending on whether we solve the cost function and then check whether it satisfies the constraint bounds or vice versa.

*3) FOCal Underdetermined System Solver (FOCUSS):* FOCUSS is a non-parametric algorithm that consists of two parts [203]. It starts by finding a low resolution estimation of the sparse signal, and then pruning this solution to a sparser signal representation through several iterations. The solution at each iteration step is found by taking the pseudo-inverse of a modified weighted matrix. The pseudo-inverse of the modified weighted matrix is defined by $(\mathbf{A}\mathbf{W})^+ = (\mathbf{A}\mathbf{W})^H (\mathbf{A}\mathbf{W} \cdot (\mathbf{A}\mathbf{W})^H)^{-1}$. This iterative algorithm is the solution of the following optimization problem:

$$\text{Find } \mathbf{s} = \mathbf{W}\mathbf{q}, \text{ where: } \min \|\mathbf{q}\|_{\ell_2} \text{ s.t. } \mathbf{x} = \mathbf{A}\mathbf{W}\mathbf{q} \qquad (77)$$

Description of this algorithm is given in Table X and an extended version is discussed in [203].

TABLE X
FOCUSS (BASIC)

- Step 1: $\mathbf{W}_{p_i} = diag(\mathbf{s}_{i-1})$
- Step 2: $\mathbf{q}_i = (\mathbf{A}\mathbf{W}_{p_i})^+ \mathbf{x}$
- Step 3: $\mathbf{s}_i = \mathbf{W}_{p_i} \cdot \mathbf{q}_i$

TABLE XI
IDE STEPS

- Detection Step: Find indices of inactive sources:

$$I = \{1 \le i \le m: \ |\mathbf{a}_i^T \cdot \mathbf{x} - \sum_{j \ne i}^{m} \hat{s}_j^l \mathbf{a}_i^T \cdot \mathbf{a}_j| < \epsilon^l\}$$

- Estimation Step: Find the following projection as the new estimate:

$$\mathbf{s}^{l+1} = \mathrm{argmin}_{\mathbf{s}} \sum_{i \in I}^{l} s_i^2 \ \text{s.t.} \ \mathbf{x}(t) = \mathbf{A} \cdot \mathbf{s}(t)$$

The solution is derived from Karush-Kuhn-Tucker system of equations. At the $l + 1^{th}$ iteration:

$$\mathbf{s}_i = \mathbf{A}_i^T \cdot \mathbf{P}(\mathbf{x} - \mathbf{A}_a \cdot \mathbf{s}_a)$$
$$\mathbf{s}_a = (\mathbf{A}_a^T \mathbf{P} \mathbf{A}_a)^{-1} \mathbf{A}_a^T \mathbf{P} \cdot \mathbf{x}$$

where the matrices and vectors are partitioned into inactive/active parts as $\mathbf{A}_i, \mathbf{A}_a, \mathbf{s}_i, \mathbf{s}_a$ and $\mathbf{P} = (\mathbf{A}_i \mathbf{A}_i^T)^{-1}$

- stop after a fixed number of iterations.

*4) Iterative Detection and Estimation (IDE):* The idea behind this method is based on a geometrical interpretation of the sparsity. Consider the elements of vector $\mathbf{s}$ are i.i.d. random variables. By plotting a sample distribution of vector $\mathbf{s}$, which is obtained by plotting a large number of samples in the $S$-space, it is observed that the points tend to concentrate first around the origin, then along the coordinate axes, then across the coordinate planes. The algorithm used in IDE is given in Table XI. In this table, $\mathbf{s}_i$'s are the inactive sources, $\mathbf{s}_a$'s are the active sources, $\mathbf{A}_i$ is the column of $\mathbf{A}$ corresponding to the inactive $\mathbf{s}_i$ and $\mathbf{A}_a$ is the column of A corresponding to the active $\mathbf{s}_a$. Notice that IDE has some resemblances to the RDE method discussed in Sec. III-A.2, IMAT mentioned in Sec. III-A.2, and MIMAT explained in Sec. VII-A.2.

*5) Smoothed $\ell_0$-norm (SL0) Method:* As discussed earlier, the criterion for sparsity is the $\ell_0$-norm; thus our minimization is

$$\min \|\mathbf{s}\|_{\ell_0} \quad \text{s.t.} \quad \mathbf{A} \cdot \mathbf{s} = \mathbf{x} \tag{78}$$

The $\ell_0$-norm has two major drawbacks: the need for a combinatorial search, and its sensitivity to noise. These problems arise from the fact that the $\ell_0$-norm is discontinuous. The idea of SL0 is to approximate the $\ell_0$-norm with functions of type [204]:

$$f_\sigma(s) \triangleq e^{-\frac{s^2}{2\sigma^2}} \tag{79}$$

TABLE XII
SL0 STEPS

- Initialization:
  1) Set $\hat{\mathbf{s}}_0$ equal to the minimum $\ell_2$-norm solution of $\mathbf{A}\mathbf{s} = \mathbf{x}$, obtained by pseudo-inverse of $\mathbf{A}$.
  2) Choose a suitable decreasing sequence for $\sigma$, $[\sigma_1, \ldots, \sigma_K]$.
- For $i = 1, \ldots, K$:
  1) Set $\sigma = \sigma_i$.
  2) Maximize the function $F_\sigma$ on the feasible set $\mathcal{S} = \{\mathbf{s} | \mathbf{A}\mathbf{s} = \mathbf{x}\}$ using $L$ iterations of the steepest ascent algorithm (followed by projection onto the feasible set):
     - Initialization: $\mathbf{s} = \hat{\mathbf{s}}_{i-1}$.
     - for $j = 1, \ldots, L$ (loop $L$ times):
       a) Let: $\Delta\mathbf{s} = [s_1 e^{-\frac{s_1^2}{2\sigma^2}}, \ldots, s_n e^{-\frac{s_n^2}{2\sigma^2}}]^T$.
       b) Set $\mathbf{s} \leftarrow \mathbf{s} - \mu\Delta\mathbf{s}$ (where $\mu$ is a small positive constant).
       c) Project $\mathbf{s}$ back onto the feasible set $\mathcal{S}$:
          $$\mathbf{s} \leftarrow \mathbf{s} - \mathbf{A}^T (\mathbf{A}\mathbf{A}^T)^{-1} (\mathbf{A}\mathbf{s} - \mathbf{x})$$
  3) Set $\hat{\mathbf{s}}_i = \mathbf{s}$.
- Final answer is $\hat{\mathbf{s}} = \hat{\mathbf{s}}_K$

where $\sigma$ is a parameter which determines the quality of the approximation. Note that we have

$$\lim_{\sigma \to 0} f_\sigma(s) = \begin{cases} 1 & \text{if } s = 0 \\ 0 & \text{if } s \ne 0 \end{cases} \tag{80}$$

For the vector $\mathbf{s}$, we have $\|\mathbf{s}\|_0 \approx n - F_\sigma(s)$, where $F_\sigma(s) = \sum_{i=1}^{n} f_\sigma(s_i)$. Now minimizing $\|\mathbf{s}\|_0$ is equivalent to maximizing $F_\sigma(s)$ for some appropriate values of $\sigma$. For small values of $\sigma$, $F_\sigma(s)$ is highly non-smooth and contains many local maxima, and therefore its maximization over $\mathbf{A} \cdot \mathbf{s} = \mathbf{x}$ may not be global. On the other hand, for larger values of $\sigma$, $F_\sigma(s)$ is a smoother function and contains fewer local maxima, and its maximization may be possible (in fact there is no local maxima for large values of $\sigma$ [204]). Hence we use a decreasing sequence for $\sigma$ in the steepest ascent algorithm and may escape from getting trapped into local maxima and reach the actual maximum for small values of $\sigma$, which gives the minimum $\ell_0$-norm solution. The algorithm is summarized in Table XII.

*6) Comparison of Different Techniques:* The above techniques have been simulated and the results are depicted in Fig. 31. In order to compare the efficiency and computational complexity of these methods; we use a fixed synthetic mixing matrix and source vectors. The elements of the mixing matrix are obtained from zero mean independent Gaussian random variables with variance $\sigma = 1$. Sparse sources have been artificially generated using a Bernoulli-Gaussian model: $s_i = p \ N(0, \sigma_{on}) + (1 - p) \ N(0, \sigma_{off})$. We set $\sigma_{off} = 0.01$, $\sigma_{on} = 1$ and $p = 0.1$. Then, we compute the noisy mixture vector $\mathbf{x}$ from $\mathbf{x} = \mathbf{A}\mathbf{s} + \boldsymbol{\nu}$, where $\boldsymbol{\nu}$ is the noise vector. The elements of the vector $\boldsymbol{\nu}$ are generated according to independent zero mean Gaussian random variables with variance $\sigma_\nu^2$. We use Orthogonal Matching Pursuit (OMP) which is a variant of Matching Pursuit [200]. OMP has a better performance in estimating the source vector in comparison to

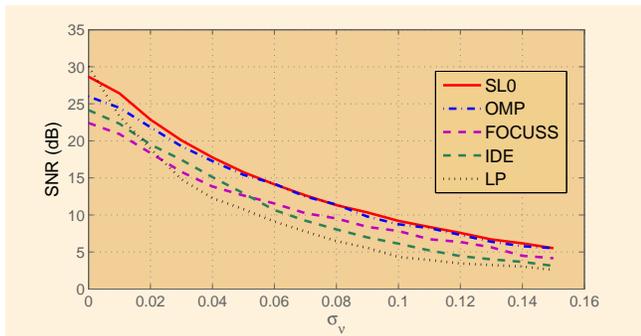

Fig. 31. Performance of various methods with respect to the standard deviation.

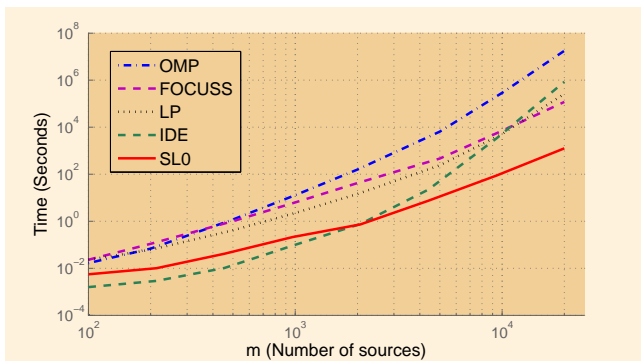

Fig. 32. Computational time (Complexity) versus the number of sources.

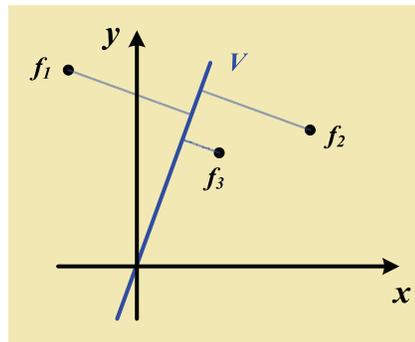

Fig. 33. Linear least squares: $e = d^2(f_1, V) + d^2(f_2, V) + d^2(f_3, V)$.

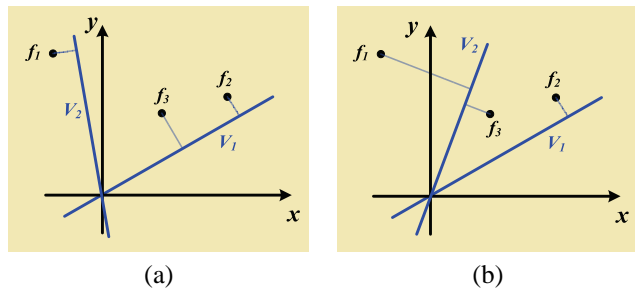

Fig. 34. Objective function: (a) $e = d^2(f_1, V_2) + d^2(f_2, V_1) + d^2(f_3, V_1)$ and (b) $e = d^2(f_1, V_2) + d^2(f_2, V_2) + d^2(f_3, V_1)$. Configuration of $V_1, V_2$ in (a) creates the partition $P_1 = \{f_1\}$ and $P_2 = \{f_2, f_3\}$ while the configuration in (b) causes the partition $P_1 = \{f_1, f_2\}$ and $P_2 = \{f_3\}$.

Matching Pursuit. Fig. 32 demonstrates the time needed for each algorithm to estimate the vector **s** with respect to the number of sources. This figure shows that IDE and SL0 have the lowest complexity.

### E. Sparse Dictionary Representation (SDR) and Signal Modeling

A signal $\mathbf{x} \in \mathbb{R}^n$ may be sparse in a given basis but not sparse in a different basis. For example, an image may be sparse in a wavelet basis (i.e., most of the wavelet coefficients are small) even though the image itself may not be sparse (i.e., many of the gray values of the image are relatively large). Thus, given a class $\mathcal{S} \subset \mathbb{R}^n$, an important problem is to find a basis or a frame in which all signals in $\mathcal{S}$ can be represented sparsely. More specifically, given a class of signals $\mathcal{S} \subset \mathbb{R}^n$, it is important to find a basis (or a frame) $D = \{w_j\}_{j=1}^d$ (if it exists) for $\mathbb{R}^n$ such that every data vector $\mathbf{x} \in \mathcal{S}$ can be represented by at most $k \ll n$ linear combinations of elements of $D$. The dictionary design problem has been addressed in [19], [20], [21], [81], [84], [95], [208]. A related problem is the signal modeling problem in which the class $\mathcal{S}$ is to be modeled by a union of subspaces $\mathcal{M} = \bigcup_{i=1}^l V_i$ where each $V_i$ is a subspace of $\mathbb{R}^n$ with a dimension of $V_i \leq k$ where $k \ll n$ [43]. If the subspaces $V_i$ are known, then it is possible to pick a basis $E^i = \{e_j^i\}_j$ for each $V_i$ and construct a dictionary $D = \bigcup_{i=1}^l E^i$ in which every signal of $\mathcal{S}$ has sparsity $k$ (or is almost $k$ sparse). The model $\mathcal{M} = \bigcup_{i=1}^l V_i$ can be found from an observed set of data $F = \{f_1, \ldots, f_m\} \subset \mathcal{S}$ by solving (if possible) the following non-linear least squares problem:

Find subspaces $V_1, \ldots, V_l$ of $\mathbb{R}^n$ that minimize the expression

$$e\big(F, \{V_1, \ldots, V_l\}\big) = \sum_{i=1}^m \min_{1 \leq j \leq l} d^2(f_i, V_j) \quad (81)$$

over all possible choices of $l$ subspaces with $\dim V_i \leq k < N$. Here $d$ denotes the Euclidian distance in $\mathbb{R}^n$ and $k$ is an integer with $1 \leq k < n$ for $i = 1, \ldots, l$. Note that $e\big(F, \{V_1, \ldots, V_l\}\big)$ is calculated as follows: for each $f_i \in F$ and fixed $\{V_1, \ldots, V_l\}$, the subspace $V_i \in \{V_1, \ldots, V_l\}$ closest to $f_i$ is found and the distance $d^2(f_i, V_j)$ is computed. This process is repeated for all $f_i \in F$ and the squares of the distances are added together to find $e(F, \{V_1, \ldots, V_l\})$. The optimal model is then obtained as the union $\mathcal{M} = \bigcup_i V_i^o$, where $\{V_1^o, \ldots, V_l^o\}$ minimize the expression (81). When $l = 1$ this problem reduces to the classical least squares problem. However, when $l > 1$ the set $\bigcup_i V_i$ is a nonlinear set and the problem is fully non-linear (see Figs. 34 and 35). A more general nonlinear least squares problem has been studied for finite and infinite Hilbert spaces [43]. In that general setting, the existence of solutions is proved and a meta-algorithm for searching for the solution is described.

For the special finite dimensional case of $\mathbb{R}^n$ in (81), the search algorithm is an iterative algorithm that alternates between data partition and the optimization of a simpler least squares problem. The search algorithm can be summarized as in Table XIII.

The above algorithm, which is similar to k-means, consists of two parts [209]: 1) an initialization; and 2) an iterative

TABLE XIII
Search Algorithm

- **Input**:
  - initial partition $\{F_l^1, \ldots, F_l^1\}$
  - Data set $\mathcal{F}$
- **Iterations**:
  1) Use the SVD to find $\{V_1^1, \ldots, V_l^1\}$ by minimizing $e(F_i^1, V_i)$ for each $i$, and compute $\Gamma_1 = \sum_i e(F_i^1, V_i^1)$;
  2) Set $j = 1$;
  3) **While** $\Gamma_j = \sum_i e(F_i^j, V_i^j) > e(\mathcal{F}, \{V_1^j, \ldots, V_l^j\})$
  4) Choose a new partition $\{F_1^{j+1}, \ldots, F_l^{j+1}\}$ that satisfies, $f \in \mathcal{F}_k^{j+1}$ implies that $d(f, V_k^j) \leq d(f, V_h^j), h = 1, \ldots, l$;
  5) Use SVD to find and choose $\{V_1^{j+1}, \ldots, V_l^{j+1}\}$, by minimizing $e(F_i^{j+1}, V_i)$ for each $i$, and compute $\Gamma_{j+1} = \sum_i e(F_i^{j+1}, V_i^{j+1})$;
  6) Increment $j$ by 1, i.e., $j \to j + 1$;
  7) **End while**
- **Output**:
  - $P_j$ and $V_{P_j}$.

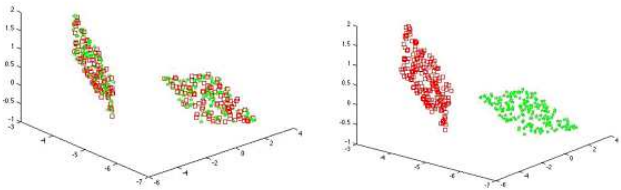

Fig. 35. Data $F$ belongs to two planes in $\mathbb{R}^3$. The algorithm uses a random initial partition in left hand side, and produces the final partition and optimal subspaces in the right hand side.

search algorithm. The initialization is a Hough-like transform that partitions the data set $F$ into $l$ classes $\{F_1^1, \ldots, F_l^1\}$, such that each class $F_i^1$ consists of vectors that are close to the same (yet unknown) subspace $V_i^1$. For each partition $F_i^1$ we find the subspace $V_i^1$ closest to the vectors in $F_i^1$ among all subspaces $V \subset \mathbb{R}^n$ of dimension dim$V \leq k$. Finding the subspace $V_i^1$ closets to the vectors in $F_i^1$ for a fixed $i$ is the classical linear least squares problem and can be found exactly using a Singular Value Decomposition. This initialization process produces a first approximation $\{V_1^1, \ldots, V_l^1\}$ of the optimal spaces $\{V_1^o, \ldots, V_l^o\}$ minimizing (81). We now use the subspaces $\{V_1^1, \ldots, V_l^1\}$ to find a new partition of the data $F_1^2, \ldots, F_l^2$ in such a way that the vectors in $F_i^2$ consist of all the vectors in $F$ that are closest to $V_i^1$, $i = 1, \ldots, l$. Using this new partition $F_1^2, \ldots, F_l^2$ of $F$, we find (for each $i$) the subspace $V_i^2$ closest to the vectors in $F_i^2$ using SVD as before. We continue this process until the value of $e$ is unchanged between two consecutive iterations. This is a local minimum which is also likely to be a global one (see Fig. 35).

## VII. Multipath Channel Estimation

In wireless systems, channel estimation is required for the compensation of channel distortions. The transmitted signal bounces off different objects and arrives at the receiver from multiple paths. This phenomenon causes the received signal to be a mixture of reflected and scattered versions of the transmitted signal. The mobility of the transmitter, receiver, and scattering objects results in rapid changes in the channel response, and thus the channel estimation process becomes more complicated. Due to the sparse distribution of scattering objects, a multipath channel is sparse in the time domain as shown in Fig. 36. By taking sparsity into consideration, channel estimation can be simplified and/or made more accurate. The sparse time varying multipath channel is modeled as:

$$h(t, \tau) = \sum_{l=0}^{k-1} \alpha_l(t) \delta(t - \tau) \tag{82}$$

where $k$ is the number of taps, $\alpha_l$ is the $l^{th}$ complex path gain, and $\tau_l$ is the corresponding path delay. At time $t$, the transfer function is given by:

$$H(t, f) = \int_{-\infty}^{+\infty} h(t, \tau) e^{-j2\pi f \tau} d\tau \tag{83}$$

The estimation of the multipath channel impulse response is very much similar to the determination of analog epochs and amplitudes of discontinuity for finite rate of innovation as shown in section II-B.4 Fig. 8 and (19). Essentially, if a known train of impulses is transmitted and the received signal from the multipath channel is filtered and sampled (information domain as shown in Fig. 8 -rate of innovation), the channel impulse response can be estimated from these samples using an annihilating filter (the Prony method [28]) defined in the $\mathcal{Z}$-transform and a pseudo-inverse matrix inversion, in principle[18]. Once the channel impulse response is estimated, its effect is compensated; this process can be repeated according to the dynamics of the time varying channel.

A special case of multipath channel is an OFDM channel, which is widely used in ADSL, DAB, DVB, WLAN, WMAN, and WIMAX[19]. OFDM is a digital multi-carrier transmission technique where a single data stream is distributed over several sub-carrier frequencies to achieve robustness against multipath channels besides many other advantages. Channel estimation in OFDM can be easier than for other modulation schemes; the channel impulse response is now quantized[20] and instead of an annihilating filter defined in the $\mathcal{Z}$-transform, we can use DFT and ELP of section III-A. Also, instead of a known train of impulses, some of the available sub-carriers of OFDM in each transmitting block are assigned to predetermined patterns, which are usually called comb-type pilots. These pilot tones help the receiver to extract some of the DFT samples of the discrete time varying channel (82) at the respective frequencies in each transmitting block. These characteristics make the OFDM channel estimation similar to unknown sparse signal recovery of section II-A.1 and the impulsive noise removal of section III-A.2. Because of these advantages, our main example and simulations are related to OFDM channel estimation.

---

[18]Similar to Pisarenko method for spectral estimation [28].

[19]These acronyms are defined in Table XV at the end of the paper.

[20]For simplicity, we assume a time-quantized channel impulse response which works quite well. For an accurate channel estimation of OFDM channel with AWGN, we can use fractional time delays.

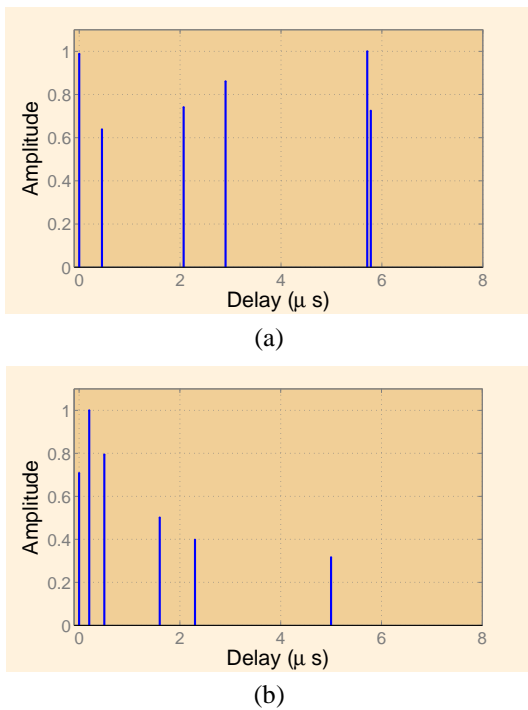

(a)

(b)

Fig. 36. The impulse response of two typical multipath channels; (a) Brazil-D and (b) TU6 channel profiles.

### A. OFDM Channel Estimation

For OFDM, the discrete version of the time varying channel of (83) in the frequency domain becomes

$$H[r,i] \triangleq H(rT_f, i\Delta f) = \sum_{l=0}^{k-1} h[r,l]e^{-\frac{j2\pi il}{n}} \tag{84}$$

where

$$h[r,l] = h(rT_f, lT_s) \tag{85}$$

where $T_f$ and $n$ are the symbol length and number of sub-carriers in each OFDM symbol, respectively. $\Delta f$ is the sub-carrier spacing, and $T_s = \frac{1}{\Delta f}$ is the sample interval. The above equation shows that for the $r^{th}$ OFDM symbol, $H[r,i]$ is the DFT of $h[r,l]$.

Two major methods are used in the equalization process: 1) zero forcing and 2) MMSE. In the zero forcing method, regardless of the noise variance, equalization is obtained by dividing the received OFDM symbol by the estimated channel frequency response; while in the MMSE method, the approximation is chosen such that the MSE of the transmitted data vector $\left( E\left[ \|\mathbf{X} - \hat{\mathbf{X}}\|^2 \right] \right)$ is minimized, which introduces the noise variance in the equations.

*1) Statement of the Problem:* The goal of the channel estimation process is to obtain the channel impulse response from the noisy values of the channel transfer function in the pilot positions. This is equivalent to solving the following equation for $\mathbf{h}$ which is also shown graphically in Fig. 37.

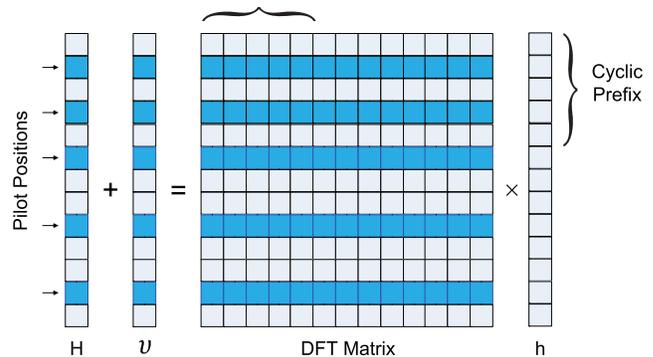

Fig. 37. Graphical representation of the sparse problem involved in OFDM channel estimation

$$\mathbf{H}_{i_p} = \mathbf{F}_{i_p}\mathbf{h} + \boldsymbol{\nu}_{i_p} \tag{86}$$

where $i_p$ is an index vector denoting the pilot positions in the frequency spectrum, $\mathbf{H}_{i_p}$ is a vector containing the value of the channel frequency spectrum in these pilot positions and $\mathbf{F}_{i_p}$ denotes the matrix obtained from taking the rows of the DFT matrix pertaining to the pilot positions. $\boldsymbol{\nu}_{i_p}$ is the additive noise on the pilot points in the frequency domain. Thus, the channel estimation problem is equivalent to finding the sparse vector $\mathbf{h}$ from the above set of equations for a set of pilots. Various channel estimation methods [210] have been used with the usual tradeoffs of optimality and complexity. The Least Square [210], ML [211], [212], Minimum Mean Square Error (MMSE) [213], [214], [215], and LMMSE [213], [211], [216] techniques are among some of these methods. But none of these techniques use the inherent sparsity of the multipath channel $\mathbf{h}$, and thus, they are not as accurate.

*2) Sparse OFDM Channel Estimation:* In the following, we present two methods that utilize this sparsity to enhance the channel estimation process.

*CS Based Channel Estimation:* Recently the idea of using time-domain sparsity in OFDM channel estimation has been proposed by [217], [218], [219]. The use of sparsity decreases the channel estimation error and hence the number of required pilots (overhead), thus increasing the bandwidth efficiency. In [217], the authors proposed to use CS for OFDM channel estimation and proved that, in case of uniform pilot insertion, the OFDM channel estimation problem satisfies the Restricted Isometric Property (RIP) described in section II-B, and thus LP-based algorithms similar to the ones discussed in section VI-D.2 can be used for channel estimation. Simulation results show that this method works effectively even in fast time varying channels. Furthermore, from (7), a lower bound on the number of pilots can be obtained for a given number of channel taps (sparsity).

However, the authors of [217] did not consider zero padding at the endpoints of the OFDM bandwidth in their scenario which is an essential part of the current standards based on OFDM transmission. This assumption causes the matrix $\mathbf{F}_{i_p}$ defined in (86) to be ill-conditioned and thus, the RIP condition defined in (9) is not satisfied. Also we do not have any pilots in the zero padded parts which complicates the channel estimation techniques. In [24], we propose a method that exploits the inherent sparsity and also solves the zero padding problem. This algorithm is briefly discussed in the following.

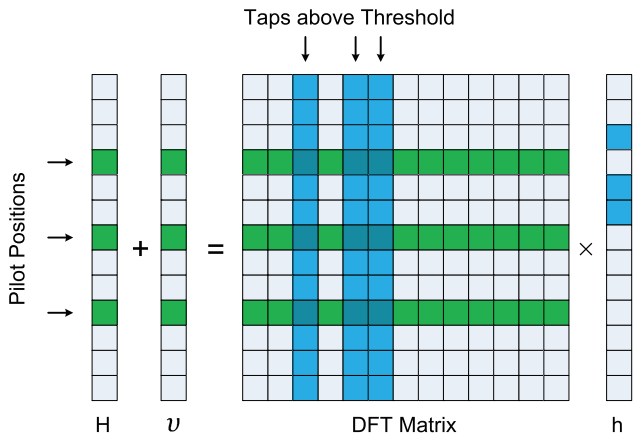

Fig. 38. Graphical representation of the equation involved in obtaining the tap values above the threshold

*Modified IMAT (MIMAT) for OFDM Channel Estimation:* In this method, the spectrum of the channel is initially estimated using a simple interpolation such as linear interpolation between pilot sub-carriers. This initial estimate is further improved in a series of iterations between time (sparse) and frequency (information) domains to find the sparsest channel impulse response by using an adaptive thresholding scheme; in each iteration, after finding the location of the taps (locations with previously estimated amplitudes higher than the threshold), their respective amplitudes are again found using the MMSE criterion. In each iteration, due to thresholding, some of the false taps that are noise samples with amplitudes above the threshold are discarded. Thus, the new iteration starts with a lower number of false taps. Moreover, because of the MMSE estimator, the valid taps approach their actual values in each new iteration. In the last iteration, the actual taps are detected and the MMSE estimator gives their respective values. This method is similar to RDE and IDE methods discussed in section III-A.2 and VI-D.4.

Table XIV summarizes the steps in the MIMAT algorithm. In the threshold of the MIMAT algorithm, $\alpha$ and $\beta$ are constants which depend on the number of taps and initial powers of noise and channel impulses. In the first iteration, the threshold is a small number and with each iteration it is gradually increased. Intuitively, this gradual increase of the threshold with the iteration number, results in a gradual reduction of false taps (taps that are created due to noise). In each iteration, the tap values are obtained from:

$$\hat{\mathbf{H}}_{LSi_p} = \mathbf{H}_{i_p} + \boldsymbol{\nu}_{i_p} = \tilde{\mathbf{F}} \cdot \mathbf{h}_t \qquad (87)$$

where $t$ denotes the index of nonzero impulses obtained from the previous step and $\tilde{\mathbf{F}}$ is obtained from $\mathbf{F}_{i_p}$ by keeping the columns determined by $t$. A graphical representation of the above equation is given in Fig. 38. The amplitudes of

- **Initialization**:
  - Find an initial estimate of the time domain channel using linear interpolation: $\hat{h}^{(0)} = \hat{h}_{linear}$
- **Iterations**:
  1) Set Threshold=$\beta e^{\alpha i}$.
  2) Using the threshold from the previous step, find the locations of the taps $t$ by thresholding the time domain channel from the previous iteration($\tilde{h}^{(i-1)}$).
  3) Solve for the value of the non-zero impulses using MMSE:
  $$\hat{h}_t = SNR \cdot \tilde{\mathbf{F}}^H (\tilde{\mathbf{F}} \cdot SNR \cdot \tilde{\mathbf{F}}^H + \mathbf{I})^{-1} \quad (88)$$
  4) Find the new estimate of the channel ($\hat{h}^{(i)}$) by substituting the taps in their detected positions.
  5) Stop if the estimated channel is the same as the previous iteration or when a maximum number of iterations is reached.

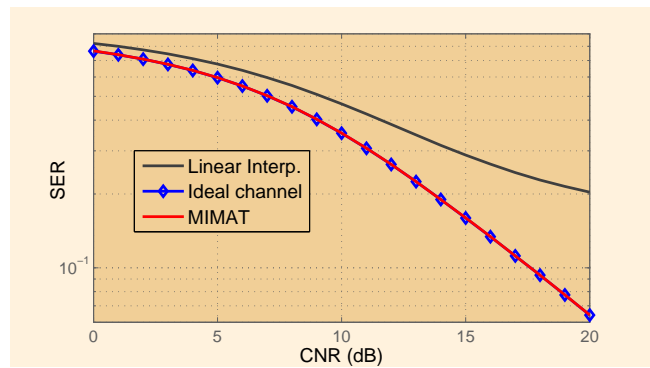

Fig. 39. SER (Symbol Error Rate) vs. CNR (Carrier to Noise Ratio) for the ideal channel, linear interpolation, and the MIMAT for the Brazil channel.

nonzero impulses can be obtained from simple iterations, pseudo-inverse, or the MMSE equation (88) of Table XIV that yields better results under additive noise environments.

The equation that has to be solved in (87) is usually overdetermined which helps the suppression of the noise in each iteration step. Note that the solution presented in (88), represents a variant of the MMSE solution when the location of discrete impulses are known. If further statistical knowledge is available, this solution can be modified and a better estimation is obtained; however, this makes the approximation process more complex. This algorithm does not need many steps of iterations; the positions of the non-zero impulses are perfectly detected in 3 or 4 iterations for most types of channels.

### B. Simulation Results and Discussions

For OFDM simulations, the DVB-H standard was used with the 16-QAM constellation in the 2K mode ($2^{11}$ FFT size). The channel profile was the Brazil channel D. Fig. 39 shows the Symbol Error Rate (SER) versus the Carrier-to-Noise Ratio (CNR) after equalizing using the proposed MIMAT algorithm and the standard linear interpolation in the frequency domain using the noisy pilot samples. As can be seen in Fig. 39, the SER obtained from the MIMAT algorithm coincides with the one obtained from the hypothetical ideal channel (where the exact channel frequency response is used for equalization). Thus, in this sense the proposed channel estimation is perfect in time invariant channels. Also, this figure shows that the standard linear interpolation method performs poorly compared to MIMAT.

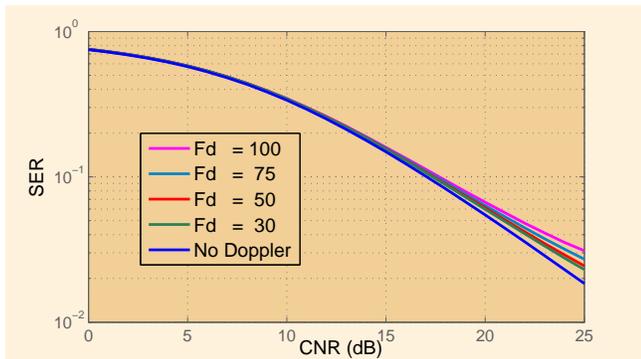

Fig. 40. SER (Symbol Error Rate) vs. CNR (Carrier to Noise Ratio) of MIMAT method for the Brazil channel with various Doppler frequencies.

This estimation technique is highly robust in rapidly changing channels and shows only minor performance degradation as the Doppler frequency increases as shown in Fig. 40.

## VIII. CONCLUSION

A unified view of sparse signal processing has been presented in tutorial form. The sparsity in the key areas of sampling, coding, spectral estimation, array processing, component analysis, and channel estimation has been carefully exploited. Some form of uniform or random sampling has been shown to underpin the associated sparse processing methods used in each of these fields. The reconstruction methods used in each application domain have been introduced and the interconnections among them have been highlighted.

This development has revealed; for example, that the iterative methods developed for random sampling can be applied to real-field block and convolutional channel coding for impulsive noise (salt-and-pepper noise in the case of images) removal, sparse component analysis, and channel estimation for orthogonal frequency division multiplexing systems. These iterative reconstruction methods have been shown to be naturally extendable to spectral estimation and sparse array processing due to their similarity to channel coding in terms of mathematical models. Conversely, the minimum description length method developed for spectral estimation and array processing has potential for application in other areas. The error locator polynomial method developed for channel coding has, moreover, been shown to be a discrete version of the annihilating filter used in sampling with a finite rate of innovation and the Prony method in spectral estimation; the Pisarenko and MUSIC methods are further improvements of the Prony method.

Linkages with emergent areas such as compressive sensing and sensor networks have also been considered. In addition, it has been suggested that the linear programming methods developed for compressive sensing and sparse component analysis can be applied to other applications with possible reduction of sampling rate. As such, this tutorial has provided the route for new applications of sparse signal processing to emerge, which can potentially reduce computational complexity and improve performance quality.

## APPENDIX I
## ACCELERATION METHODS: CHEBYSHEV AND CONJUGATE GRADIENT (CG) [65]

### A. Chebyshev Algorithm

- Initialization:

$$
\begin{aligned}
x_0[i] &= 0 \\
x_1[i] &= \frac{2}{A+B}\mathcal{PS}\{x[i]\} \\
\rho &= \frac{B-A}{B+A} \\
\lambda_1 &= 2
\end{aligned}
\tag{89}
$$

- For $n = 2, \ldots, N$:

$$
\begin{aligned}
\lambda_n &= \left(1 - \frac{\rho^2}{4}\lambda_{n-1}\right)^{-1} \\
x_{n+1}[i] &= x_{n-1}[i] + \lambda_n\Big(x_n[i] - x_{n-1}[i] \\
&\quad + \frac{2}{A+B}\mathcal{PS}\{x[i] - x_n[i]\}\Big)
\end{aligned}
\tag{90}
$$

where $\mathcal{S}$ and $\mathcal{P}$ are the sampling and filtering operators, respectively. Also, $A$ and $B$ are frame bound parameters and $N$ is the number of iterations.

### B. Conjugate Gradient Algorithm

- Initialization:

$$
\begin{aligned}
x_0[i] &= 0 \\
r_0[i] &= p_0[i] = \mathcal{PS}\{x[i]\} \\
\lambda_1 &= 2
\end{aligned}
\tag{91}
$$

- For $n = 2, \ldots, N$:

$$
\begin{aligned}
\lambda_n &= \frac{\langle r_n[i], p_n[i]\rangle}{\langle p_n[i], \mathcal{PS}\{p_n[i]\}\rangle} \\
x_{n+1}[i] &= x_n[i] + \lambda_n p_n[i] \\
r_{n+1}[i] &= r_n[i] - \lambda_n \mathcal{PS}\{p_n[i]\} \\
\lambda'_n &= \frac{\langle r_{n+1}[i], \mathcal{PS}\{p_n[i]\}\rangle}{\langle p_n[i], \mathcal{PS}\{p_n[i]\}\rangle} \\
p_{n+1}[i] &= r_{n+1}[i] - \lambda'_n p_n[i]
\end{aligned}
\tag{92}
$$

where $\mathcal{S}$ and $\mathcal{P}$ are the sampling and filtering operators, respectively. $N$ is the number of iterations and $\langle x[i], y[i]\rangle$ denotes the inner product of the two functions $x[i]$ and $y[i]$.

## Appendix II
## ELP Decoding for Erasure Channels [53]

For lost samples, the polynomial locator for the erasure samples is

$$H(z_i) = \prod_{m=1}^{k} \left( z_i - e^{j\frac{2\pi \cdot i_m}{n}} \right) = \sum_{t=0}^{k} h_t \ z^{k-t}, \quad (93)$$

,

$$H(z_{i_m}) = 0, \quad m = 1, 2, \ldots, k \quad (94)$$

where $z_i = e^{j\frac{2\pi \cdot i}{n}}$. The polynomial coefficients $h_t$, $t = 0, \ldots, k$ can be found from the product in (93); it is easier to find $h_t$ by obtaining the inverse FFT of $H(z)$. By multiplying (94) by $e[i_m] \cdot \left( z_{i_m} \right)^r$ (where $r$ is an integer) and summing over $m$, we get

$$\sum_{t=0}^{k} h_t \cdot \sum_{m=1}^{k} \left[ e[i_m] \cdot (z_{i_m})^{k+r-t} \right] = 0 \quad (95)$$

Since the inner summation is the DFT of the missing samples $e[i_m]$, we get

$$\sum_{t=0}^{k} h_t \cdot E[k + r - t] = 0 \quad (96)$$

where $E[.]$ is the DFT of $e[i]$. The received samples, $d[i]$, can be thought of as the original over-sampled signal, $x[i]$, minus the missing signal $e[i_m]$. The error signal, $e[i]$, is the difference between the corrupted and the original over-sampled signal and hence is equal to the values of the missing samples for $i = i_m$ and is equal to zero otherwise. In the frequency domain, we have

$$E[j] = X[j] - D[j], \quad j = 1, \ldots, n \quad (97)$$

Since $X[j] = 0$ for $j \in \Theta$ (see the footnote on page 10), then

$$E[j] = -D[j], \quad j \in \Theta \quad (98)$$

The remaining values of $E[j]$ can be found from (96), by the following recursion:

$$E[r] = \frac{-1}{h_k} \sum_{t=1}^{k} h_{k-t} E[r + t] \quad (99)$$

where $r \notin \Theta$ and the index additions are in $mod(n)$.

## Appendix III
## ELP Decoding for Impulsive Noise Channels [32], [96]

For all integer values of $r$ such that $r \in \Theta$ and $r + k \in \Theta$, we obtain a system of $k$ equations with $k + 1$ unknowns ($h_t$ coefficients). These equations yield a unique solution for the polynomial with the additional condition that the first nonzero $h_t$ is equal to one. After finding the coefficients, we need to determine the roots of the polynomial (93). Since the roots of $H(z)$ are of the form $e^{j\frac{2\pi \cdot i_m}{n}}$, the inverse DFT (IDFT) of the $\{h_m\}_{m=0}^{k}$ can be used. Before performing IDFT, we have to pad $n - 1 - k$ zeros at the end of the $\{h_m\}_{m=0}^{k}$ sequence to obtain an $n$-point signal. We refer to the new signal (after IDFT) as $\{H_i\}_{i=0}^{n-1}$. Each zero in $\{H_i\}$ represents an error in $r[i]$ at the same location.

## Appendix IV
## Proofs for MDL formulas

Proof of (58): The eigenvectors of $\mathbf{R}_{ML}^{-1}$ are the same as the ones for $\hat{\mathbf{R}}$ and its eigenvalues are the reciprocals of the eigenvalues of $R_{ML}$ and we know the eigenvalues constitute an orthonormal set. Thus we have:

$$
\begin{aligned}
tr(\mathbf{R}_{ML}^{-1}\hat{\mathbf{R}}) &= tr\big((\sum_{i=1}^{k} \hat{\lambda}_i^{-1}\hat{\mathbf{v}}_i\hat{\mathbf{v}}_i^H + \hat{\sigma}_{ML}^{-2} \sum_{i=k+1}^{n} \hat{\mathbf{v}}_i\hat{\mathbf{v}}_i^H) \\
&\quad \cdot (\sum_{i=1}^{n} \hat{\lambda}_i\hat{\mathbf{v}}_i\hat{\mathbf{v}}_i^H)\big) \\
&= tr\big(\sum_{i=1}^{k} \hat{\mathbf{v}}_i\hat{\mathbf{v}}_i^H + \sum_{i=k+1}^{n} \frac{\hat{\lambda}_i}{\hat{\sigma}_{ML}^2}\hat{\mathbf{v}}_i\hat{\mathbf{v}}_i^H\big)
\end{aligned}
\quad (100)
$$

where $tr(.)$ represents the *trace* operator on matrices. We know that if both $\mathbf{AB}$ and $\mathbf{BA}$ are defined, $tr(\mathbf{AB}) = tr(\mathbf{BA})$. Thus:

$$tr\big(\hat{\mathbf{v}}_i\hat{\mathbf{v}}_i^H\big) = tr\big(\hat{\mathbf{v}}_i^H\hat{\mathbf{v}}_i\big) \quad (101)$$

and since $\hat{\mathbf{v}}_i^H\hat{\mathbf{v}}_i = 1$ for $i = 1, \ldots, n$, we have:

$$
\begin{aligned}
tr(\mathbf{R}_{ML}^{-1}\hat{\mathbf{R}}) &= \sum_{i=1}^{k} tr\big(\hat{\mathbf{v}}_i^H\hat{\mathbf{v}}_i\big) + \sum_{i=k+1}^{n} \frac{\hat{\lambda}_i}{\hat{\sigma}_{ML}^2} tr\big(\hat{\mathbf{v}}_i^H\hat{\mathbf{v}}_i\big) \\
&= k + \hat{\sigma}_{ML}^{-2} \sum_{i=k+1}^{n} \hat{\lambda}_i
\end{aligned}
\quad (102)
$$

We also have $\hat{\sigma}_{ML}^2 = \frac{1}{n-k} \sum_{i=k+1}^{n} \hat{\lambda}_i$, which results in:

$$tr(\mathbf{R}_{ML}^{-1}\hat{\mathbf{R}}) = k + n - k = n \quad (103)$$

Proof of (60): The term $\frac{\kappa}{2} \log m$ is the penalty function and we have:

$$
\begin{aligned}
-\log f(x; \mathbf{R}_{ML}) &= -\log \left( \frac{1}{|\pi \mathbf{R}_{ML}|^m} e^{-tr(\mathbf{R}_{ML}^{-1}\hat{\mathbf{R}})} \right) \\
&= m\log(\pi) + m\log(|\mathbf{R}_{ML}|) \\
&\quad + tr(\mathbf{R}_{ML}^{-1}\hat{\mathbf{R}}) \\
&= m\log|\mathbf{R}_{ML}| \\
&\quad + \underbrace{n + m\log(\pi)}_{C(m)}
\end{aligned}
\quad (104)
$$

The first term of the above equation can be written as:

$$
\begin{aligned}
m\log|\mathbf{R}_{ML}| &= m\log\big((\prod_{i=1}^{k} \hat{\lambda}_i)((\hat{\sigma}_{ML}^2)^{n-k})\big) \\
&= m\sum_{i=1}^{k} \log(\hat{\lambda}_i) \\
&\quad + m(n-k)\log\left( \frac{\sum_{i=k+1}^{n} \hat{\lambda}_i}{n-k} \right)
\end{aligned}
\quad (105)
$$

therefore,

$$
\begin{aligned}
-\log f(x; \mathbf{R}_{ML}) &= m\sum_{i=1}^{n} \log(\hat{\lambda}_i) \\
&\quad + m(n-k)\log\left( \frac{1}{n-k} \sum_{i=k+1}^{n} \hat{\lambda}_i \right) \\
&\quad + C(m)
\end{aligned}
\quad (106)
$$

where $C(m)$ only depends on the parameter $m$ and not $k$. Thus we can ignore this term in the MDL criterion.

## Appendix V
## Sparse Matrix Manipulations [34], [35], [36]

Many physical phenomena and engineering problems are represented by a mathematical model very often in the form of ordinary, or partial differential equations. The explicit solutions of these equations are rarely available, unless for special cases. It is customary to try to find approximate solutions to these problems by making them discrete and linear. The resulting linear system involves very large matrices, which hopefully contain many zeros. A matrix, with very few non-zero elements is called a sparse matrix.

Solving large linear systems are difficult as it is very time consuming, costly and laborious, unless the matrices involved are sparse and have many zero elements. The sparse property of the matrix may reduce the storage and computing time considerably, crucial to solving the linear systems that represent such matrices. This is achieved by storing and computing only non-zero entries of the matrix. Various techniques can be applied to solve such a linear system [220].

There are two main approaches in finding admissible solutions for sparse linear systems, 1) Direct Methods [35], and 2) Iterative Methods [34], [221], [222]:

1) Direct Methods give the exact solutions in a finite number of elementary operations, provided that there are no rounding errors. Direct methods for a sparse linear system fall into three categories: (i) Gaussian elimination techniques, (ii) Triangular factorization, in particular using decomposition techniques, such as, LU, Incomplete Lower and Upper (ILU), and Cholesky factorization, (iii) Householder reduction to upper triangular form. Direct methods, in real applications, have an advantage of providing solutions with robust and predictable behavior. Detailed analysis and discussion of these methods have been well established, and can be found in [35], [36], and references therein.

2) Iterative Methods consist of Jacobi, Gauss-Siedel, and Successive Over-Relaxation (SOR). The most suitable method for the sparse linear system is the projection method, and in particular the Krylov subspace method. With the new development in iterative methods for approximate solution of linear systems, it has been realized that new iterative techniques of projection methods, and in particular the combination of preconditioning and projection onto Krylov subspace iterations, can be a simple and efficient solver of sparse linear systems, and can compete with the direct methods in its applications [34], [221].

Each of the above two methods is suitable for different classes of matrices. Direct methods are best applied to those classes that produce only a few fill-ins - a fill-in being a new non-zero entry created at a position where the original matrix contains a zero entry after a linear algebraic operation is performed. Preconditioned iterative methods are the preferred choice for the class of matrices that produce many fill-ins in the process.

There are two types of sparse matrices: matrices with regularly patterned non-zeros (group I), and matrices with irregularly structured matrices (group II). Such a distinction between matrices is particularly important in the iterative solution methods since algebraic operations for the group I of matrices can be significantly reduced using a computer. There is a large number of different sparse matrices archived and accessible on the website managed by T. Davis, The University of Florida Sparse Matrix Collection, http://www.cise.ufl.edu/research/sparse/matrices.

### A. Solution of Linear Systems

Consider the linear system

$$\mathbf{A}\mathbf{x} = \mathbf{b} \tag{107}$$

where $\mathbf{A}$ is an $m \times n$ matrix, $\mathbf{x}$ is the unknown, and $\mathbf{b}$ is an $m \times 1$ vector. For the exact solution of this system, there are three cases:

1) The square matrix $\mathbf{A}$ ($m = n$) is invertible (non-singular): there is a unique solution $\mathbf{x} = \mathbf{A}^{-1}\mathbf{b}$.
2) The matrix $\mathbf{A}$ is singular and $\mathbf{b} \in \mathcal{R}(\mathbf{A})$ ($\mathbf{b}$ belongs to the *Range* of $\mathbf{A}$): there are infinitely many solutions $\mathbf{x} = \mathbf{x}_0 + \mathbf{v}$, for all $\mathbf{v} \in Ker(\mathbf{A})$ (*Null* space of $\mathbf{A}$), where $\mathbf{x}_0$ is a particular solution of $\mathbf{A}\mathbf{x} = \mathbf{b}$.
3) The matrix $\mathbf{A}$ is singular and $\mathbf{b} \notin \mathcal{R}(\mathbf{A})$: there is no solution.

However, for large $n$, calculating the inverse of $\mathbf{A}$ is a complex task. Further, if approximate solutions are found, it may be difficult to estimate how accurate they are. This will depend on the entries of matrix $\mathbf{A}$. Further, for the case 2 above, where $n \leq m$ and $rank(\mathbf{A}) = n$, denote the pseudo inverse by $\mathbf{A}^+ = (\mathbf{A}^T\mathbf{A})^{-1}\mathbf{A}^T$, the problem is converted to finding the vector $\mathbf{x} = \mathbf{A}^+\mathbf{b}$ (an approximate solution) subject to the following conditions:

- $\mathbf{x}$ satisfies the least square solution. Namely, vector $\mathbf{x}$ minimizes the norm of $\|\mathbf{r}\|$, where the residual vector $\mathbf{r} = \mathbf{b} - \mathbf{A}\mathbf{x}$.
- $\mathbf{x}$ is the solution of the system $\mathbf{A}\mathbf{x} = \mathbf{b} - \mathbf{r}$, when $\mathbf{A}^T\mathbf{r} = 0$.

Note that in the case that $\mathbf{A}$ is the product of certain special matrices (diagonal, orthogonal, triangular, and other invertible matrices), the solution can be found by various direct methods [34], [35], [36].

### B. Iterative Methods

Consider an $n \times n$ real coefficient matrix $\mathbf{A}$ and a real $n$-vector $\mathbf{b}$ in linear system (107), the decomposition of $\mathbf{A}$ as:

$$\mathbf{A} = \mathbf{L} + \mathbf{D} + \mathbf{U} \tag{108}$$

where $\mathbf{D}$ is the diagonal matrix of $\mathbf{A}$ with all non-zero entries, and $\mathbf{L}$ and $\mathbf{U}$ are the strict lower and upper matrices of $\mathbf{A}$, respectively. The iterative solution vector $\mathbf{x}_{k+1}$ is given by

$$
\begin{aligned}
\mathbf{x}_{k+1} &= -(\mathbf{D} + \mathbf{L})^{-1}\left[\mathbf{U}\mathbf{x}_k - \mathbf{b}\right] \quad \text{(Gauss-Seidel)} \\
\mathbf{x}_{k+1} &= -\mathbf{D}^{-1}\left[(\mathbf{L} + \mathbf{U})\mathbf{x}_k - \mathbf{b}\right] \quad \text{(Jacobi)}
\end{aligned} \tag{109}
$$

or in a general form,

$$\mathbf{x}_{k+1} = \mathbf{G}\mathbf{x}_k + \mathbf{f} \tag{110}$$

where in Jacobi iterations $\mathbf{G} = \mathbf{G}_J(\mathbf{A}) = \mathbf{I} - \mathbf{D}^{-1}\mathbf{A}$, $\mathbf{f} = \mathbf{D}^{-1}\mathbf{b}$, and in Guass-Seidel iteration $\mathbf{G} = \mathbf{G}_{GS}(\mathbf{A}) = \mathbf{I} - (\mathbf{D} + \mathbf{L})^{-1}\mathbf{A}$, $\mathbf{f} = (\mathbf{D} + \mathbf{L})^{-1}\mathbf{b}$. This iteration can be viewed as a technique for solving the linear system:

$$\begin{aligned}(\mathbf{I} - \mathbf{G})\mathbf{x} &= \mathbf{f}, \\ \text{or } \mathbf{M}^{-1}\mathbf{A}\mathbf{x} &= \mathbf{M}^{-1}\mathbf{b}\end{aligned} \tag{111}$$

where the precondition matrix $\mathbf{M}$ is given by $\mathbf{M}_J = \mathbf{D}$, and $\mathbf{M}_{GS} = \mathbf{D} + \mathbf{L}$. The iterations given above converge and the limit is a solution of the original linear system (for details see [36], [221], [222]).

*1) Krylov Subspace Methods:* Let $\mathbf{x}_0$ be an initial approximation to the solution of (107), $\mathbf{r}_0 = \mathbf{b} - \mathbf{A}\mathbf{x}_0$ be the initial residual, and let $\mathbb{K}_m(\mathbf{A}, \mathbf{r}_0)$ be the Krylov subspace of dimension $m$ defined by

$$\mathbb{K}_m(\mathbf{A}, \mathbf{r}_0) = Span\{\mathbf{r}_0, \mathbf{A}\mathbf{r}_0, \dots, \mathbf{A}^{m-1}\mathbf{r}_0\} \tag{112}$$

where these subspaces are nested, i.e., $\mathbb{K}_m \subseteq \mathbb{K}_{m+1}$, ($m = 1, 2, 3, \dots$). Krylov subspace methods are iterative methods in which $\mathbf{x}_m$, an approximation to the solution of (107), at the $m^{th}$ step, is found in $\mathbf{x}_0 + \mathbb{K}_m$, namely the approximation is of the form

$$\mathbf{x} = \mathbf{x}_0 + q_{m-1}(\mathbf{A})\mathbf{r}_0 \tag{113}$$

where, $q_{m-1}$ is a polynomial of degree at most $m - 1$. If the system is real, then coefficients of $q_{m-1}$ are real. This natural expression implies that the residual $\mathbf{r}_m = \mathbf{b} - \mathbf{A}\mathbf{x}_m$ is associated with the so-called residual polynomial $p_m$ of degree at most $m$ with $p_m(0) = 1$ because

$$\begin{aligned}\mathbf{r}_m = \mathbf{b} - \mathbf{A}\mathbf{x}_m &= \mathbf{r}_0 - \mathbf{A}p_m(\mathbf{A})\mathbf{r}_0 \\ &= p_m(\mathbf{A})\mathbf{r}_0\end{aligned} \tag{114}$$

The error satisfies $\mathbf{e} = \mathbf{x}_m - \mathbf{x}_* = p_m(\mathbf{A})(\mathbf{x}_0 - \mathbf{x}_*)$, where $\mathbf{x}_*$ is the exact solution of (107). Let us denote by $\mathcal{P}_m$ the set of all polynomials $p$ of degree at most $m$ such that $p(0) = 1$. The approximation $\mathbf{x}_m \in \mathbf{x}_0 + \mathbb{K}_m$ is often found by requiring $\mathbf{x}_m$ to be the minimizer of some functional. There are different methods depending on the characteristics of the matrix and implementation. Thus, each method defines implicitly a different polynomial $p_m \in \mathcal{P}_m$ (for details see [223]).

The extra condition imposed for convergence and completeness are

- $\mathbf{r}_{rm}$ is orthogonal to $\mathbb{K}_m$ (Galerkin condition: $\mathbf{r}_m \perp \mathbb{K}_m$),
- Minimum residual condition:

$$\mathbf{r}_m = \min_{\mathbf{x} \in \mathbf{x}_0 + \mathbb{K}_m} \|\mathbf{b} - \mathbf{A}\mathbf{x}\| \tag{115}$$

We note that the nested property of the Krylov subspaces imply that any method for which one of the conditions (115) holds will terminate in at most $n$ steps. The desired methods are those which produce a good approximation to the solution of (107) in many fewer than $n$ iterations. An important ingredient that makes Krylov subspace methods work is the use of preconditioners, a matrix or operator $\mathbf{M}$ used to convert equation (107) into (111). There is no one method which is recommended for all problems.

Some of the applications on Sparse matrices besides the one discussed in the main body of the paper are:

### C. Applications in Photonics and Electromagnetics

Numerical simulation for the development of photonics and electromagnetic CAD software packages has been the subject of intensive research in the past decade. The most widely used simulation method is the Finite-Difference Time-Domain (FDTD) for time-domain analysis of Maxwell's equations. The more recently introduced schemes would require the solution of large sparse linear systems at each unit of time [224]. A popular method for time-domain problems is the time-domain beam propagation method [225], which necessitates a multiplication of the input field vector with a very sparse matrix. This sparse multiplication is advantageous in that the method can become very efficient and highly parallel [226]. Another method for the time-domain simulation of Maxwell's equations is the Finite-Element Time-Domain (FETD) technique [224]. The FETD leads to an almost sparse linear system; the computational complexity of this method can be considerably reduced by inverting the sparse matrix [227].

### D. Applications in Genomic Signal Processing [228]

Micro-arrays (DNA and protein) are parallel biosensors capable of detecting a large number of different genomic particles simultaneously. DNA micro-arrays that use tens of thousands of probe spots detect multiple targets in a single experiment. This is a wasteful use of the sensing resources in comparative DNA micro-array experiments. Generally, only a fraction of the total number of genes with respect to a reference sample is differentially expressed, and, thus, a vast number of probe spots may not provide any useful information. An alternative design is the so-called compressed micro-arrays; this translates to significantly lower costs, simpler image acquisition and processing, and smaller amount of genomic material needed for experiments. To recover signals from compressed micro-array measurements, ideas from CS (Sec. II-B) can be employed. For sparse measurement matrices, sparse algorithms can be used to lower the computational complexity than the widely used linear-programming-based methods [92], [228].

## Acknowledgments

We would like to sincerely thank our colleagues for their specific contributions in various sections in this paper. Especially, Dr. M. Nouri Moghadam from Math. Dept. of George Washington University, who wrote Appendix V; Dr. H. Saeedi from University of Massachusetts who contributed to LDPC codes, and Dr. K. Mehrany from EE Dept. of Sharif University who contributed to section V-C and also M. Valiollahzadeh who edited and contributed to the SCA section. We are especially indebted to Prof. B. Sankur from Bogazici University in Turkey for his careful review and comments. The contribution of some of the sparse array examples by Dr. A. Austeng from University of Oslo is also appreciated. We are also thankful to the students of the Multimedia Lab and members of ACRI at Sharif University for their invaluable help and simulations. We are specifically indebted to V. Montazer Hodjat, S. Jafarzadeh, A. Salemi, M. Soltanalian, E. Azizi and H. Firuzi.

## REFERENCES

[1] P. J. S. G. Ferreira, "Mathematics for multimedia signal processing II: Discrete finite frames and signal reconstruction," *Signal Proc. for Multimedia, IOS press*, vol. 174, pp. 35–54, Dec. 1999.

[2] F. Marvasti, *Nonuniform Sampling: Theory and Practice*. Springer, formerly Kluwer Academic/Plenum Publishers, 2001.

[3] R. G. Baraniuk, "A lecture on compressive sensing," *IEEE Signal Proc. Magazine*, vol. 24, no. 4, pp. 118–121, July 2007.

[4] M. Vetterli, P. Marziliano, and T. Blu, "Sampling signals with finite rate of innovation," *IEEE Trans. on Signal Proc.*, vol. 50, no. 6, pp. 1417–1428, June 2002.

[5] S. Lin and D. J. Costello, *Error control coding*. Prentice-Hall, Englewood Cliffs, NJ, 1983.

[6] T. Richardson and R. Urbanke, *Modern Coding Theory*. Cambridge University Press, 2008.

[7] F. Marvasti, M. Hung, and M. R. Nakhai, "The application of walsh transform for forward error correction," *Proc. of IEEE Int. Conf. on Acoustics, Speech and Signal Proc., ICASSP'99*, vol. 5, pp. 2459–2462, 1999.

[8] S. L. Marple, *Digital Spectral Analysis*. Prentice-Hall, Englewood Cliffs, NJ, 1987.

[9] S. M. Kay and S. L. Marple, "Spectrum analysis-a modern perspective," *Proc. IEEE, Reprinted by IEEE Press, (Modern Spectrum Analysis II)*, vol. 69, no. 11, pp. 1380–1419, Nov. 1981.

[10] S. M. Kay, *Modern Spectral Estimation: Theory and Application*. Prentice-Hall, Englewood Cliffs, N.J., Jan. 1988.

[11] P. Stoica and R. L. Moses, *Introduction to Spectral Analysis*. Upper Saddle River, NJ: Prentice-Hall, 1997.

[12] P. Stoica and A. Nehorai, "Music, maximum likelihood, and Cramer-Rao bound," *IEEE Trans. on ASSP*, vol. 37, no. 5, pp. 720–741, May 1989.

[13] ——, "Performance study of conditional and unconditional direction-of-arrival estimation," *IEEE Trans. on ASSP*, vol. 38, no. 10, pp. 1783–1795, October 1990.

[14] S. Holm, A. Austeng, K. Iranpour, and J. F. Hopperstad, "Sparse sampling in array processing," in *Nonuniform Sampling: Theory and Practice*, F. Marvasti, Ed. Springer, formerly Kluwer Academic/Plenum Publishers, 2001, pp. 787–833.

[15] S. Aeron, M. Zhao, and V. Saligrama, "Fundamental tradeoffs between sparsity, sensing diversity and sensing capacity," *Asilomar Conf. on Signals, Systems and Computers, ACSSC '06, Pacific Grove, CA*, pp. 295–299, Oct.-Nov. 2006.

[16] P. Bofill and M. Zibulevsky, "Underdetermined blind source separation using sparse representations," *Signal Proc., Elsevier*, vol. 81, no. 11, pp. 2353–2362, Nov. 2001.

[17] M. A. Girolami and J. G. Taylor, *Self-Organising Neural Networks: Independent Component Analysis and Blind Source Separation*. Springer Verlag, London, 1999.

[18] P. Georgiev, F. Theis, and A. Cichocki, "Sparse component analysis and blind source separation of underdetermined mixtures," *IEEE Trans. on Neural Networks*, vol. 16, no. 4, pp. 992–996, July 2005.

[19] M. Aharon, M. Elad, and A. M. Bruckstein, "The k-svd: An algorithm for designing of overcomplete dictionaries for sparse representation," *IEEE Trans. on Signal Proc.*, vol. 54, no. 11, pp. 4311–4322, Nov. 2006.

[20] ——, "On the uniqueness of overcomplete dictionaries, and a practical way to retrieve them," *Linear Algebra and Applications*, vol. 416, no. 1, pp. 48–67, July 2006.

[21] R. Gribonval and M. Nielsen, "Sparse representations in unions of bases," *IEEE Trans. on Info. Theory*, vol. 49, no. 12, pp. 3320–3325, Dec. 2003.

[22] P. Fertl and G. Matz, "Efficient OFDM channel estimation in mobile environments based on irregular sampling," *in Proc. Asilomar Conf. Signals, Systems, Computers, Pacific Grove, CA, Okt*, pp. 1777–1781, Nov. 2006.

[23] O. Ureten and N. Serinken, "Decision directed iterative equalization of OFDM symbols using non-uniform interpolation," *Proc. IEEE VTC'06 (fall), Montreal, Canada.*

[24] M. Soltanolkotabi, A. Amini, and F. Marvasti, "OFDM channel estimation based on adaptive thresholding for sparse signal detection," *arXive 0901.3948*, Jan. 2009.

[25] J. L. Brown, "Sampling extentions for multiband signals," *IEEE Trans. on Acoust. Speech, Signal Proc.*, vol. 33, pp. 312–315, Feb. 1985.

[26] O. G. Guleryuz, "Nonlinear approximation based image recovery using adaptive sparse reconstructions and iterated denoising, parts I and II," *IEEE Trans. on Image Proc.*, vol. 15, no. 3, pp. 539–571, March 2006.

[27] H. Rauhut, "On the impossibility of uniform sparse reconstruction using greedy methods," *STSIP*, vol. 7, no. 2, pp. 197–215, May 2008.

[28] T. Blu, P. Dragotti, M. Vetterli, P. Marziliano, and L. Coulot, "Sparse sampling of signal innovations: Theory, algorithms, and performance bounds," *IEEE Signal Proc. Magazine*, vol. 25, no. 2, March 2008.

[29] F. Marvasti, "Guest editor's comments on special issue on nonuniform sampling," *STSIP*, vol. 7, no. 2, pp. 109–112, May 2008.

[30] D. Donoho, "Compressed sensing," *IEEE Trans. on Info. Theory*, vol. 52, no. 4, pp. 1289–1306, Apr. 2006.

[31] E. J. Candes, "Compressive sampling," *Proc. of the Int. Congress of Mathematics, Madrid, Spain*, vol. 3, pp. 1433–1452, 2006.

[32] S. Zahedpour, S. Feizi, A. Amini, and F. Marvasti, "Impulsive noise cancellation based on soft decision and recursion," *IEEE Trans. on Instrumentation and Measurement*, vol. 67, no. 12, Dec. 2008.

[33] S. Feizi, S. Zahedpour, M. Soltanolkotabi, A. Amini, and F. Marvasti, "Salt and pepper noise removal for images," *IEEE Proc. on Int. Conf. on Telecommunications ICT'08, St. Petersburg, Russia*, pp. 1–5, June 2008.

[34] Y. Saad, *Iterative Methods for Sparse Linear Systems*. SIAM publications, 2nd edition, 2003.

[35] T. A. Davis, *Direct Methods for Sparse Linear Systems (Fundamentals of Algorithms)*. SIAM Publications, 2006.

[36] J. S. Duff, A. M. Erisman, and J. K. Reid, *Direct Methods for Sparse matrices*. Oxford University Press, 1986.

[37] P. D. Grunwald, I. J. Myung, and M. A. Pitt, *Advances in Minimum Description Length: Theory and Applications*. MIT Press, April 2005.

[38] A. Kumar, P. Ishwar, and K. Ramchandran, "On distributed sampling of bandlimited and non-bandlimited sensor fields," *IEEE Int. Conf. on Acoustics, Speech and Signal Proc., ICASSP'04, Montreal, Canada*, vol. 3, pp. 925–928, May 2004.

[39] P. Fertl and G. Matz, "Multi-user channel estimation in OFDMA uplink systems based on irregular sampling and reduced pilot overhead," *Proc. IEEE Int. Conf. on Acoustics, Speech and Signal Proc. ICASSP'07, Hawaii*, vol. 3, pp. 297–300, April 2007.

[40] F. Marvasti, "Spectral analysis of random sampling and error free recovery by an iterative method," *Trans. of IECE of Japan*, vol. 69, no. 2, pp. 79–82, Feb. 1986.

[41] R. A. DeVore, "Deterministic constructions of compressed sensing matrices," *Journal of Complexity*, vol. 23, no. 4-6, pp. 918–925, Aug. 2007.

[42] F. Marvasti and A. K. Jain, "Zero crossings, bandwidth compression, and restoration of nonlinearly distorted band-limited signals," *Journal of Opt. Soc. of America*, vol. 3, no. 5, pp. 651–654, 1986.

[43] A. Aldroubi, C. Cabrelli, and U. Molter, "Optimal non-linear models for sparsity and sampling," *to be published in the Journal of Fourier Analysis and Applications, Special Issue on Compressed Sampling*, vol. 14, no. 6, pp. 48–67, July 2008.

[44] A. Amini and F. Marvasti, "Convergence analysis of an iterative method for the reconstruction of multi-band signals from their uniform and periodic nonuniform samples," *STSIP*, vol. 7, no. 2, pp. 109–112, May 2008.

[45] P. J. S. G. Ferreira, "Iterative and non-iterative recovery of missing samples for 1-D band-limited signals," in *Nonuniform Sampling: Theory and Practice*, F. Marvasti, Ed. Springer, formerly Kluwer Academic/Plenum Publishers, 2001, pp. 235–281.

[46] ——, "The stability of a procedure for the recovery of lost samples in band-limited signals," *IEEE Trans. on Signal Proc.*, vol. 40, no. 3, pp. 195–205, Dec. 1994.

[47] F. Marvasti, *A Unified Approach to Zero-crossings and Nonuniform Sampling of Single and Multi-dimensional Signals and Systems*. Nonuniform Publication, Oak Park, ILL, 1987.

[48] ——, *Nonuniform Sampling*, R. J. Marks, Ed. New York: Springer-Verlag, 1993.

[49] A. I. Zayed and P. L. Butzer, "Lagrange interpolation and sampling theorems," in *Nonuniform Sampling: Theory and Practice*, F. Marvasti,

Ed. Springer, formerly Kluwer Academic/Plenum Publishers, 2001, pp. 123–168.

[50] F. Marvasti, P. Clarkson, M. Dokic, U. Goenchanart, and C. Liu, "Reconstruction of speech signals with lost samples," *IEEE Trans. on Signal Proc.*, vol. 40, no. 12, pp. 2897–2903, Dec. 1992.

[51] M. Unser, "Sampling-50 years after Shannon," *Proc. of the IEEE*, vol. 88, no. 4, pp. 569–587, Apr. 2000.

[52] F. Marvasti, "Random topics in nonuniform sampling," in *Nonuniform Sampling: Theory and Practice*, F. Marvasti, Ed. Springer, formerly Kluwer Academic/Plenum Publishers, 2001, pp. 169–234.

[53] F. Marvasti, M. Hasan, M. Eckhart, and S. Talebi, "Efficient algorithms for burst error recovery using FFT and other transform kernels," *IEEE Trans. on Signal Proc.*, vol. 47, no. 4, April 1999.

[54] F. Marvasti, M. Analoui, and M. Gamshadzahi, "Recovery of signals from nonuniform samples using iterative methods," *IEEE Trans. on ASSP*, vol. 39, no. 4, pp. 872–878, April 1991.

[55] H. Feichtinger and K. Gröchenig, "Theory and practice of irregular sampling," in *Wavelets- Mathematics and Applications*, J. J. Benedetto and et. al., Eds. CRC Publications, 1994, pp. 305–363.

[56] P. J. S. G. Ferreira, "Noniterative and fast iterative methods for interpolation and extrapolation," *IEEE Trans. on Signal Proc*, vol. 42, no. 11, pp. 3278–3282, Nov. 1994.

[57] A. Aldroubi and K. Gröchenig, "Non-uniform sampling and reconstruction in shift-invariant spaces," *SIAM Review*, vol. 43, no. 4, pp. 585–620, 2001.

[58] A. Aldroubi, "Non-uniform weighted average sampling and exact reconstruction in shift-invariant and wavelet spaces," *Applied and Computational Harmonic Analysis*, vol. 13, no. 2, pp. 151–161, sept. 2002.

[59] A. Papoulis and C. Chamzas, "Detection of hidden periodicities by adaptive extrapolation," *IEEE Trans. on Acoustics, Speech, and Signal Proc.*, vol. 27, no. 5, pp. 492–500, Oct. 1979.

[60] C. Chamzas and W. Y. Xu, "An improved version of Papoulis-Gerchberg algorithm on band-limited extrapolation," *IEEE Trans. on Acoustics, Speech, and Signal Proc.*, vol. 32, no. 2, pp. 437–440, April 1984.

[61] P. J. S. G. Ferreira, "Interpolation and the discrete Papoulis-Gerchberg algorithm," *IEEE Trans. on Signal Proc.*, vol. 42, no. 10, Oct. 1994.

[62] K. Gröchenig and T. Strohmer, "Numerical and theoretical aspects of non-uniform sampling of band-limited images," in *Nonuniform Sampling: Theory and Practice*, F. Marvasti, Ed. Springer, formerly Kluwer Academic/Plenum Publishers, 2001, pp. 283–324.

[63] D. C. Youla, "Generalized image restoration by the method of alternating orthogonal projections," *IEEE Trans. Circuits Syst.*, vol. 25, no. 9, pp. 694–702, Sept. 1978.

[64] D. C. Youla and H. Webb, "Image restoration by the method of convex projections: Part 1 -theory," *IEEE Trans. Med. Imag.*, vol. 1, no. 2, pp. 81–94, Oct. 1982.

[65] K. Gröchenig, "Acceleration of the frame algorithm," *IEEE Trans. on Signal Proc.*, vol. 41, no. 12, pp. 3331–3340, Dec. 1993.

[66] A. Ali-Amini, M. Babaie-Zadeh, and C. Jutten, "A new approach for sparse decomposition and sparse source separation," *EUSIPCO2006, Florence*, Sept. 2006.

[67] F. Marvasti, "Applications to error correction codes," in *Nonuniform Sampling: Theory and Practice*, F. Marvasti, Ed. Springer, formerly Kluwer Academic/Plenum Publishers, 2001, pp. 689–738.

[68] Y. Tsaig and D. Donoho, "Extensions of compressed sensing," *Signal Proc.*, vol. 86, no. 3, pp. 549–571, March 2006.

[69] E. Candes and T. Tao, "Near-optimal signal recovery from random projections: Universal encoding strategies," *IEEE Trans. on Info. Theory*, vol. 52, no. 12, pp. 5406– 5425, Dec. 2006.

[70] A. J. Jerri, "The Shannon sampling theorem-its various extension and applications: a tutorial review," *Proc. of IEEE*, vol. 65, no. 11, pp. 1565–1596, April 1977.

[71] E. Candes and M. Wakin, "An introduction to compressive sampling," *IEEE Signal Proc. Magazine*, vol. 25, no. 2, pp. 21–30, March 2008.

[72] Y. Eldar, "Compressed sensing of analog signals," *Preprint*, 2008.

[73] O. Christensen, *An introduction to frames and Riesz basis*, ser. Applied and Numerical Harmonic Analysis. Birkhauser, 2003.

[74] E. Candes and J. Romberg, "Sparsity and incoherence in compressive sampling," *Inverse Problems*, vol. 23, pp. 969–985, April 2007.

[75] D. Donoho and X. Hou, "Uncertainty principle and ideal atomic decomposition," *IEEE Trans. on Info. Theory*, vol. 47, no. 7, pp. 2845–2862, Nov. 2001.

[76] R. G. Baraniuk, M. Davenport, R. DeVore, and M. B. Wakin, "A simple proof of the restricted isometry property for random matrices," *Preprint*, 2007.

[77] A. C. Gilbert, M. J. Strauss, J. A. Tropp, and R. Vershynin, "Algorithmic linear dimension reduction in the $\ell_1$ norm for sparse vectors," preprint, 2006.

[78] D. L. Donoho, "For most large underdetermined systems of linear equations the minimal $\ell_1$-norm solution is also the sparsest solution," *Comm. Pure Appl. Math*, vol. 59, pp. 797–829, 2004.

[79] J. A. Tropp, "Recovery of short linear combinations via $\ell_1$ minimization," *IEEE Trans. on Info. Theory*, vol. 90, no. 4, pp. 1568–1570, July 2005.

[80] E. Candes and T. Tao, "Decoding by linear programming," *IEEE Trans. on Info. Theory*, vol. 51, pp. 4203–4215, Dec. 2005.

[81] J. A. Tropp, "Greed is good: Algorithmic results for sparse approximation," *IEEE Trans. on Info. Theory*, vol. 50, no. 10, pp. 2231–2242, Oct. 2004.

[82] J. Tropp and A. Gilbert, "Signal recovery from partial information via orthogonal matching pursuit," *IEEE Trans. on Info. Theory*, vol. 53, no. 12, pp. 4655–4666, Dec. 2007.

[83] E. Candes and J. Romberg, "Quantitative robust uncertainty principles and optimally sparse decompositions," *Foundations of Comput. Math.*, vol. 6, no. 2, pp. 227–254, Apr. 2006.

[84] E. Candes, J. Romberg, and T. Tao, "Robust uncertainty principles: Exact signal reconstruction from highly incomplete frequency information," *IEEE Trans. on Info. Theory*, vol. 52, no. 2, pp. 489–509, Feb. 2006.

[85] V. Saligrama, "Deterministic designs with deterministic gaurantees: Toepliz compressed sensing matrices, sequence design and system identification," *arXiv:0806.4958*, July 2008.

[86] R. Berinde, A. C. Gilbert, P. Indyk, H. Karloff, and M. J. Strauss, *Combining geometry and combinatorics: A unified approach to sparse signal recovery*. preprint 2008.

[87] M. F. Duarte, M. B. Wakin, and R. G. Baraniuk, "Fast reconstruction of piecewise smooth signals from random projections," *In Proc. SPARS05, Rennes, France*, vol. 1, pp. 1064–1070, Nov. 2005.

[88] A. C. Gilbert, Y. Kotidis, S. Muthukrishnan, and M. J. Strauss, "One-pass wavelet decompositions of data streams," *IEEE Trans. Knowl. Data Eng.*, vol. 15, no. 3, pp. 541–554, May/June 2003.

[89] A. C. Gilbert, M. J. Strauss, J. A. Tropp, and R. Vershynin, "One sketch for all: fast algorithms for compressed sensing," *In ACM STOC 2007*, pp. 237–246, 2007.

[90] S. Sarvotham, D. Baron, and R. G. Baraniuk, "Compressed sensing reconstruction via belief propagation," *Technical Report ECE-0601, ECE Dept., Rice University*, July. 2006.

[91] ——, "Sudocodes - fast measurement and reconstruction of sparse signals," *IEEE ISIT*, 2006.

[92] W. Xu and B. Hassibi, "Efficient compressive sensing with deterministic guarantees using expander graphs," *IEEE Info. Theory Workshop, ITW'07*, pp. 414–419, Sep. 2007.

[93] A. Aldroubi, H. Wang, and K. Zaringhalam, *Sequential compressed sampling via Huffman codes*. preprint 2008.

[94] P. L. Dragotti, M. Vetterli, and T. Blu, "Sampling moments and reconstructing signals of finite rate of innovation: Shannon meets strang-fix," *IEEE Trans. on Signal Proc.*, vol. 55, no. 5, pp. 1741–1757, May 2007.

[95] I. Maravic and M. Vetterli, "Sampling and reconstruction of signals with finite rate of innovation in the presence of noise," *IEEE Trans. on Signal Proc.*, vol. 53, no. 8, pp. 2788–2805, Aug. 2005.

[96] P. Azmi and F. Marvasti, "Robust decoding of DFT-based error-control codes for impulsive and additive white gaussian noise channels," *IEE Proc. Comm.*, vol. 152, no. 3, pp. 265–271, June 2005.

[97] F. Marvasti and M. Nafie, "Sampling theorem: A unified outlook on information theory, block and convolutional codes," *Special Issue on Info. Theory and its Applications, IEICE Trans. on Fundamentals of Electronics, Comm. and Computer Sciences, Section E*, vol. 76, no. 9, pp. 1383–1391, Sept. 1993.

[98] J. Wolf, "Redundancy, the discrete Fourier transform, and impulse noise cancellation," *IEEE Trans. on Comm.*, vol. 31, no. 3, pp. 458–461, Mar 1983.

[99] R. E. Blahut, "Transform techniques for error control codes," *IBM Journal of Research and Development*, vol. 23, no. 3, pp. 299–315, May 1979.

[100] T. G. M. Jr., "Coding of real-number sequences for error correction: A digital signal processing problem," *IEEE Trans. on Selected Areas in Comm.*, vol. 2, no. 2, March 1984.

[101] C. Berrou, A. Glavieux, and P. Thitimajshima, "Near Shannon limit error- correcting coding and decoding: Turbo codes," *Geneva, May 1993*, vol. 2, pp. 1064–1070, Aug. 1993.

[102] C. N. Hadjicostis and G. C. Verghese, "Coding approaches to fault tolerance in linear dynamic systems," *IEEE Trans. on Info. Theory*, vol. 51, no. 1, pp. 210–228, Jan. 2005.

[103] C. J. Anfinson and F. T. Luk, "A linear algebraic model of algorithm-based fault tolerance," *IEEE Trans. on Comput.*, vol. 37, no. 12, pp. 1599–1604, Dec. 1988.

[104] V. S. S. Nair and J. A. Abraham, "Real-number codes for fault-tolerant matrix operations on processor arrays," *IEEE Trans. on Computer*, vol. 39, no. 4, pp. 426–435, Apr. 1990.

[105] A. L. N. Reddy and P. Banerjee, "Algorithm-based fault detection for signal processing applications," *IEEE Trans. on Computer*, vol. 39, no. 10, pp. 1304–1308, Oct. 1990.

[106] J. M. N. Vieira and P. J. S. G. Ferreira, "Interpolation, spectrum analysis, error-control coding, and fault tolerant computing," *In Proc. of ICASSP'97, Munich, Germany*, vol. 3, pp. 1831–1834, Apr. 1997.

[107] F. Marvasti, "Error concealment of speech, image and video signals," *United Stated Patents 6,601,206*, July 2003.

[108] M. Nafie and F. Marvasti, "Implementation of recovery of speech with missing samples on a DSP chip," *Electronics Letters*, vol. 30, no. 1, pp. 12–13, Jan. 1994.

[109] A. Momenai and S. Talebi, "Improving the stability of the DFT error recovery codes by using the vandermonde fast decoding," *Proc. of ICASSP'06*, 2006.

[110] ——, "Improving the stability of DFT error recovery codes by using sparse oversampling patterns," *2007 - Elsevier*, vol. 87, no. 6, pp. 1448–1461, June 2007.

[111] C. Wong, F. Marvasti, and W. Chambers, "Implementation of recovery of speech with impulsive noise on a DSP chip," *Electronics Letters*, vol. 31, no. 17, pp. 1412–1413, Aug. 1995.

[112] D. Mandelbaum, "On decoding of Reed-Solomon codes," *IEEE Trans. on Info. Theory*, vol. 17, no. 6, pp. 707–712, Nov. 1971.

[113] R. G. Gallager, "Low-density parity-check codes," *IRE Trans. on Info. Theory*, vol. 8, pp. 21–28, Jan. 1962.

[114] D. J. C. MacKay and R. M. Neal, "Near Shannon-limit performance of low-density parity-check codes," *Electronics Letters*, vol. 32, pp. 1645–1646, Aug. 1996.

[115] J. Hagenauer, E. Offer, and L. Papke, "Iterative decoding of binary block and convolutional codes," *IEEE Trans. on Info. Theory*, vol. 42, pp. 429–445, March 1996.

[116] T. J. Richardson, M. A. Shokrollahi, and R. L. Urbanke, "Design of capacity approaching irregular low-density parity-check codes," *IEEE Trans. on Info. Theory*, vol. 47, no. 2, pp. 619–637, Feb. 2001.

[117] R. M. Tanner, "A recursive approach to low complexity codes," *IEEE Trans. on Info. Theory*, vol. 27, no. 5, pp. 533–547, Sept. 1981.

[118] J. Feldman, "Decoding error-correcting codes via linear programming," Ph.D. dissertation, Massachusetts Institute of Technology, Cambridge, MA, 2003.

[119] J. Feldman, M. J. Wainwright, and D. R. Karger, "Using linear programming to decode binary linear codes," *IEEE Trans. on Info. Theory*, vol. 51, no. 3, pp. 954–972, March 2005.

[120] P. O. Vontobel and R. Koetter, "On low-complexity linear-programming decoding of LDPC codes," *European Trans. on Telecommunications*, vol. 18, no. 5, pp. 509–517, 2007.

[121] S. K. Chilappagari and M. Chertkov, "Provably efficient instanton search algorithm for LP decoding of LDPC codes over the BSC," *submitted to IEEE Trans. on Info. Theory, arXiv:0808.2515v2*, Sept. 2008.

[122] B. G. R. de Prony, "Essai éxperimental et analytique: sur les lois de la dilatabilité de fluides élastique et sur celles de la force expansive de la vapeur de lalkool, á différentes températures," *Journal de lÉcole Polytechnique*, vol. 1, pp. 24–76, 1795.

[123] J. J. Fuchs, "Extension of the Pisarenko method to sparse linear arrays," *IEEE Trans. on Signal Proc.*, vol. 45, no. 10, pp. 2413–2421, Oct. 1997.

[124] R. Schmidt, "Multiple emitter location and signal parameter estimation," *IEEE Trans. on Antennas and Propagation*, vol. 34, no. 3, pp. 276–280, March 1986.

[125] H. Krim and M. Viberg, "Two decades of array signal processing research: the parametric approach," *IEEE Signal Proc. Magazine*, vol. 13, no. 4, pp. 67–94, July 1996.

[126] B. D. V. Veen and K. M. Buckley, "Beamforming: a versatile approach to spatial filtering," *IEEE ASSP Magazine*, vol. 5, no. 2, pp. 4–24, Apr. 1988.

[127] S. Valaee and P. Kabal, "An information theoretic approach to source enumeration in array signal processing," *IEEE Trans. on Signal Proc.*, vol. 52, no. 5, pp. 1171–1178, May 2004.

[128] R. Roy and T. Kailath, "Esprit-estimation of signal parameters via rotational invariance techniques," *IEEE Trans. on ASSP*, vol. 37, no. 7, pp. 984–995, July 1989.

[129] M. Wax and T. Kailath, "Detection of signals by information theoretic criteria," *IEEE Trans. on ASSP*, vol. 33, no. 2, pp. 387–392, April 1985.

[130] I. Ziskind and M. Wax, "Maximum likelihood localization of multiple sources by alternating projection," *IEEE Trans. on ASSP*, vol. 36, no. 10, pp. 1553–1560, October 1988.

[131] M. Viberg and B. Ottersten, "Sensor array processing based on subspace fitting," *IEEE Trans. on Signal Proc.*, vol. 39, no. 5, pp. 1110–1121, May 1991.

[132] S. Shahbazpanahi, S. Valaee, and A. B. Gershman, "A covariance fitting approach to parametric localization of multiple incoherently distributed sources," *IEEE Trans. on Signal Proc.*, vol. 52, no. 3, pp. 592–600, March 2004.

[133] J. Rissanen, "Modeling by shortest data description," *Automatica*, vol. 14, pp. 465–471, 1978.

[134] H. Akaike, "A new look on the statistical model identification," *IEEE Trans. on Automatic Control*, vol. 19, no. 6, pp. 716–723, Dec. 1974.

[135] M. Kaveh, H. Wang, and H. Hung, "On the theoretical performance of a class of estimators of the number of narrow-band sources," *IEEE Trans. on ASSP*, vol. 35, no. 9, pp. 1350–1352, Sept. 1987.

[136] Q. T. Zhang, K. M. Wong, P. C. Yip, and J. P. Reilly, "Statistical analysis of the performance of information theoretic criteria in the detection of the number of signals in array processing," *IEEE Trans. on ASSP*, vol. 37, no. 10, pp. 1557–1567, Oct. 1989.

[137] H. V. Poor, *An Introduction to Signal Detection Estimation.* Springer, 1994.

[138] T. M. Cover and J. A. Thomas, *Elements of Information Theory, 2nd Ed.* John Wiley & Sons, 2006.

[139] J. Rissanen, "A universal prior for integers and estimation by minimum description length," *Annals of Statistics*, vol. 11, no. 2, pp. 416–431, June 1983.

[140] T. W. Anderson and T. Wilbur, *An Introduction to Multivariate Statistical Analysis, 2nd Ed.* Wiley, 1958.

[141] F. Haddadi, M. R. M. Mohammadi, M. M. Nayebi, and M. R. Aref, "Statistical performance analysis of detection of signals by information theoretic criteria," *submitted to IEEE Trans. on Signal Proc.*, 2008.

[142] R. M. Leahy and B. D. Jeffs, "On the design of maximally sparse beamforming arrays," *IEEE Trans. on Antennas and Propagation*, vol. 39, no. 8, pp. 1178–1187, Aug. 1991.

[143] A. Austeng, J. E. Kirkebo, and S. Holm, "A flexible algorithm for layout-optimized sparse cmut arrays," *Proc. IEEE Ultrasonics Symp., (Montreal, Canada)*, vol. 2, pp. 1266–1269, Aug. 2004.

[144] S. Holm, A. Austeng, and J. E. Kirkebo, "Experience with sparse arrays at the university of Oslo," *in Proc. 29th ESA Antenna Workshop on Multiple Beams and Reconfigurable Antennas, Noordwijk, Netherlands*, pp. 215–218, Apr. 2007.

[145] H. S. r Jacobsen and K. Madsen, "Synthesis of nonuniformly spaced arrays using a general nonlinear minimax optimization method," *IEEE Trans. on Antennas and Propagation*, vol. 24, no. 4, pp. 501–506, July 1976.

[146] S. Holm, B. Elgetun, and G. Dahl, "Properties of the beampattern of weight- and layout-optimized sparse arrays," *IEEE Trans. on Ultrasonics, Ferroelectrics and Frequency Control*, vol. 44, no. 5, pp. 983–991, Sept. 1997.

[147] W. J. Hendricks, "The totally random versus the bin approach for random arrays," *IEEE Trans. Antennas and Propagation*, vol. 39, no. 12, pp. 1757–1762, Dec. 1991.

[148] G. R. Lockwood and F. S. Foster, "Optimizing the radiation pattern of sparse periodic two-dimensional arrays," *IEEE Trans. on Ultrasonic, Ferroelectrics and Frequency Control*, vol. 43, pp. 15–19, Jan. 1996.

[149] A. Austeng and S. Holm, "Sparse 2-D arrays for 3-D phased array imaging - design methods," *IEEE Trans. on Ultrasonic, Ferroelectrics and Frequency Control*, vol. 49, no. 8, pp. 1073–1086, Aug. 2002.

[150] ——, "Sparse 2-D arrays for 3-D phased array imaging - experimental validation," *IEEE Trans. on Ultrasonic, Ferroelectrics and Frequency Control*, vol. 49, no. 8, pp. 1087–1093, Aug. 2002.

[151] J. E. Kirkebo and A. Austeng, "Improved beamforming using curved sparse 2-D arrays in ultrasound," *IEEE Trans. Ultrasonic, Ferroelectrics and Frequency Control*, vol. 46, no. 2, pp. 119–128, May 2007.

[152] J. E. Kirkebo, A. Austeng, and S. Holm, "Layout-optimized cylindrical sonar arrays," *Proc. IEEE OCEANS, Kobe, Japan*, vol. 2, pp. 598–602, Nov. 2004.

[153] M. Gastpar and M. Vetterli, "Source-channel communication in sensor networks," *in Lecture Notes in Computer Science, Springer, New York, NY*, pp. 162–177, April 2003.

[154] A. M. Sayeed, "A statistical signal modeling framework for wireless sensor networks," *in Proc. 2nd Int. Workshop on Info. Proc. in Sensor Networks, IPSN'03, UW Tech. Rep. ECE-1-04*, pp. 162–177, Feb. 2004.

[155] K. Liu and A. M. Sayeed, "Optimal distributed detection strategies for wireless sensor networks," *in Proc. 42nd Annual Allerton Conf. on Comm., Control and Comp.*, Oct. 2004.

[156] A. D'Costa, V. Ramachandran, and A. Sayeed, "Distributed classification of gaussian space-time sources in wireless sensor networks," *IEEE Journal on Selected Areas in Comm.*, vol. 22, no. 6, pp. 1026–1036, Aug. 2004.

[157] W. U. Bajwa, J. Haupt, A. M. Sayeed, and R. Nowak, "Compressive wireless sensing," *in Proc. Int. Symposium on Info. Proc. in Sensor Networks, IPSN'06, Nashville, TN*, pp. 134–142, Apr. 2006.

[158] R. Rangarajan, R. Raich, and A. Hero, "Sequential design of experiments for a rayleigh inverse scattering problem," *in IEEE Workshop on Statistical Signal Proc., Bordeaux, France*, July 2005.

[159] Y. Yang and R. S. Blum, "Radar waveform design using minimum mean-square error and mutual information," *Fourth IEEE Workshop on Sensor Array and Multichannel Proc.*, vol. 12, no. 14, pp. 234–238, July 2006.

[160] A. Kumar, P. Ishwar, and K. Ramchandran, "On distributed sampling of smooth non-bandlimited fields," *Int. Simp. On Info. Proc. In Sensor Networks (ISPN2004)*, pp. 89–98, April 2004.

[161] E. Meijering, "A chronology of interpolation: From ancient astronomy to modern signal and image processing," *In Proc. of IEE*, vol. 90, no. 3, pp. 319–342, March 2002.

[162] D. Ganesan, S. Ratnasamy, H. Wang, and D. Estrin, "Coping with irregular spatio-temporal sampling in sensor networks," *ACM SIGCOMM Computer Communication Review*, vol. 34, no. 1, pp. 125–130, Jan. 2004.

[163] R. Wagner, R. Baraniuk, S. Du, D. Johnson, and A. Cohen, "An architecture for distributed wavelet analysis and processing in sensor networks," *in Proc. of Int. Workshop Info. Proc. in Sensor Networks, IPSN'06, Nashville, TN*, pp. 243–250, Apr. 2006.

[164] R. Wagner, H. Choi, R. Baraniuk, and V. Delouille, "Distributed wavelet transform for irregular sensor network grids," *In Proc. IEEE Stat. Signal Proc. Workshop (SSP)*, July 2005.

[165] S. S. Pradhan, J. Kusuma, and K. Ramchandran, "Distributed compression in a dense microsensor network," *IEEE Signal Proc. Magazine*, vol. 19, no. 2, pp. 51–60, March 2002.

[166] M. Gastpar and M. Vetterli, "Power, spatio-temporal bandwidth, and distortion in large sensor networks," *IEEE Journal on Selected Areas in Comm.*, vol. 23, no. 4, pp. 745–754, Apr. 2005.

[167] W. U. Bajwa, A. M. Sayeed, and R. Nowak, "Matched source-channel communication for field estimation in wireless sensor networks," *In Proc. 4th Int. Symposium on Info. Proc. in Sensor Networks, IPSN'05, Los Angeles, CA*, no. 44, pp. 332–339, Apr. 2005.

[168] R. Mudumbai, J. Hespanha, U. Madhow, and G. Barriac, "Scalable feedback control for distributed beamforming in sensor networks," *Proc. of the Int. Symposium on Info. Theory, ISIT'05*, Sept. 2005.

[169] J. Haupt and R. Nowak, "Signal reconstruction from noisy random projections," *IEEE Trans. on Info. Theory*, vol. 52, no. 9, pp. 4036–4068, Aug. 2006.

[170] D. Baron, M. B. Wakin, M. F. Duarte, S. Sarvotham, and R. G. Baraniuk, "Distributed compressed sensing," *submitted for publication, preprint, http://www.ece.rice.edu/drorb/pdf/DCS112005.pdf*, Nov. 2005.

[171] W. Wang, M. Garofalakis, and K. Ramchandran, "Distributed sparse random projections for refinable approximation," *in Proc. IPSN'07, Cambridge, MA*, pp. 331–339, Apr. 2007.

[172] M. F. Duarte, M. B. Wakin, D. Baron, and R. G. Baraniuk, "Universal distributed sensing via random projections," *in Proc. Int. Workshop on Info. Proc. in Sensor Networks, IPSN'06*, pp. 177–185, Apr. 2006.

[173] J. Haupt, W. U. Bajwa, M. Rabbat, and R. Nowak, "Compressed sensing for networked data," *IEEE Signal Proc. Magazine*, pp. 92–101, March 2008.

[174] M. Crovella and E. Kolaczyk, "Graph wavelets for spatial traffic analysis," *INFOCOM 2003. Twenty-Second Annual Joint Conf. of the IEEE Computer and Communications Societies. IEEE*, vol. 3, pp. 1848–1857, Mar. 2003.

[175] R. Coifman and M. Maggioni, "Diffusion wavelets," *Applied Computational and Harmonic Analysis*, vol. 21, no. 1, pp. 53–94, June 2006.

[176] W. U. Bajwa, J. Haupt, A. M. Sayeed, and R. Nowak, "Joint source-channel communication for distributed estimation in sensor networks,"

[177] *IEEE Trans. on Info. Theory*, vol. 53, no. 10, pp. 3629–3653, Oct. 2007.

[177] S. Boyd, A. Ghosh, B. Prabhakar, and D. Shah, "Randomized gossip algorithms," *IEEE Trans. on Info. Theory and IEEE/ACM Trans. on Networking*, vol. 52, no. 6, pp. 2508–2530, June 2006.

[178] M. Rabbat, J. Haupt, A. Singh, and R. Nowak, "Decentralized compression and predistribution via randomized gossiping," *in Proc. IPSN'06, Nashville, TN*, pp. 51–59, Apr. 2006.

[179] M. A. T. Figueiredo, R. D. Nowak, and S. J. Wright, "Gradient projection for sparse reconstruction: Application to compressed sensing and other inverse problems," *IEEE Journal on Selected Topics in Signal Proc.*, vol. 1, no. 4, pp. 586–597, Dec. 2007.

[180] S. J. Kim, K. Koh, M. Lustig, S. Boyd, and D. Gorinevsky, "A method for large-scale $\ell_1$-regularized least squares problems with applications in signal processing and statistics," *IEEE Journal on Selected Topics in Signal Proc.*, vol. 1, no. 4, pp. 606–617, Dec. 2007.

[181] Y. Rachlin, R. Negi, and P. Khosla, "Sensing capacity for discrete sensor network applications," *Proc. of the Fourth Int. Symposium on Info. Proc. in Sensor Networks*, pp. 126–132, Apr. 2005.

[182] S. Aeron, M. Zhao, and V. Saligrama, "Information theoretic bounds to sensing capacity of sensor networks under fixed SNR," *IEEE Info. Theory Workshop, ITW'07, Lake Tahoe, CA*, pp. 84–89, Sept. 2007.

[183] ——, "Fundamental limits on sensing capacity for sensor networks and compressed sensing," *submitted to IEEE Trans. on Info. Theory*, no. 2, Apr. 2008.

[184] S. Sanei and J. Chambers, *EEG Signal Processing*.  John Wiley.

[185] J. Herault and C. Jutten, "Space or time adaptive signal processing by neural network models," *Proc. of American Institute of Physics (AIP) Conf.: Neural Networks for Computing*, pp. 206–211, 1986.

[186] A. M. Aibinu, A. R. Najeeb, M. J. E. Salami, and A. A. Shafie, "Optimal model order selection for transient error autoregressive moving average (TERA) MRI reconstruction method," *Proc. World Academy of Science, Engineering and Technology*, vol. 32, pp. 191–195, Aug. 2008.

[187] M. S. Pedersen, J. Larsen, U. Kjems, and L. C. Parra, *A Survey of Convolutive Blind Source Separation Methods*, 2007.

[188] P. Comon, "Independent component analysis: A new concept," *Signal Proc.*, vol. 36, pp. 287–314, 1994.

[189] M. Zibulevsky and B. A. Pearlmutter, "Blind source separation by sparse decomposition in a signal dictionary," *Neural Comp.*, vol. 13, no. 4, 2001.

[190] ——, "Blind source separation by sparse decomposition," *Univ. New Mexico Tech. Rep. CS99-1*, July 1999.

[191] Y. Luo, J. A. Chambers, S. Lambotharan, and I. Proudler, "Exploitation of source non-stationarity in underdetermined blind source separation with advanced clustering techniques," *IEEE Trans. on Signal Proc.*, vol. 54, no. 6, 2006.

[192] Y. Li, S. Amari, A. Cichocki, D. W. C. Ho, and S. Xie, "Underdetermined blind source separation based on sparse representation," *IEEE Trans. on Signal Proc.*, vol. 54, no. 2, pp. 423–437, 2006.

[193] A. Jourjine, S. Rickard, and O. Yilmaz, "Blind separation of disjoint orthogonal signals: demixing n sources from 2 mixtures," *Proc. of IEEE Conf. on Acoustic, Speech, and Signal Proc. ICASSP'2000*, vol. 5, pp. 2985–2988, 2000.

[194] L. Vielva, D. Erdogmus, C. Pantaleon, I. Santamaria, J. Pereda, and J. C. Principe, "Underdetermined blind source separation in a time-varying environment," *Proc. of IEEE Conf. On Acoustic, Speech, and Signal Proc. ICASSP'02*, vol. 3, pp. 3049–3052, 2002.

[195] I. Takigawa and et. al., "On the minimum $\ell_1$-norm signal recovery in underdetermined source separation," *Proc. of 5th Int. Conf. on Independent Component Analysis*, pp. 22–24, 2004.

[196] C. C. Took, S. Sanei, and J. Chambers, "A filtering approach to underdetermined BSS with application to temporomandibular disorders," *in Proc IEEE ICASSP'06, France*, 2006.

[197] T. Melia and S. Rickard, "Underdetermined blind source separation in echoic environment using DESPRIT," *EURASIP Journal on Advances in Signal Proc., Article No. 86484*, 2007.

[198] K. Nazarpour, S. Sanei, L. Shoker, and J. A. Chambers, "Parallel space-time-frequency decomposition of EEG signals for brain computer interfacing," *Proc. of EUSIPCO 2006, Florence, Italy*, 2006.

[199] R. Gribonval and S. Lesage, "A survey of sparse component analysis for blind source separation: principles, perspectives, and new challenges," *in Proc. of ESANN 2006*, pp. 323–330, Apr. 2006.

[200] S. G. Mallat and Z. Zhang, "Matching pursuits with time-frequency dictionaries," *IEEE Trans. on Signal Proc.*, vol. 41, no. 12, pp. 3397–3415, Nov. 1993.

[201] S. S. Chen, D. L. Donoho, and M. A. Saunders, "Atomic decomposition by basis pursuit," *SIAM Journal on Scientific Computing*, vol. 20, pp. 33–61, 1998.

[202] E. Candes, "$\ell_1$-magic: Recovery of sparse signals," *http://www.acm.caltech.edu/l1magic/*. [Online]. Available: http://www.acm.caltech.edu/l1magic/

[203] I. F. Gorodnitsky and B. D. Rao, "Sparse signal reconstruction from limited data using FOCUSS, a re-weighted minimum norm algorithm," *IEEE Trans. on Signal Proc.*, vol. 45, no. 3, pp. 600–616, March 1997.

[204] G. H. Mohimani, M. Babaie-Zadeh, and C. Jutten, "Fast sparse representation using smoothed $\ell^0$-norm," *to appear in IEEE Trans. on Signal Proc.*

[205] R. Gribonval and P. Vandergheynst, "On the exponential convergence of matching pursuit in quasi-incoherent dictionaries," *IEEE Trans. on Info. Theory*, vol. 52, no. 1, pp. 255–261, Jan. 2006.

[206] D. L. Donoho, M. Elad, and V. Temlyakov, "Stable recovery of sparse overcomplete representations in the presence of noise," *IEEE Trans. on Info. Theory*, vol. 52, no. 1, pp. 6–18, Jan. 2006.

[207] M. S. Bazaraa, J. J. Jarvis, and H. D. Sherali, *Linear programming and network flows*. John Wiley & Sons, Inc. New York, NY, USA, 1990.

[208] J. J. Fuchs, "On sparse representations in arbitrary redundant bases," *IEEE Tran. on Info. Theory*, vol. 50, no. 6, pp. 1341–1344, June 2004.

[209] A. Aldroubi and K. Zaringhalam, *Non-linear Least square in $\mathbb{R}^N$*. Acta Applicanda Mathematicae, in press.

[210] J. J. V. de Beek, O. Edfors, M. Sandell, S. K. Wilson, and P. Borjesson, "On channel estimation in OFDM systems," *Proc. 45th IEEE Vehicular Technology Conf., Chicago*, vol. 2, pp. 815–819, Jul. 1995.

[211] M. Morelli and U. Mengali, "A comparison of pilot-aided channel estimation methods for OFDM systems," *IEEE Trans. on Signal Proc.*, vol. 49, no. 12, pp. 3065–3073, Dec. 2001.

[212] E. G. Larsson and et. al., "Joint symbol timing and channel estimation for OFDM based WLANs," *IEEE Comm. Letters*, vol. 5, no. 8, pp. 325–327, Aug. 2001.

[213] O. Edfors, M. Sandell, J.-J. V. de Beek, and S. K. Wilson, "Ofdm channel estimation by singular value decomposition," *IEEE Trans. on Comm.*, vol. 46, no. 7, pp. 931–939, July 1998.

[214] P. Strobach, "Low-rank adaptive filters," *IEEE Trans. on Signal Proc.*, vol. 44, no. 2, pp. 2932–2947, Dec. 1996.

[215] S. Coleri, M. Ergen, A. Puri, and A. Bahai, "Channel estimation techniques based on pilot arrangement in OFDM systems," *IEEE Trans. on Broadcasting*, vol. 48, no. 3, pp. 223–229, Sept. 2002.

[216] S. G. Kang, Y. M. Ha, and E. K. Joo, "A comparative investigation on channel estimation algorithms for OFDM in mobile communications," *IEEE Trans. on Broadcasting*, vol. 49, no. 2, pp. 142–149, June 2003.

[217] G. Tauböck and F. Hlawatsch, "A compressed sensing technique for OFDM channel estimation in mobile environments: exploiting channel sparsity for reducing pilots," *Proc. ICASSP'08, Las Vegas*, pp. 2885–2888, March. 2008.

[218] M. R. Raghavendra and K. Giridhar, "Improving channel estimation in OFDM systems for sparse multipath channels," *IEEE Signal Proc. Letters*, vol. 12, no. 1, pp. 52–55, Jan. 2005.

[219] T. Kang and R. A. Iltis, "Matching pursuits channel estimation for an underwater acoustic ofdm modem," *IEEE Int. Conf. on Acoustics, Speech and Signal Proc., ICASSP'08, Las Vegas*, pp. 5296–5299, March 2008.

[220] G. H. Golub and C. V. Loan, *Matrix Computations, 3rd edition*. The John Hopkins University Press, Baltimore, 1996.

[221] Z. Zelatev, *Computational Methods for General Sparse Matrices*, ser. Mathematics and its application series. Springer, 1991.

[222] D. J. Evans, *Sparsity and its Applications*. Cambridge University Press, 1985.

[223] V. Simoncini and D. B. Szyld, "Recent computational developments in Krylov subspace methods for linear systems," *Numer. Linear Algebra Appl.*, vol. 14, pp. 1–59, Apr. 2007.

[224] F. L. Teixeira, "FDTD/FETD methods: a review on some advances and selected applications," *Journal of Microwave and Optoelectronics*, vol. 6, no. 1, pp. 83–95, June 2007.

[225] H. M. Masoudi, M. A. AlSunaidi, and J. M. Arnold, "Time-domain finite-difference beam propagation method," *IEEE Photonic Tech. Lett.*, vol. 11, no. 10, pp. 1274–1276, Oct. 1999.

[226] H. M. Masoudi and J. M. Arnold, "Parallel three-dimensional finite-difference beam propagation methods," *Int. J. Num. Mod*, vol. 8, pp. 95–107, March 1995.

[227] B. He and F. L. Teixeira, "Sparse and explicit FETD via sparse approximation of hodge matrix," *IEEE Microwave and Wireless Components Lett.*, vol. 16, no. 6, pp. 348–356, April 2006.

[228] F. Parvaresh, H. Vikalo, H. Misra, and B. Hassibi, "Recovering sparse signals using sparse measurement matrices in compressed DNA microarrays," *IEEE J. Selec. Topics in Signal Proc.*, vol. 2, no. 3, pp. 275–285, June 2008.

TABLE XV

LIST OF ACRONYMS

| | | | |
|---|---|---|---|
| **ADSL**: | Asynchronous Digital Subscriber Line | **AIC**: | Akaike Information Criterion |
| **ARMA**: | Auto-Regressive Moving Average | **AR**: | Auto-Regressive |
| **BW**: | BandWidth | **BSS**: | Blind Source Separation |
| **CFAR**: | Constant False Alarm Rate | **CAD**: | Computer Aided Design |
| **CS**: | Compressed Sensing | **CG**: | Conjugate Gradient |
| **DAB**: | Digital Audio Broadcasting | **CT**: | Computer Tomography |
| **DCT**: | Discrete Cosine Transform | **DC**: | Direct Current: Zero-Frequency Coefficient |
| **DFT**: | Discrete Fourier Transform | **DHT**: | Discrete Hartley Transform |
| **DOA**: | Direction Of Arrival | **DST**: | Discrete Sine Transform |
| **DT**: | Discrete Transform | **DVB**: | Digital Video Broadcasting |
| **DWT**: | Discrete Wavelet Transform | **EEG**: | ElectroEncephaloGraphy |
| **ELP**: | Error Locator Polynomial | **ESPRIT**: | Estimation of Signal Parameters via Rotational Invariance Techniques |
| **FDTD**: | Finite-Difference Time-Domain | | |
| **FETD**: | Finite-Element Time-Domain | **FOCUSS**: | FOCal Under-determined System Solver |
| **FPE**: | Final Prediction Error | | |
| **GA**: | Genetic Algorithm | **HNQ**: | Hannan and Quinn method |
| **ICA**: | Independent Component Analysis | **IDE**: | Iterative Detection and Estimation |
| **IDT**: | Inverse Discrete Transform | **IMAT**: | Iterative Methods with Adaptive Thresholding |
| **KLT**: | Karhunen Loeve Transform | | |
| $l_1$: | Absolute Summable Discrete Signals | $l_2$: | Finite Energy Discrete Signals |
| **LDPC**: | Low Density Parity Check | **LP**: | Linear Programming |
| **MA**: | Moving Average | **MAP**: | Maximum A Posteriori probability |
| **MDL**: | Minimum Description Length | | |
| **MIMAT**: | Modified IMAT | **ML**: | Maximum Likelihood |
| **MMSE**: | Minimum Mean Squared Error | **MSL**: | Multi-Source Location |
| **MUSIC**: | MUltiple SIgnal Classification | **NP**: | Non-Polynomial time |
| **OCT**: | Optical Coherence Tomography | **OFDM**: | Orthogonal Frequency Division Multiplex |
| **OFDMA**: | Orthogonal Frequency Division Multiple Access | | |
| | | **OMP**: | Orthogonal Matching Pursuit |
| **OSR**: | Over Sampling Ratio | **PCA**: | Principle Component Analysis |
| **PDF**: | Probability Density Function | **PHD**: | Pisarenko Harmonic Decomposition |
| **POCS**: | Projection Onto Convex Sets | **PPM**: | Pulse-Position Modulation |
| **RDE**: | Recursive Detection and Estimation | **RIP**: | Restricted Isometry Property |
| **RS**: | Reed-Solomon | **RV**: | Residual Variance |
| **SA**: | Simulated Annealing | **SCA**: | Sparse Component Analysis |
| **SDCT**: | Sorted DCT | **SDFT**: | Sorted DFT |
| **SDR**: | Sparse Dictionary Representation | **SER**: | Symbol Error Rate |
| **SI**: | Shift Invariant | **SL0**: | Smoothed $\ell_0$-norm |
| **SNR**: | Signal-to-Noise Ratio | **ULA**: | Uniform Linear Array |
| **UWB**: | Ultra Wide Band | **WIMAX**: | Worldwide Inter-operability for Microwave Access |
| **WLAN**: | Wireless Local Area Network | | |
| **WMAN**: | Wireless Metropolitan Area Network | | |